\documentclass[11pt, a4paper]{article}
\usepackage[utf8]{inputenc}
\usepackage[english]{babel}
\usepackage[a4paper,top=2cm,bottom=2.5cm,left=2cm,right=2cm,marginparwidth=1.75cm]{geometry}
\usepackage{amsmath,amsfonts}
\usepackage{amssymb}
\usepackage{bbold}
\usepackage{caption,adjustbox}
\usepackage{subcaption}
\usepackage{graphicx}
\usepackage[colorlinks=true,linkcolor=Black,citecolor=Black,urlcolor=Blue]{hyperref}
\usepackage[usenames,dvipsnames]{xcolor}
\usepackage{physics, tensor, cancel}
\usepackage[title,toc,titletoc]{appendix}
\usepackage{tikz}
\usepackage[affil-sl]{authblk}

\usepackage[symbol]{footmisc}

\usepackage{comment}
\usepackage{csquotes}
\usepackage{hyperref}
\usepackage[normalem]{ulem}
\usepackage{fancyhdr}
\usepackage{academicons}
\usepackage[dvipsnames]{xcolor}
\usepackage{colortbl}
\usepackage{multicol}
\usepackage{pdflscape}
\usepackage{multirow}
\usepackage{booktabs}    
\usepackage{calc}        
\usepackage{placeins}
\usepackage{mathtools}
\usepackage{hhline}
\usepackage{makecell}

\usepackage{booktabs}
\usepackage{wrapfig}

\usepackage[style=numeric-comp,sorting=none,maxbibnames=9]{biblatex}
\usepackage{subcaption}

\usepackage{tikz-cd}
\usetikzlibrary{decorations.pathmorphing}
\usepackage{biblatex}
\usepackage{ytableau}
\usepackage{youngtab}
\usepackage{graphicx}
\usepackage{listings}
 \DeclareUnicodeCharacter{0301}{\'{e}}

\newcounter{countCode}
\lstnewenvironment{code} [1][caption=Ponme caption, label=default]{%
	 
	\setcounter{lstlisting}{\value{countCode}} 
	\lstset{ %
	language=java,
	basicstyle=\ttfamily\footnotesize,       
	numbers=left,                   
	numberstyle=\sc,      
	stepnumber=1,                   
	numbersep=5pt,                 
	numberstyle=\color{red!50!blue},
    	backgroundcolor=\color{lightgray!20},
	rulecolor=\color{blue},
	keywordstyle=\color{red}\bfseries,
	showspaces=false,               
	showstringspaces=false,         
	showtabs=false,                 
	frame=single,                   
	framexleftmargin=0mm,
	numberblanklines=false,
	xleftmargin=5pt,
	breaklines=true,
	breakatwhitespace=true,
	breakautoindent=true,
	captionpos=t,
	texcl=true,
	tabsize=2,                      
	extendedchars=true,
	inputencoding=utf8, 
	escapechar=\%,
	morekeywords={print, println, size, background, strokeWeight, fill, line, rect, ellipse, triangle, arc, save, PI, HALF_PI, QUARTER_PI, TAU, TWO_PI, width, height,},
	emph=[1]{print,println,}, emphstyle=[1]{\color{blue}}, 
	emph=[2]{width,height,}, emphstyle=[2]{\bf\color{violet}}, 
	emph=[3]{PI, HALF_PI, QUARTER_PI, TAU, TWO_PI}, emphstyle=[3]\color{orange!50!violet}, 
	emph=[4]{line, rect, ellipse, triangle, arc,}, emphstyle=[4]\color{green!70!black}, 
	#1}}{\addtocounter{countCode}{1}}

\DeclareMathOperator\arctanh{arctanh}

\definecolor{rossoCP3}{cmyk}{0,.88,.77,.40}
\title{\LARGE \bf  \textcolor{rossoCP3}{The Dilatonic Dynamics \\ of 
\\ Baryonic Crystals, Branes and Spheres} 

}
\author[1]{Jahmall Bersini\thanks{\href{mailto:jahmall.bersini@ipmu.jp}{jahmall.bersini@ipmu.jp}} }

\author[2,3,4]{Alessandra D'Alise  \thanks{\href{mailto:alessandra.dalise@unina.it}{alessandra.dalise@unina.it}}}
\author[2,3]{ Matías Torres\thanks{\href{mailto:matiasignacio.torressandoval@unina.it}{matiasignacio.torressandoval@unina.it}}}
\author[2,3,4]{Francesco Sannino\thanks{\href{mailto:sannino@qtc.sdu.dk}{sannino@qtc.sdu.dk}}}

\affil[1]{\small  Kavli IPMU (WPI), UTIAS, The University of Tokyo, Kashiwa, Chiba 277-8583, Japan}

\affil[2]{\small Dept. of Physics E. Pancini, Università di Napoli Federico II, via Cintia, 80126 Napoli, Italy}
\affil[3]{\small INFN sezione di Napoli, via Cintia, 80126 Napoli, Italy}
\affil[4]{\small Quantum  Theory Center ($\hbar$QTC) at IMADA and D-IAS, Southern Denmark Univ., Campusvej 55, 5230 Odense M, Denmark}
\date{}

\date{}
\begin{document}
 \maketitle

\begin{abstract}
We systematically analyze the impact of dilatonic dynamics on Skyrme spheres, crystals and branes. The effects of the dilatonic model parameters, encompassing different underlying near-conformal dynamics, on the macroscopic properties of Skyrmions such as their mass and radius, are discussed.  For spheres and crystals  we identify special values of the ratio of the decay constants for which the second order differential equations reduce to a solvable first order system. Additionally, in the case of the crystals, the dilaton presence spatially separates the baryon and isospin charge distributions.  For branes, we show how the dilaton smooths out their configurations. Our results are expected to have wide implications from the study of near-conformal dynamics stemming from QCD-like theories to phenomenological investigations of nuclear matter in extreme regimes.

\end{abstract}

\newpage

 \tableofcontents 

 \newpage
 
\section{Introduction}

Strong dynamics is notoriously hard to tackle both analytically and numerically. Over several decades, methodologies have been devised to access different dynamical regimes of the theory. At sufficiently low energies chiral Lagrangians have been shown to faithfully describe the dynamics of underlying specific strongly underlying theories in terms of their Goldstone bosons. This approach has landed precious information on the dynamics and spectrum of various strongly interacting models of which Quantum Chromodynamics (QCD) is the time-honored example.  The approach is also routinely employed to inform and extract information from first principle numerical simulations. Another remarkable property of chiral Lagrangians is that their non-perturbative solutions describe extended objects that, depending on the boundary conditions (BCs), can be identified with distinct physical configurations, from crystals to branes and spheres. The spherical case (the hedgehog solution) is the oldest example dating back to the pioneering work of Skyrme \cite{Skyrme:1961vq}. Once the topological sector of the theory is taken into account \cite{Adkins:1983ya, Adkins:1983hy} and the solitonic solutions properly quantized they describe half-integer states with topological charge identified as the baryon number of the underlying theory. These extended states are taken to be the nucleons of the underlying theory. Astonishingly, the same model can simultaneously describe mesonic degrees of freedom and their scattering properties as well as the spectrum and form factors of extended baryonic states including the Goldstone boson scattering off them. Therefore the same effective Lagrangian coefficients control both, the pion dynamics as well as the baryon spectrum and form factors. The operators of the chiral effective Lagrangian can be organized in the number of derivative and pion masses. The solitonic solutions can be mapped into baryon states of the underlying QCD-like dynamics at large number of colours. The dependence on the number of colours is naturally encoded in the pion decay constant and depends on the underlying fermion representation with respect to the asymptotically free gauge group. The Skyrme model has also been enriched by adding at the Lagrangian level massive vector mesons that have been shown to play an important role, not only to give a deeper understanding of the origin of the Skyrme term but also for the associated phenomenological consequences \cite{Adkins:1983nw, Zahed:1986qz,Jain:1987sz,Meissner:1987ge}. Lastly, a less explored avenue that still deserves much attention especially after the renewed experimental interest in heavy baryon spectroscopy and transitions \cite{Detmold:2015aaa,Bigi:2016mdz,HeavyFlavorAveragingGroup:2022wzx} is the study of heavy baryons as bound states of a Skyrme soliton and heavy mesons \cite{Callan:1985hy,Jaffe:1976yi,Scoccola:1988wa,Kaplan:1989fc,Kondo:1991fc,Weigel:1993zd,Eichten:1980mw,Guralnik:1992dj,Jenkins:1992ic,Oh:1994vd,Gupta:1993kd,Schechter:1995vr,Harada:1997we}. The original idea pioneered by Callan and Klebanov  \cite{Callan:1985hy} and first tailored for baryons made by one strange quark and two light quark could, however, account only for half of the physical states possible in QCD. The issue was resolved analytically in \cite{Harada:1997bk} by consistently introducing excited heavy mesons.  
When working with heavy quarks such as the charm and the bottom ones the approach hugely benefits from the marriage between large number of colours light dynamics and the celebrated heavy quark limit of QCD \cite{Voloshin:1986dir,Isgur:1989vq, Wise:1993wa,Neubert:1993mb,Luke:1992cs,Falk:1992ws,Falk:1990pz}.

Beyond the traditional phenomenological nuclear and particle physics applications  here we systematically investigate the solitonic dynamics when the underlying gauge-fermion theory is  near-conformal. The region in the number of flavours versus the number of colours, of asymptotically free gauge theories, where the infrared theory is conformal is   dubbed \emph{conformal window}. The  phase diagram structure for QCD and QCD-like theories appeared in \cite{Sannino:2004qp,Dietrich:2006cm}.    At the lower edge of the conformal window the theory undergoes a
quantum phase transition from an infrared CFT to a phase characterized by both conformal and chiral symmetry breaking \cite{Miransky:1984ef,Miransky:1996pd}. When the transition is sufficiently smooth, close to the lower end of the conformal window, the theory exhibits a near-conformal phase characterized by the existence of a region of the renormalization group (RG) flow in which the coupling remains nearly constant signaling the occurrence of \emph{walking dynamics} \cite{Holdom:1988gs,Holdom:1988gr,Cohen:1988sq}. Walking behavior lies at the core of many phenomenological models of dynamical electroweak symmetry breaking e.g. within the technicolor \cite{Weinberg:1975gm, Cohen:1988sq, Appelquist:1999dq, Duan:2000dy, Sannino:2004qp,Dietrich:2005jn,Dietrich:2006cm, Cacciapaglia:2020kgq} and (fundamental partial) composite Goldstone Higgs scenarios \cite{Kaplan:1983fs, Kaplan:1983sm, Cacciapaglia:2014uja, Gripaios:2009pe, Galloway:2010bp, Barnard:2013zea, Sannino:2016sfx, Orlando:2020yii, Bersini:2022bnx}. 

The infrared dynamics of the near-conformal theory can be modeled by augmenting the standard chiral Lagrangian via the introduction of a new light scalar degree of freedom with the same quantum numbers of the vacuum which is commonly referred to as the \emph{dilaton} or \emph{radion}. In this scenario, the dilaton is the Goldstone boson stemming from the spontaneous breaking of scale invariance whereas sources of explicit conformal breaking leading to the near-conformal phase can be encoded in the effective dilaton potential. The aforementioned phenomenological applications have motivated several investigations of the resulting dilaton effective field theory (EFT) \cite{Dietrich:2005jn,  Chacko:2012sy, Matsuzaki:2013eva, Kasai:2016ifi, Hansen:2016fri, Golterman:2016lsd, Golterman:2016cdd, Appelquist:2017wcg, Golterman:2018mfm,  Cata:2019edh, Golterman:2020tdq, Appelquist:2020bqj, Golterman:2021ohm, Appelquist:2022mjb, appelquist2020dilaton, Crewther:2013vea, Cata:2018wzl, Zwicky:2023bzk}.

In this work we present several classes of extended objects emerging as solitons of the dilaton augmented EFT. We identify crystal, brane, and sphere phases constituted respectively by ordered arrays of hadronic tubes, layers, and spherical hedgehog solitonic solutions at non-vanishing baryon number. 

Until very recently, it was considered hard to construct analytical solutions representing baryonic condensates in the low-energy limit of QCD. However, in the references \cite{Canfora:2023zmt,Alvarez:2017cjm,Ayon-Beato:2015eca,Canfora:2022jmh,Canfora:2018rdz,Barriga:2021eki,Barriga:2022izc,Canfora:2022jmh,Canfora:2023pkx} an exact method to build hadronic solitons has been introduced. These solutions disclose intriguing non-perturbative features of the finite density phase diagram. For these reasons, it is very interesting to analyze what happens when the dilatonic degree of freedom is taken into account.

Skyrmions under the effects of dilaton dynamics have already been employed to describe dense nuclear matter such as the core of neutron stars \cite{Lee:2003eg, Park:2008zg, Ma:2018xjw, Shao:2022njr, Paeng:2017qvp, Brown:2001nh, Holt:2014hma, Ma:2013ooa, Paeng:2015noa}. In this framework, conformal invariance is seen as an emergent hidden symmetry of real-world QCD at high densities. Starting from the pioneering work of G.E. Brown and M. Rho \cite{brown1991scaling}, who introduced the dilaton mode to derive a series of scaling relations among the values of decay constants and masses in vacuum and at finite density, the dilaton EFT is considered to be a key ingredient for studying Skyrmion matter in extreme conditions. These relations are shown to be modified when considering generalized EFTs  \cite{Harada:1996wb}. While earlier works considered a simplified version of the effective action, the construction has later been refined in \cite{Li:2016uzn} and since then the dilaton EFT has become one of the building blocks of the generalized nuclear effective field theory (GnEFT), which aims at the description of hadronic matter from low to compact star densities (see e.g. \cite{Ma:2023ugl, Ma:2021nuf} for recent reviews). However, to the best of our knowledge, no comprehensive analysis of the spherical Skyrmion properties in near-conformal theories has been performed.  

\vskip .1cm
\emph{The overarching goal of the present work is to combine numerical and analytical methods to perform a systematic investigation of Skyrmion solutions in the presence of dilatonic dynamics. Applications range from the dynamics of  QCD-like theories close to the lower end of the conformal window to nuclear matter in extreme regimes.}
\vskip .1cm 

We structure our work as follows.  After introducing the dilaton augmented EFT in Sec.\ref{dresscode}, we present a systematic numerical analysis in Sec.\ref{dahs}.  More specifically, we employ the standard hedgehog ansatz \cite{Adkins:1983ya} and determine relevant physical quantities such as the Skyrmion profile, mass, and size as a function of the physical parameters responsible for the deviations from conformal dynamics. Concretely, these are the anomalous dimension of the quark mass operator, the dilaton mass, and the conformal dimension of the relevant operator triggering the RG flow away from the infrared fixed point. We unveil an intricate dependence of the solitonic properties on these parameters as a consequence of the competition of dilaton and pion mass terms. We discover that the dilaton decouples rapidly from the dynamics once its mass exceeds the mass of the Goldstone modes. Moreover, we show that the dilaton decouples faster when the anomalous dimension of the chiral condensate is large.

We further show the existence of a special value of the ratio of dilaton and pion decay constants where the coupled second order field equations of the dilaton augmented EFT reduce to a first order solvable system. Therefore, the construction of analytic solitonic solutions is achieved through the identification of a novel special point in the parameter space of the theory. Interestingly, such a point does not correspond to any bound on the energy. This is a surprising result since, without the dilaton, the field equations with the spherical hedgehog ansatz are not solvable. However, we find that the so-constructed solutions, which feature a negative topological charge $B=-1$, are singular at the origin and do not carry finite energy.

In Sec.\ref{boxer}, we study the impact of the dilaton on Skyrmion crystals investigated in \cite{Canfora:2018rdz, Barriga:2021eki,Barriga:2022izc,Canfora:2023pkx}. After briefly reviewing the analytical solution in the absence of the dilaton, we show that even for the emerging hadronic tubes there exists a special parameters point where the coupled second order equation of motions (EOM)s reduce to a first order system (again without a BPS bound on the energy). The special parameter point is shown to be related to the one observed in the spherically symmetric case by mapping into each other the respective EOMs. We further notice that the presence of the dilaton spatially separates the baryon and isospin charge distributions similarly to spin-charge separation discussed in  \cite{PhysRevLett.125.190401, Lerda:1994kb}.   

We then move to study the impact of dilaton on the solitonic branes discovered in \cite{Alvarez:2017cjm,Ayon-Beato:2015eca,Canfora:2022jmh,Canfora:2023zmt,Canfora:2023pkx}. Assuming that the fields depend at most on time and a single spatial coordinate, we find the general solution to the EOMs of the dilaton-dressed Skyrme model. While the soliton profile is left unaltered by the presence of the dilaton, the latter acquires a homogeneous vacuum expectation value which is nontrivially determined by the soliton properties and the dilaton potential. For any given topological charge, we determine the lower bound on the dilaton mass such that the solution is real. Finally, we discuss how the dilaton smooths out the brane structure at low values of the dilaton mass.

We offer our conclusion in Sec.\ref{conclusion}. The appendix summarizes our numerical results for the spherical Skyrmion solutions.

\vspace{2em}

\section{Dilaton dressing of the chiral Lagrangian} \label{dresscode}
The Skyrme Lagrangian for $N_f=2$ massive flavours in four dimensions reads
\begin{align}
I[U] = \int d^4 x \sqrt{-g} \ \frac{K}{2} \operatorname{Tr}\left(\frac{1}{2}R_\mu R^\mu+\frac{\lambda}{16}F_{\mu\nu}F^{\mu\nu}+ \frac{m_\pi^2}{2} \left(U+U^{\dagger}\right)\right) \ , \\
R_\mu  =U^{-1} \nabla_\mu U=R_\mu^j t_j \ , \quad F_{\mu \nu}=\left[R_\mu, R_\nu\right] \ . \quad 
\label{NLSMaction}
  \end{align}
Here $U\in SU(2)$, $\nabla_{\mu}$ is the covariant derivative, and $t_{a}=i\sigma_{a}$ are the generators of the $SU(2)$ algebra, being $\sigma_{a}$ the Pauli matrices. Here $m_{\pi}$  stems from the pion mass while $K$ is related to the meson decay constant via $f_{\pi}=\sqrt{K}$. The above theory can describe low-energy Quantum Chromodynamics or any other underlying quantum field theory featuring the same pattern of chiral symmetry breaking.

The theory admits a conserved topological current $J_{\mu}$ given by
\begin{align}
J^{\mu}=\epsilon^{\mu\nu\lambda\rho}\operatorname{Tr}(R_{\nu}R_{\lambda}R_{\rho}) \ .
\end{align}
Integrating the zero-component of $J^{\mu}$ on a time-slice surface $\Sigma$ we obtain the topological charge
\begin{align}
    B=\frac{1}{24 \pi^2}\int_{\Sigma} \epsilon^{ijk}\operatorname{Tr}(R_{i}R_{j}R_{k})  \ .
\end{align}
This quantity corresponds to the winding number associated with $U$ and is interpreted as the Baryon number.
The classical equations of motion are obtained by performing the functional variation with respect to $U$, yielding
\begin{align}
    \nabla_{\mu}\left(R^{\mu}+\frac{\lambda}{4} [R_{\nu},F^{\mu\nu}]\right)-\frac{m_{\pi}^2}{2}\left(U-U^{\dagger}\right)=0  \ . \label{NLSMeq}
\end{align}
These correspond to a set of $3$ non-linear coupled Partial Differential Equations (PDE)s. 

In this work we explore the solitonic dynamics in a quasi-conformal regime by adding to the action a dilatonic degree of freedom. The latter is introduced to both restore conformal invariance at the action level and provide a mechanism for breaking conformality in a controllable manner. The pion mass term acts as an independent and controllable parameter for explicit conformal symmetry breaking.  We, therefore, dress every operator $O_{k}$ of mass dimension $k$ with the dilaton field $\sigma$ as \cite{coleman1988aspects, goldberger2007light}
\begin{align}
    O_{k}\rightarrow e^{(k-4)\sigma f}O_{k}  \ ,
\end{align}
where under a scale transformation $x\rightarrow e^{\alpha}x$ the field $\sigma$ transforms as 
\begin{align}
    \sigma \rightarrow \sigma -\frac{\alpha}{f} \ .
\end{align}
Here $f$ is the length related to the spontaneous breaking of scale invariance. 
Possible sources of explicit conformal breaking can be modeled by perturbing the underlying conformal theory via 
\begin{equation}
\delta L_{\mathcal{O}} = \lambda_{\mathcal{O}} \mathcal{O} \ , 
\end{equation}
 with $\mathcal{O}$ an operator with dimension $\Delta \neq 4$ and $\lambda_\mathcal{O}$  the associated coupling. The presence of such an operator  generates the following effective dilaton potential \cite{goldberger2007light, rattazzi2001comments, chacko2013effective, Appelquist:2022mjb}
 \begin{equation}\label{sigmapot}
  \,V(\sigma) =   f^{-4} \,   e^{-4\sigma f} \sum_{n=0}^\infty\ c_n \, e^{-n(\Delta-4)f \sigma }\ ,
\end{equation}
where the coefficients $c_n$ depend on the microscopic theory. In the present work we are interested in cases where the explicit conformal breaking is small which, in turn, can be realized when $\lambda_\mathcal{O} \ll 1$ and/or $\Delta \to 4$. 
In the first case, one can truncate the expansion \eqref{sigmapot} and obtain \cite{goldberger2007light, appelquist2020dilaton}
\begin{equation}\label{sigmapotDelta}
  V(\sigma) =  \frac{m_{\sigma}^2 e^{-4 f \sigma } }{4 (4-\Delta ) f^2}\left(1-\frac{4 }{\Delta }e^{-(\Delta-4)f \sigma }\right)  + \mathcal{O}( \lambda_{\mathcal{O}}^2) \ ,
\end{equation}
where we introduced the dilaton mass $m_\sigma$ and required that the ground state is realized for $\sigma = 0$. In particular, we have
\begin{equation}
\frac{c_1}{c_0} = - \frac{4}{\Delta} \ , \quad {\rm with} \quad c_0 = \frac{{f^2 \, m_{\sigma}^2   } }{4 (4-\Delta ) } \ .
\end{equation} 

The potential \eqref{sigmapotDelta} has been considered in many recent studies of QCD-like theories in the near-conformal regime \cite{appelquist2020dilaton, Appelquist:2022mjb, Bersini:2022bnx}. When $\Delta = 2$ the potential \eqref{sigmapotDelta} reduces to the usual $\phi^4$  Higgs-like potential whereas when $\Delta$ vanishes the coefficient $c_1$ becomes parametrically larger than $c_0$ and for $\Delta=0$ the potential \eqref{sigmapotDelta} diverges. 

The case of $\Delta  =0$ must be handled with care.   Here one must retain at least one extra term in \eqref{sigmapot}, namely the one with the $c_2$ coefficient.  The $c_1$ term becomes a $\sigma$ independent constant that we set to zero. Imposing the very same conditions we adopted earlier we arrive at 
\begin{equation}
   V_{\Delta \to 0}^{(1)}(\sigma) =\frac{m_\sigma^2\cosh (4 f \sigma )}{16 f^2}  \ ,
\end{equation}
where now $c_0 = c_2 = \frac{f^2 m_\sigma^2}{32}$ and $c_1= 0$. In this limit, the final dilaton potential preserves a $Z_2$ symmetry whose existence depends on the structure of specific underlying CFT and its deformation.  Alternatively, one can subtract the infinite constant appearing in the expansion of eq.\eqref{sigmapotDelta} around $\Delta=0$ and obtain the dilaton potential considered in the classic work of Coleman \cite{coleman1988aspects} 
\begin{equation} \label{coleman}
  V_{\Delta \to 0}^{(2)}(\sigma) =-\frac{m_\sigma^2}{16 f^2}  \left(1-4 f \sigma -e^{-4 f \sigma }\right)   \ .
\end{equation}
Here the $Z_2$ symmetry is removed and we shall use the more traditional Coleman potential in the following analyses.  

In the limit $\Delta\rightarrow 4$ we have that $\mathcal{O}$ becomes near marginal and it is legitimate to expand eq.\eqref{sigmapot} in powers of $\Delta - 4$ as
\begin{equation} \label{logpot}
  V(\sigma) = -\frac{m_{\sigma}^2 e^{-4 f \sigma }}{16 f^2} (1+4 f \sigma ) +\mathcal{O}\left((\Delta-4)^2\right)  \ .
\end{equation}
The form of the potential in \eqref{logpot}  agrees with the counting scheme in \cite{Golterman:2016lsd, Golterman:2016cdd, Golterman:2018mfm, Golterman:2020tdq, Golterman:2021ohm}. Moreover, it has been often employed to describe dense Skyrmion matter, see e,g, \cite{Lee:2003eg, brown1991scaling, Paeng:2017qvp, Ma:2013ooa, Park:2008zg}.


We can now also take into account the explicit breaking of conformal symmetry stemming from the presence of quark masses. We achieve this by assuming the mass term to have dimension $y=3-\gamma$ where $0<\gamma<2$ is the
anomalous dimension of the chiral condensate and its range is limited by the unitarity bound. Moreover, since for $\gamma \simeq  1$ the underlying four fermion operator becomes nearly marginal, we will focus on the interval $0<\gamma<1$.


We, finally, arrive at the following dilaton augmented chiral Lagrangian
\small{
\begin{equation}
\begin{split}
    I[U,\sigma]=& \int d^{4} x \sqrt{-g} \left[\frac{K}{2}\operatorname{Tr}\left(\frac{1}{2}R^{\mu}R_{\mu}e^{-2\sigma f}+\frac{\lambda}{16}F_{\mu\nu}F^{\mu\nu}+\frac{m_{\pi}^2}{2} \left(U+U^{\dagger}\right) e^{-y \sigma f}\right) -\left.\frac{1} {2} e^{-2 \sigma f}\nabla_\mu \sigma \nabla^\mu \sigma\right. -V(\sigma)\right]    \ .
\end{split}
\end{equation}} 
\normalsize
Although we restricted the number of flavours to two, we stress that conformality can be realized with such a small number of flavours in a number of theories, such as the ones with fermions in the two index symmetric or antisymmetric representation of the gauge group as shown first \cite{Sannino:2004qp} and generalized in \cite{Dietrich:2006cm}. In any case, here we focus on the impact of the novel dilatonic dynamics on time-honored solitons first discussed for the two and three flavour cases.  

In order to investigate these extended objects one starts with the equations of motion that read
\begin{align}
\label{EOM1}
    &\nabla_{\mu} \left(R^{\mu}e^{-2\sigma f}+\frac{\lambda}{4}[R_{\nu},F^{\mu\nu}]\right)-\frac{m_{\pi}^2}{2}\left(U-U^{\dagger}\right)e^{-y \sigma f}=0  \,, \\
    \label{EOM2}
&\nabla_{\mu}\left(e^{-2\sigma f}\nabla^{\mu}\sigma\right)+f \nabla^{\mu}\sigma\nabla_{\mu}\sigma e^{-2\sigma f}-\partial_\sigma V(\sigma)-\frac{K}{2} f\operatorname{Tr}R_{\mu}R^{\mu} e^{-2\sigma f} -\frac{K}{4} y f m_{\pi}^2 \operatorname{Tr}\left(U+U^{\dagger}\right)e^{-y \sigma f}=0  \ ,
\end{align}
while the stress-energy tensor $T_{\mu\nu}=\frac{(-2)}{\sqrt{-g}}\frac{\partial\sqrt{-g}\mathcal{L}}{\partial g^{\mu\nu}}$ is given by 
\begin{align}
    T_{\mu\nu}&=-\frac{K}{2} \operatorname{Tr}\left[\left( R_{\mu}R_{\nu}-\frac{1}{2}g_{\mu\nu}R_{\alpha}R^{\alpha}\right)e^{-2\sigma f}+\frac{\lambda}{4}\left(g^{\alpha \beta} F_{\mu \alpha} F_{\nu \beta}-\frac{1}{4} g_{\mu \nu} F_{\alpha \beta} F^{\alpha \beta}\right) \right. \nonumber\\& \left. - \frac{m_{\pi}^2}{2}  {g_{\mu\nu}\left( U+U^\dagger\right)} e^{-y\sigma f}\right]   -\left(-\nabla_{\mu}\sigma\nabla_{\nu}\sigma+\frac{1}{2}g_{\mu\nu}(\nabla\sigma)^2\right)e^{-2\sigma f} -g_{\mu\nu}V(\sigma)\ .
\end{align}
The presence of the mass term shifts the value of the minimum of the dilaton potential from $\sigma=0$ to a finite value that we indicate with $ \langle \sigma \rangle = \sigma_f$. The latter is determined by the following equation
\begin{equation} \label{bicchio}
  \partial_\sigma V(\sigma) \rvert_{\sigma=\sigma_f}+K y f m_\pi^2 e^{-y f \sigma_f} = 0 \  .
\end{equation}
This leads to new values for the decay constants and the masses of all the states appearing in the Lagrangian satisfying certain scaling relations \cite{Appelquist:2022mjb, Golterman:2020tdq, Golterman:2021ohm, appelquist2020dilaton}. Equipped with the above, we are now ready to investigate several classes of outstanding solitonic configurations starting with the celebrated hedgehog solution.

\section{Dilaton augmented hedgehog Skyrmions} \label{dahs}

\subsection{Setup}
The solitonic object with a given topological charge known as the Skyrme model  \cite{Skyrme:1961vq,Adkins:1983ya}  features a spherical hedgehog ansatz. The model describes baryons in Quantum Chromodynamics  but it has also been used, for instance, to provide a counter-example of the non-hair theorem \cite{Droz:1991cx} when coupled to general relativity. Here we investigate the modification to the hedgehog solution in the presence of dilatonic dynamics. Let us start with a general parameterization for $U(x)$ 
\begin{align}
    \begin{gathered}
U^{\pm 1}\left(x^\mu\right)=\cos (\alpha) \mathbf{1}_{2 \times 2} \pm \sin (\alpha) n^a t_a  \ , \\
n^1=\sin (\Theta) \cos (\Phi), \quad n^2=\sin (\Theta) \sin (\Phi), \quad n^3=\cos (\Theta)  \ .
\end{gathered} \label{U}
\end{align}
where $\mathbf{1}_{2 \times 2}$ is the $2 \times 2$ identity matrix and $\alpha(x)$, $\Theta(x)$ and $\Phi(x)$ are the three scalar degrees of freedom of $SU(2)$. The hedgehog ansatz for $SU(2)$ in spherical coordinates $\{r,\theta,\phi \}$reads 
\begin{align}
    \alpha=\alpha(r), \quad \Theta= \theta, \quad \Phi=\phi  \label{Ufield ANW} \ .
\end{align}
The energy and topological charge densities are respectively given by
\begin{align}
 \epsilon &= T_{00}= \frac{1}{2} e^{-2 f \sigma} \left(\frac{2 K \sin ^2(\alpha) }{r^2}+K \alpha '^2+\sigma^{\prime\ 2}\right)+\frac{\lambda  K \alpha '^2 \sin ^2(\alpha) }{r^2}\\ \nonumber&+\frac{\lambda  K \sin ^4(\alpha) }{2 r^4} -K m_\pi^2 \cos (\alpha ) e^{-f y \sigma} + V(\sigma)\ , \\   \rho_B &= J^0 = -\frac{12 \alpha ' \sin ^2(\alpha )}{r^2}  \ ,
\end{align}
where the prime denotes the derivative with respect to the radial coordinate.
The baryon mass $M_S$ and charge $B$ are respectively obtained by integrating the energy charge and  densities of the solution over the whole spacetime, that is
\begin{equation}
    M_s =4 \pi \int_0^\infty  r^2  \epsilon \ d r \,, \qquad B= \frac{1}{6 \pi} \int_0^\infty  r^2  \rho_B \ d r \,.
\end{equation}
A measure of the size of the solitonic object is given by the root mean square radius of the baryon charge $\langle r^2 \rangle^{1/2}$ defined as 
\begin{equation}
    \langle r^2 \rangle^{1/2} \equiv \left( \frac{1}{6 \pi}\int_0^\infty \rho_B \ r^4 \dd{r}\right)^{1/2} = \sqrt{- \frac{2}{\pi}  \int_0^\infty r^2 \sin^2(\alpha) \alpha^\prime d r} \ .
\end{equation}

Inserting the hedgehog ansatz in \eqref{NLSMeq} and \eqref{EOM2} we obtain the following set of coupled second order differential equations
{\small
\begin{equation}\label{skyrmerad1}
\begin{split}
 \left(2 \lambda  e^{2 f \sigma} \sin ^2(\alpha) +r^2\right)\alpha ''-2 r \alpha ' \left(f r \sigma'-1\right)+\frac{\lambda  e^{2 f \sigma} \sin (2 \alpha ) \left(r^2 \alpha '^2-\sin ^2(\alpha )\right)}{r^2}-\sin (2 \alpha ) - m_{\pi }^2 r^2 \sin (\alpha) e^{-f (y-2) \sigma}=0  \ ,
\end{split}
\end{equation}
}
{\small
\begin{equation}\label{skyrmerad2}
   \sigma''-f\sigma'^2+\frac{2}{r}\sigma'-e^{2f\sigma}\partial_\sigma V(\sigma)
  +f K\alpha'^2+\frac{2}{r^2}f K\sin^2(\alpha) -K f  m_{\pi }^2 y \cos (\alpha ) e^{f (2-y) \sigma }=0  \ .
\end{equation}
}
In the next section, we will numerically solve the above equations to investigate the dependence of the Skyrmion properties on the parameters $\Delta, m_\sigma, y$ encoding the explicit breaking of conformal invariance.

\subsection{Numerical solutions} \label{numeri}
To study the impact of dilatonic dynamics, we numerically solve the EOMs \eqref{skyrmerad1} and \eqref{skyrmerad2} in the sector of topological charge $B=1$. We measure all the dimensionful quantities in units of $\sqrt{K}$ (in QCD $\sqrt{K} = 93 \text{MeV}$) and as a starting point we consider the following  QCD-inspired values \cite{appelquist2020dilaton, Adkins:1983ya, Lee:2003eg}: 
\begin{equation}
f=0.29 \,, \qquad \lambda=\frac{1}{4.75^2} \,, \qquad m_\pi=\frac{140}{93} \ , 
\end{equation}
in units of $\sqrt{K}$. 
We then solve the EOMs for different values of the parameters $\Delta$, $m_\sigma$, $y$. In particular, we consider $m_\sigma=1,2, \dots, 10$ (except $m_\sigma =6$ due to numerical instabilities) and $\Delta=1, 1.2, 1.4, \dots, 3.8$. Moreover, we study the limiting cases $\Delta \to 4$ and $\Delta \to 0$  corresponding to the dilaton potentials \eqref{logpot} and \eqref{coleman}, respectively. For every value of $\Delta$ and $m_\sigma$, we determine the profile functions $\alpha(r)$ and $\sigma(r)$, the baryon mass $M_S$, and the root mean square radius of the baryon charge $\langle r^2 \rangle^{1/2}$. Since the topologically trivial vacuum exhibits a divergent zero-point energy, we define $M_S$ as the energy difference between the solutions with $B=1$ and $B=0$. To simplify the presentation we momentarily set $y=5/2$ and postpone to Sec.\ref{yrole} the discussion of the dependence on the anomalous dimension of the chiral condensate. However, as we shall see, all the conclusions below apply for any value of $2<y<3$.

To set the boundary conditions we impose that for $r \to \infty$ the fields approach the topologically trivial vacuum. We recall that at fixed values of time, the field $U(r)$ defines a map from the spatial manifold $\mathbb{R}^3$ to the isospin manifold $S^3$ with the boundary condition that $U(r)$ goes to its trivial vacuum for asymptotically large distances. In other words, we have 
\begin{equation}
    U(r\to\infty)=\mathbb{1}\implies \alpha(r\to \infty)=0 \,, \qquad   U(0)=-\mathbb{1}\implies \alpha(0)=\pi\ ,
\end{equation}
where the latter condition fixes the baryon charge to unity \cite{Adkins:1983ya}. Analogously, we impose $\sigma(r \to \infty) \to \sigma_f$ with $\sigma_f$ given in eq.\eqref{bicchio}. Since $\sigma_f$ depends on the values of the parameters we vary, this can be seen as a fixed BC for the normalized variable 
\begin{equation}
    \chi(r) \equiv e^{- f (\sigma(r)-\sigma_f)} \,,
\end{equation} 
namely $\chi(r \to \infty) \to 1$. Accordingly, we keep fixed the value of $\chi(r)$ at the origin in order to have the same boundary conditions for the variable $\chi$ regardless of the value of the parameters. We note that one of the main differences with respect to earlier numerical studies \cite{Lee:2003eg, Park:2008zg} is that we do consider the dependence of $\sigma_f$ on the presence of the pion mass. 

We study the dilaton potential \eqref{sigmapotDelta} for $1 \le \Delta < 4$ and its $\Delta \to 4$ limit \eqref{logpot} for $\Delta = 4$. We do not consider values of $\Delta$ smaller than $1$ being the potential \eqref{sigmapotDelta} ill-defined in the $\Delta \to 0 $ limit.  Finally, we choose $\chi(r=0)=2/5$ as the initial condition for the dilaton. 

\begin{figure}[t!]
\centering
\includegraphics[width=0.49\textwidth]{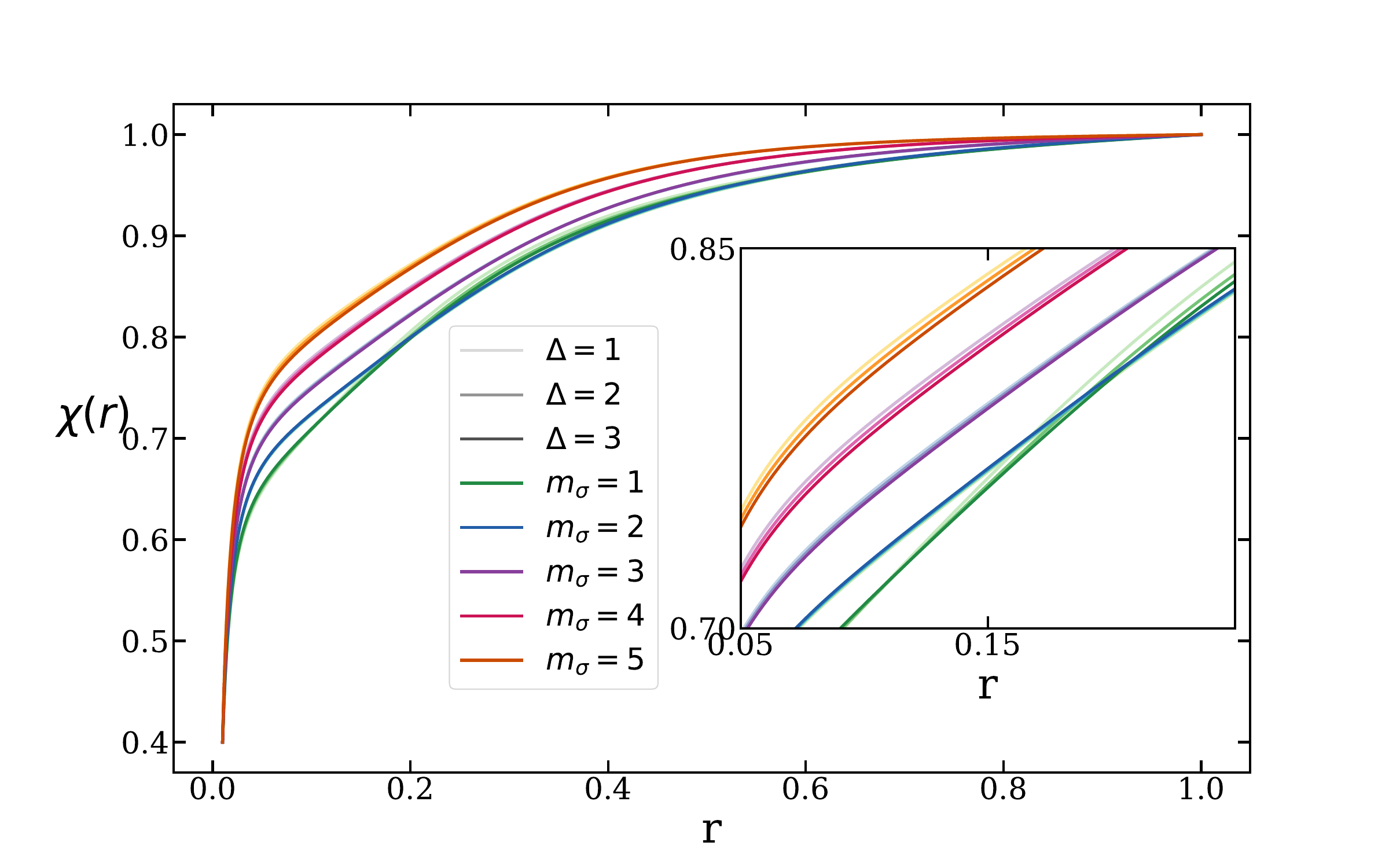} \includegraphics[width=0.49\textwidth]{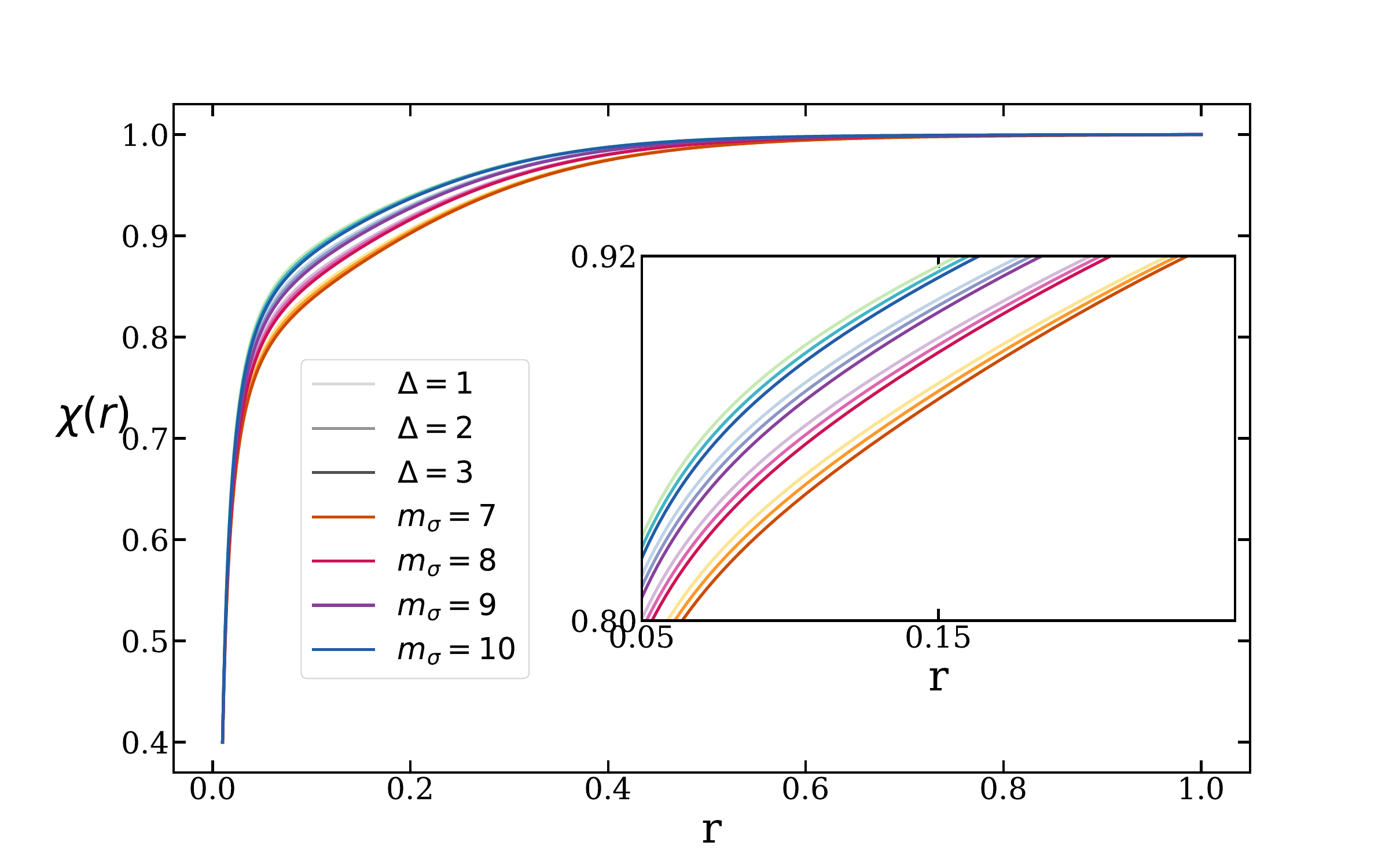} 
	\caption{Dilaton profile $\chi(r)$ for $\Delta=1,2,3$ and $m_\sigma=1,2,3,4,5$ (\emph{left panel}) and  $m_\sigma=7,8,9,10$ (\emph{right panel}). The initial condition for the dilaton is $\chi(0)=2/5$.} 
	\label{fig:appchi1}
\end{figure}

\begin{figure}[t!]
\centering
\includegraphics[width=0.49\textwidth]{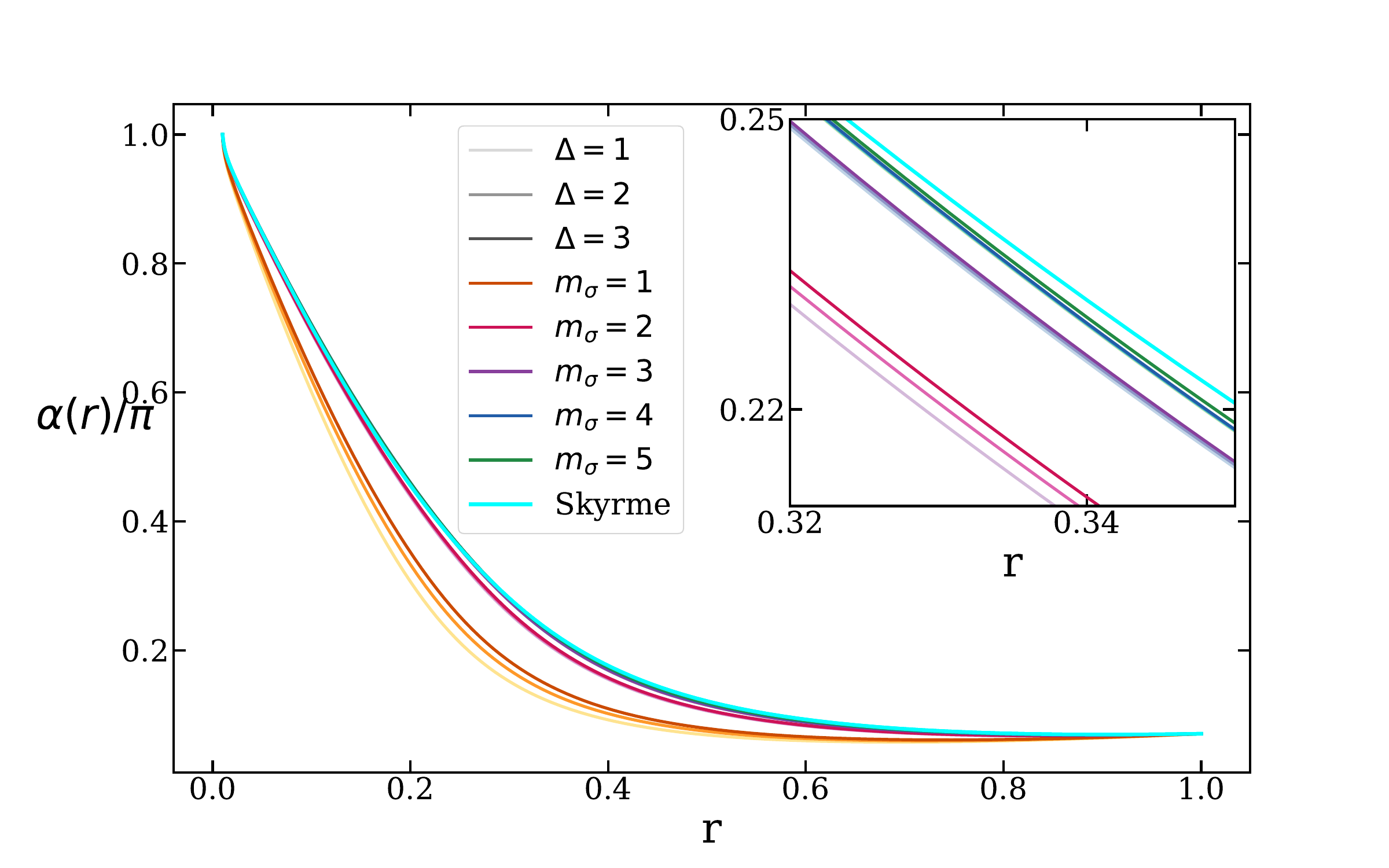} \includegraphics[width=0.49\textwidth]{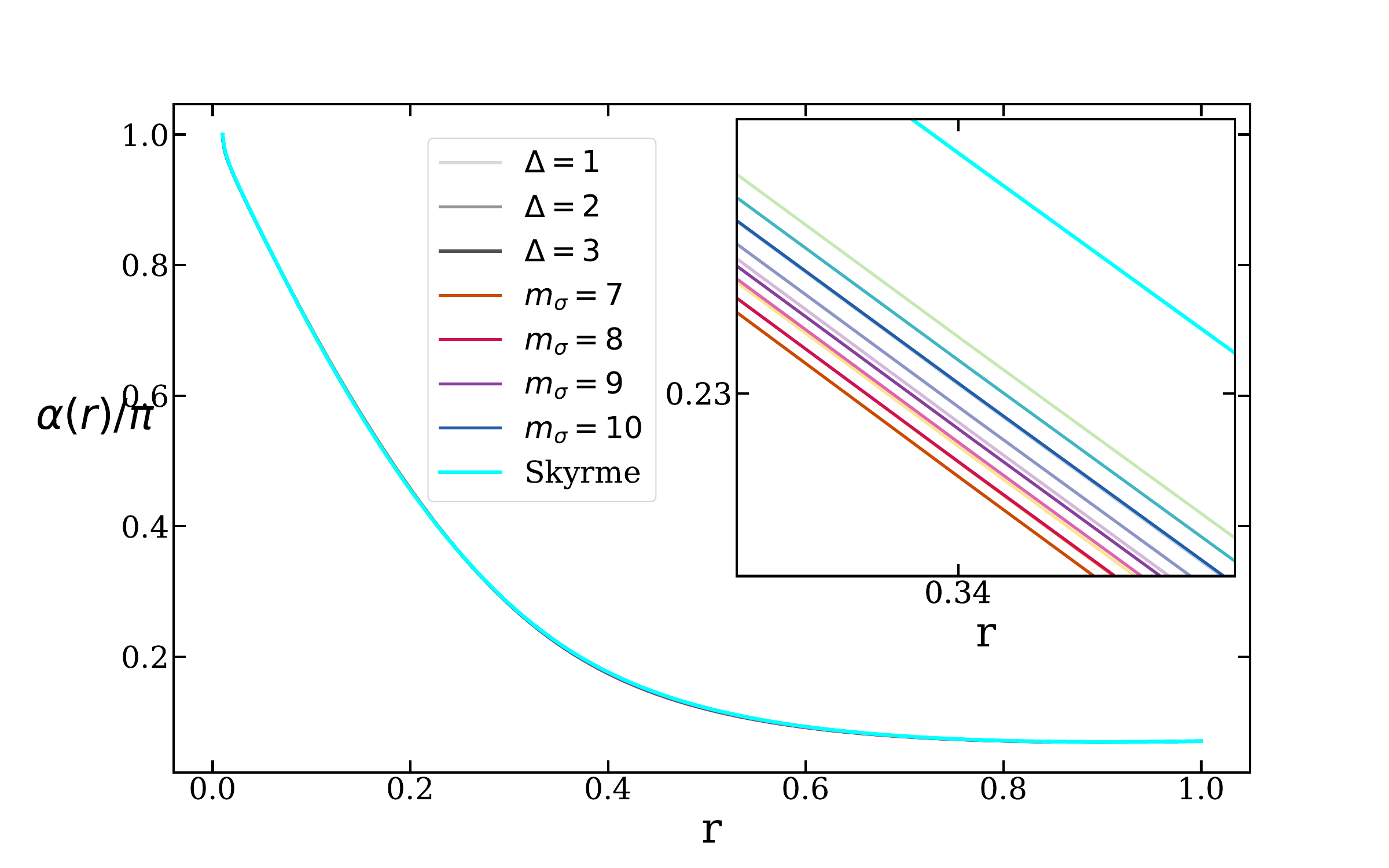} 
	\caption{Skyrmion profile $\alpha(r)$ for $\Delta=1,2,3$ and $m_\sigma=1,2,3,4,5$ (\emph{left panel}) and  $m_\sigma=7,8,9,10$ (\emph{right panel}). The initial condition for the dilaton is $\chi(0)=2/5$.} 
	\label{fig:appelquist alpha}
\end{figure}

Fig.\ref{fig:appchi1} shows the dilaton profile $\chi(r)$ for $\Delta=1,\ 2,\  3$ and, respectively, $m_\sigma=7,\ 8\  9\ 10$ and $m_\sigma=1,\ 2,\ 3, \, 5$. The general trend displayed in the figures can be summarized as follows.
\begin{enumerate}
    \item At fixed $\Delta$ the dilaton profile flattens towards its asymptotic value $\chi=1$ as $m_\sigma$ increases.
    \item  At fixed $m_\sigma$ the dilaton profile flattens towards its asymptotic value $\chi=1$ as $\Delta$ decreases.
\end{enumerate}
The profile $\alpha(r)/\pi$ is depicted in Fig.\ref{fig:appelquist alpha} for the same choice of the parameters. As $m_\sigma$ increases the dilaton decouples from the dynamics and $\alpha(r)$ converges rapidly to the solution obtained in the absence of the dilaton. In general, the solutions exhibit little dependence on $\Delta$. 

\begin{figure}[t!]
\centering
\includegraphics[width=0.5\textwidth]{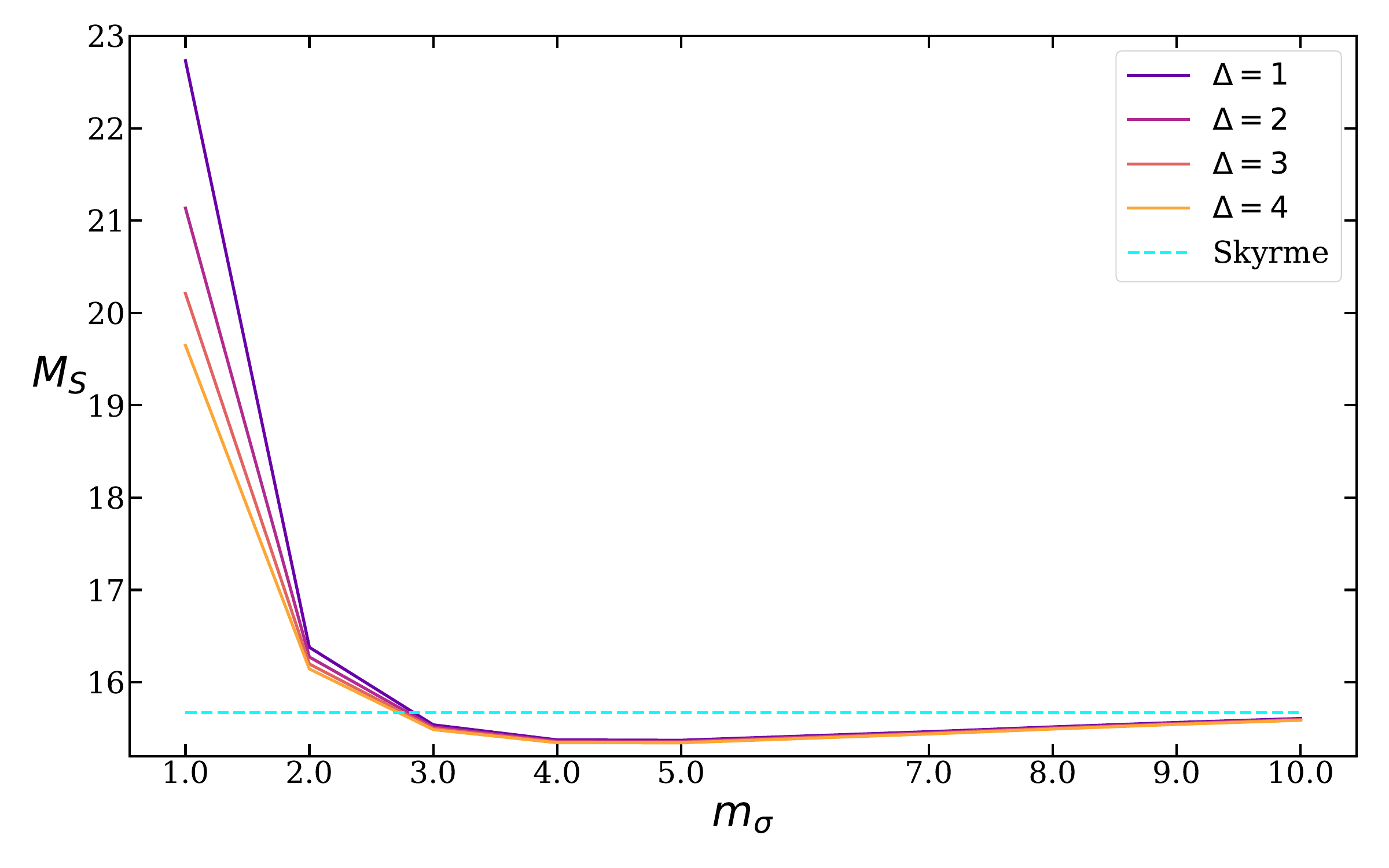}
	\caption{The Skyrmion mass $M_S$ as a function of the dilaton mass $m_\sigma$ for $\Delta=1$ (\emph{purple}), $\Delta=2$ (\emph{red}), $\Delta=3$ (\emph{orange}), and $\Delta=4$ (\emph{yellow}). The dashed line marks the value of the Skyrmion mass in the absence of the dilaton ($M_\text{Skyrme}= 15.67$ in units of $\sqrt{K}$).} 
	\label{massmass}
\end{figure}

Our results for $M_S$ are collected in appendix \ref{app1} and illustrated in Fig.\ref{msigma35710Msrquadro} for $m_\sigma=3,5,7,10$. For all the values of the free parameters, the relative numerical error is less than $0.5\%$ where such a value should be seen as a conservative upper limit. One can see that at fixed values of $m_\sigma$ the baryon mass gets smaller as $\Delta$ increases. On the other hand, the dependence of $M_S$ on $m_\sigma$ at fixed $\Delta$ is not monotonic. In fact, $M_S$ first decreases with increasing $m_\sigma$ till it reaches its minimum for $m_\sigma \sim 6$ after which it climbs towards its   Skyrme model value $M_\text{Skyrme} = 15.67$. This behavior is illustrated in Fig.\ref{massmass}. A change in the behavior of the solutions around $m_\sigma = 6$ can be also seen by analyzing the values of $\langle r^2 \rangle^{1/2}$ which are given in Tab.\ref{tab:bigtable2b} and Tab.\ref{tab:bigtable1} in App.\ref{app1}. In fact,  $\langle r^2 \rangle^{1/2}$ increases with increasing $m_\sigma$ till $m_\sigma \sim 6$ and then starts getting smaller for higher values of the dilaton mass. Moreover, while $\langle r^2 \rangle^{1/2}$ grows monotonically with $\Delta$ for $m_\sigma=1,2, \dots, 5$, the opposite holds for $m_\sigma=7,8,9,10$ as can be seen from Fig.\ref{msigma35710Msrquadro}.

Finally, it is important to stress that the dependence of the Skyrmion properties on both $\Delta$ and $m_\sigma$ becomes quickly milder as we increase $m_\sigma$ implying that \emph{the dilaton decouples rapidly from the dynamics once its mass exceeds $m_\pi$}. For instance, while for $m_\sigma=1$ both $M_S$ and $\langle r^2 \rangle^{1/2}$ differ more than $20\%$ from their values in the absence of the dilaton, such a difference reduces to less than $3\%$ for $m_\sigma \ge 3$.

\begin{figure}[t!]
\centering
\includegraphics[width=0.48\textwidth]{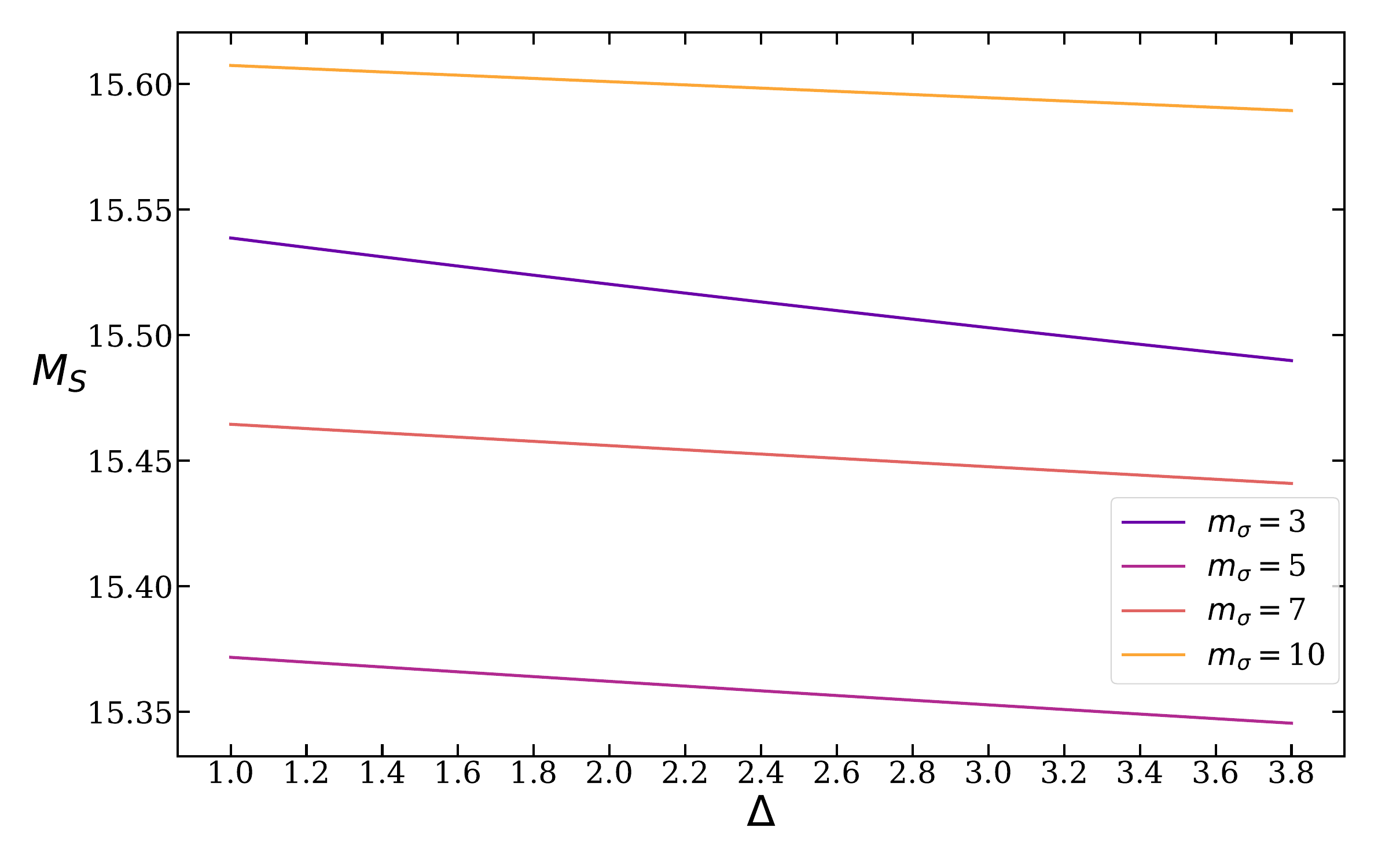} \includegraphics[width=0.48\textwidth]{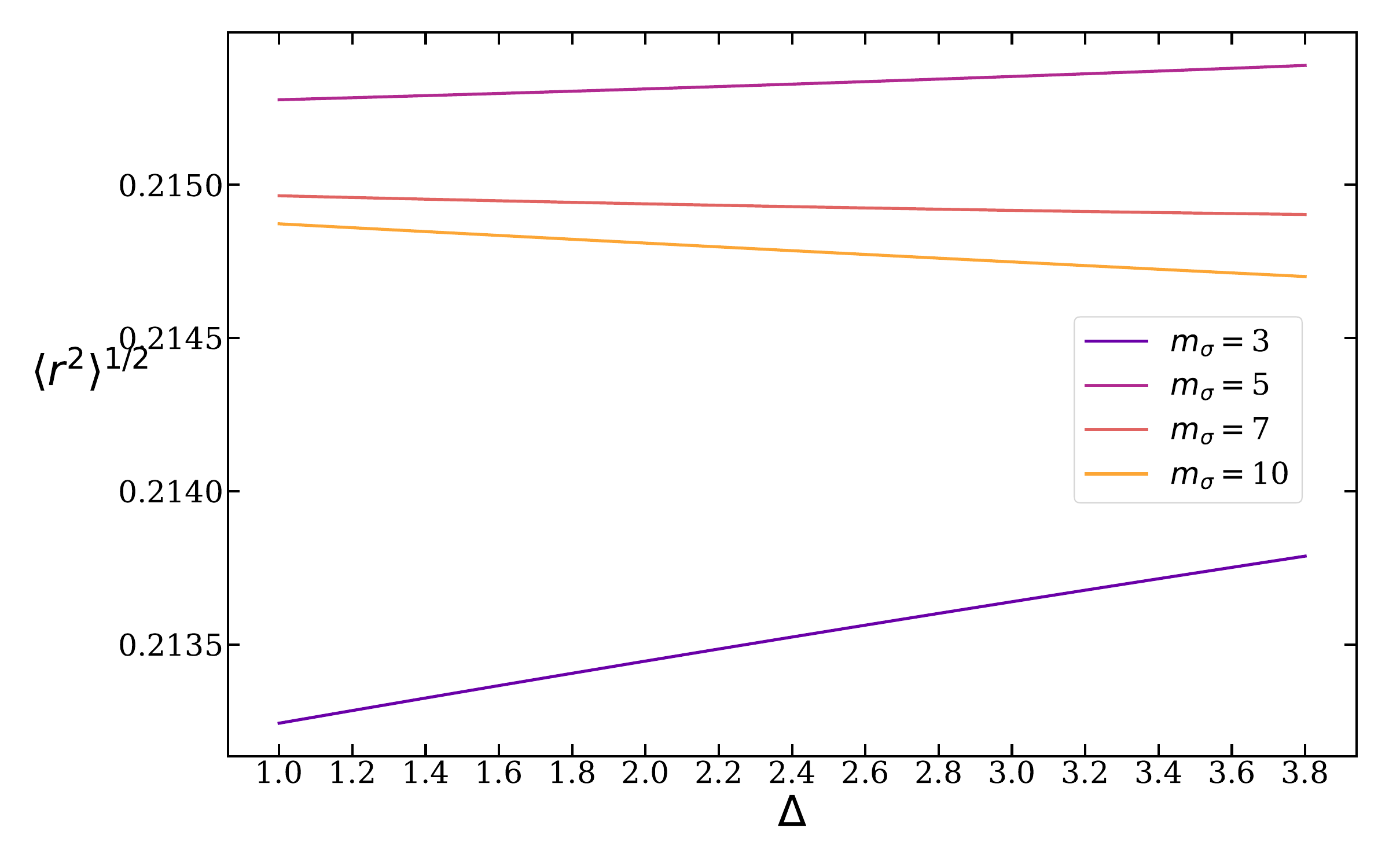} 
	\caption{$M_S$ (\emph{left}) and $\langle r^2 \rangle^{1/2}$ (\emph{right}) as a function of $\Delta$ $m_\sigma=3, 5, 7, 10$.} 
	\label{msigma35710Msrquadro}
\end{figure}


\subsubsection{Skyrmions in the $\Delta \to 0$ limit}

As previously discussed, the potential \eqref{sigmapotDelta} diverges in the $\Delta \to 0$ limit. For completeness, we now discuss the solitonic solution in the presence of the potential \eqref{coleman}, which can be seen as a regularized version of the $\Delta=0$ case. We use the values of the parameters considered in the previous section except the initial condition for the dilaton field that we take to be $\chi(0)=9/20$ since it leads to higher numerical precision.

\begin{figure}[t!]
\centering
\includegraphics[width=0.48\textwidth]{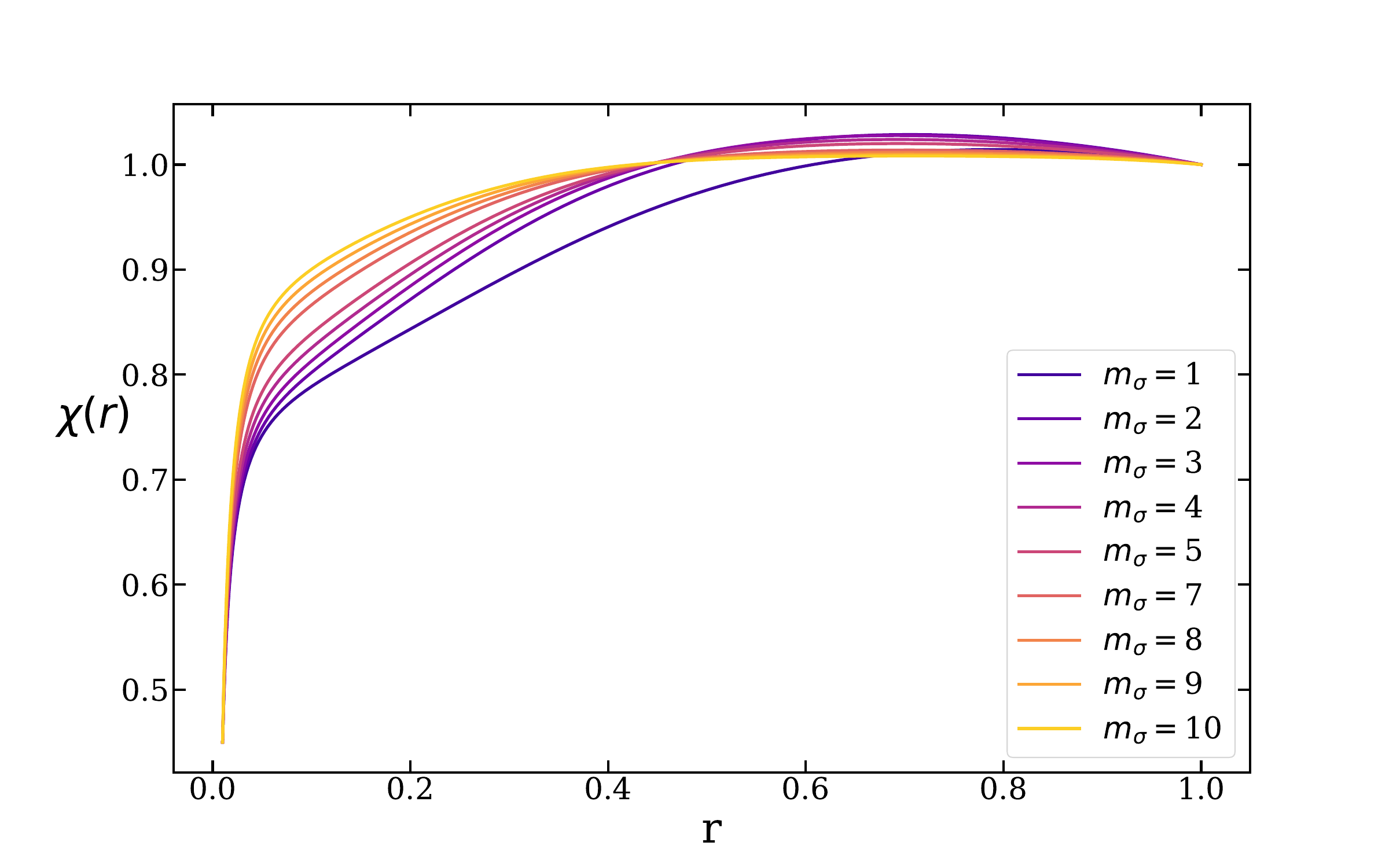} \includegraphics[width=0.48\textwidth]{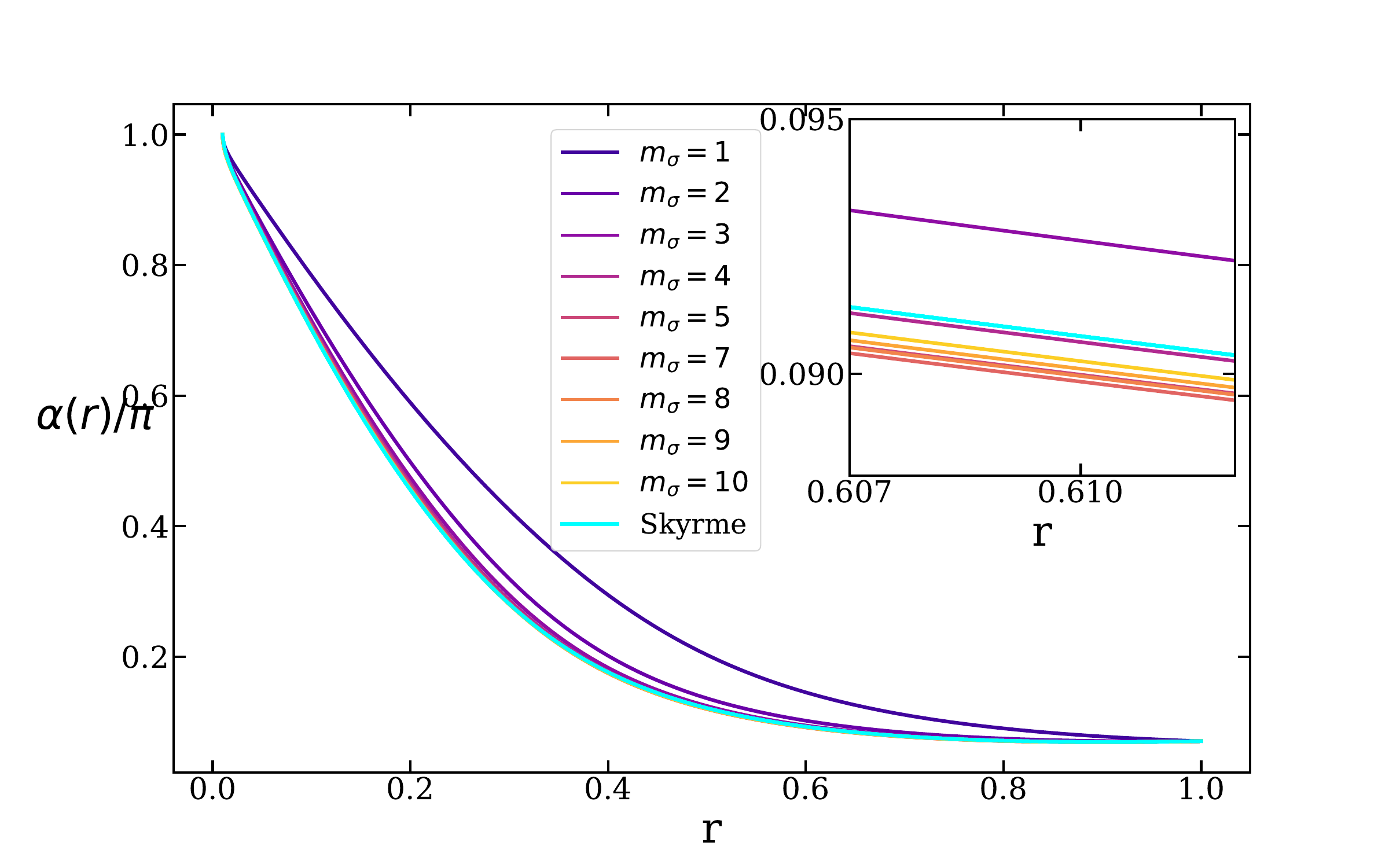} 
	\caption{Dilaton profile $\chi(r)$ (\emph{left}) and $\alpha(r)/\pi$ (\emph{right}) for $m_\sigma=1,2,3,4,5,7,8,9,10$. The initial condition for the dilaton is $\chi(0)=9/20$.} 
	\label{profilecoleman}
\end{figure}

\begin{figure}[t!]
\centering
\includegraphics[width=0.48\textwidth]{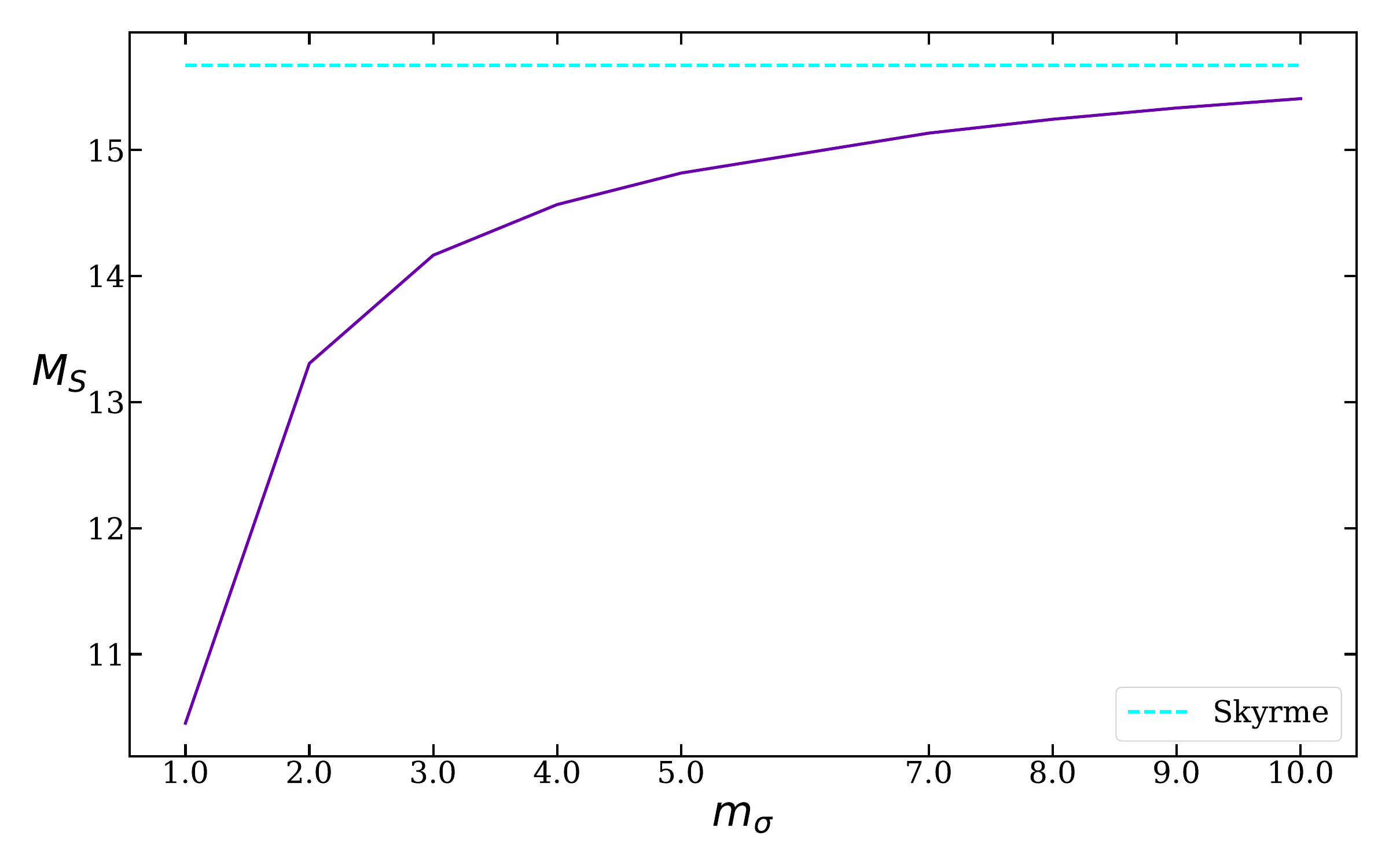} 
\includegraphics[width=0.48\textwidth]{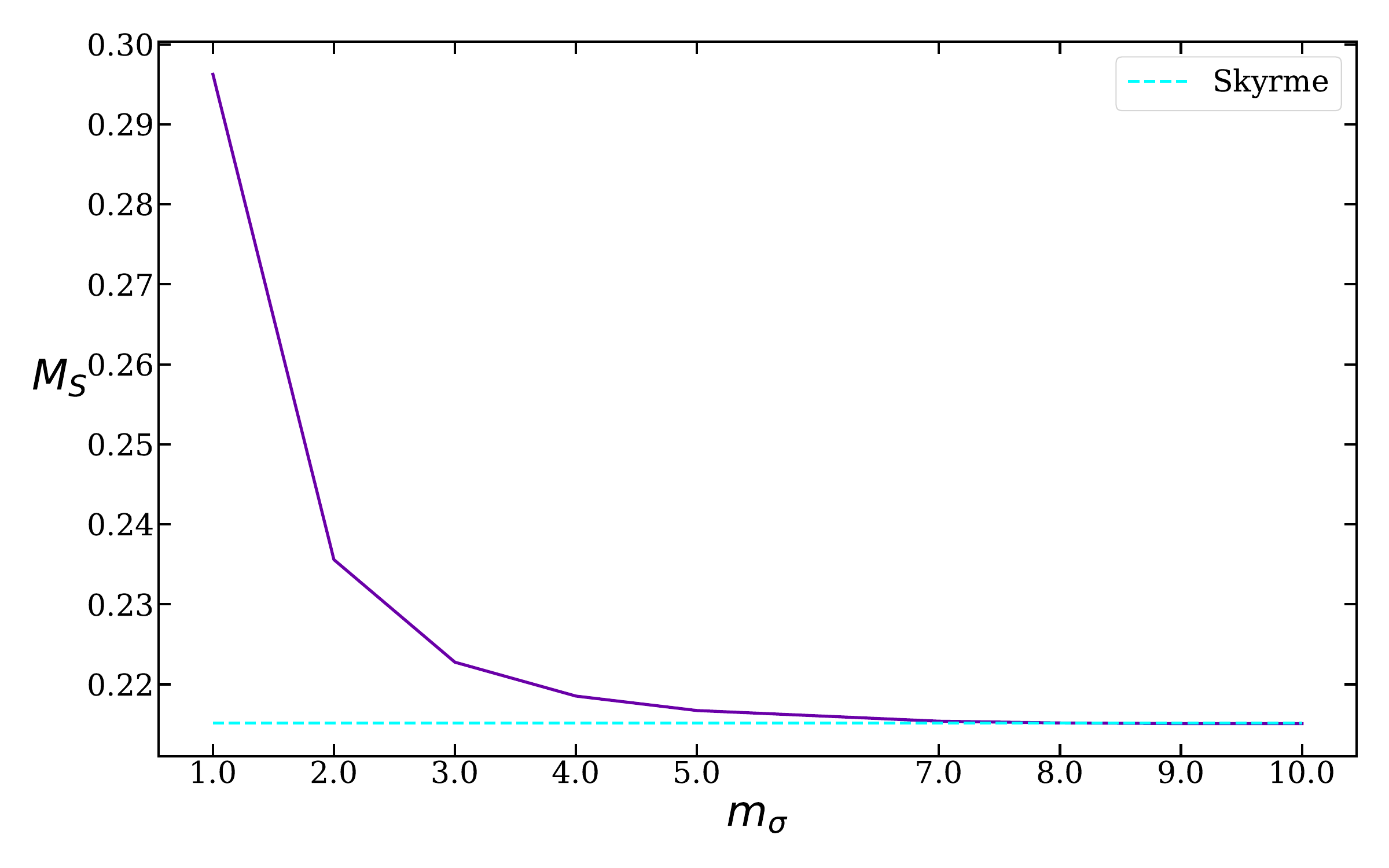}
	\caption{The Skyrmion mass $M_S$ (\emph{left}) and $\langle r^2 \rangle^{1/2}$ (\emph{right}) as a function of the dilaton mass $m_\sigma$. The dashed line marks the values obtained in the absence of the dilaton $M_{\text{Skyrme}}=15.67$ and  $\langle r^2 \rangle^{1/2}_\text{Skyrme}=0.215$.} 
	\label{MSmsigmacoleman}
\end{figure}

The profiles of the fields are displayed in Fig.\ref{profilecoleman}. Their behavior as we vary $m_\sigma$ is very similar to the one discussed in the $1 \le \Delta \le 4$ case with the dilaton profile becoming more flat and converging faster to $\chi=1$  as $m_\sigma$ increases while $\alpha(r)$ converges quickly to the Skyrme model profile. However, differences can be seen in the behavior of the Skyrmion mass and the root mean square radius as a function of $m_\sigma$. For example, differently from the cases studied above, $M_S$ is always smaller than  $M_\text{Skyrme}$  for any value of $m_\sigma$. Additionally, as illustrated in the left panel of Fig.\ref{MSmsigmacoleman}, when $m_\sigma$ increases $M_S$ grows monotonically (in the previous examples this behavior was not monotonic) converging quickly to $M_\text{Skyrme}$. We also observe that $\langle r^2 \rangle^{1/2}$ is larger than its value in the dilaton-less case  $\langle r^2 \rangle^{1/2}_\text{Skyrme}=0.049$ for any value of $m_\sigma$ and decreases monotonically for increasing values of $m_\sigma$ as shown in the right panel of Fig.\ref{MSmsigmacoleman}. Our results for $M_S$ and  $\langle r^2 \rangle^{1/2}$ are listed in App.\ref{app2}.

\subsubsection{The role of the anomalous dimension of the chiral condensate} \label{yrole}

\begin{figure}[t!]
\centering
\includegraphics[width=0.49\textwidth]{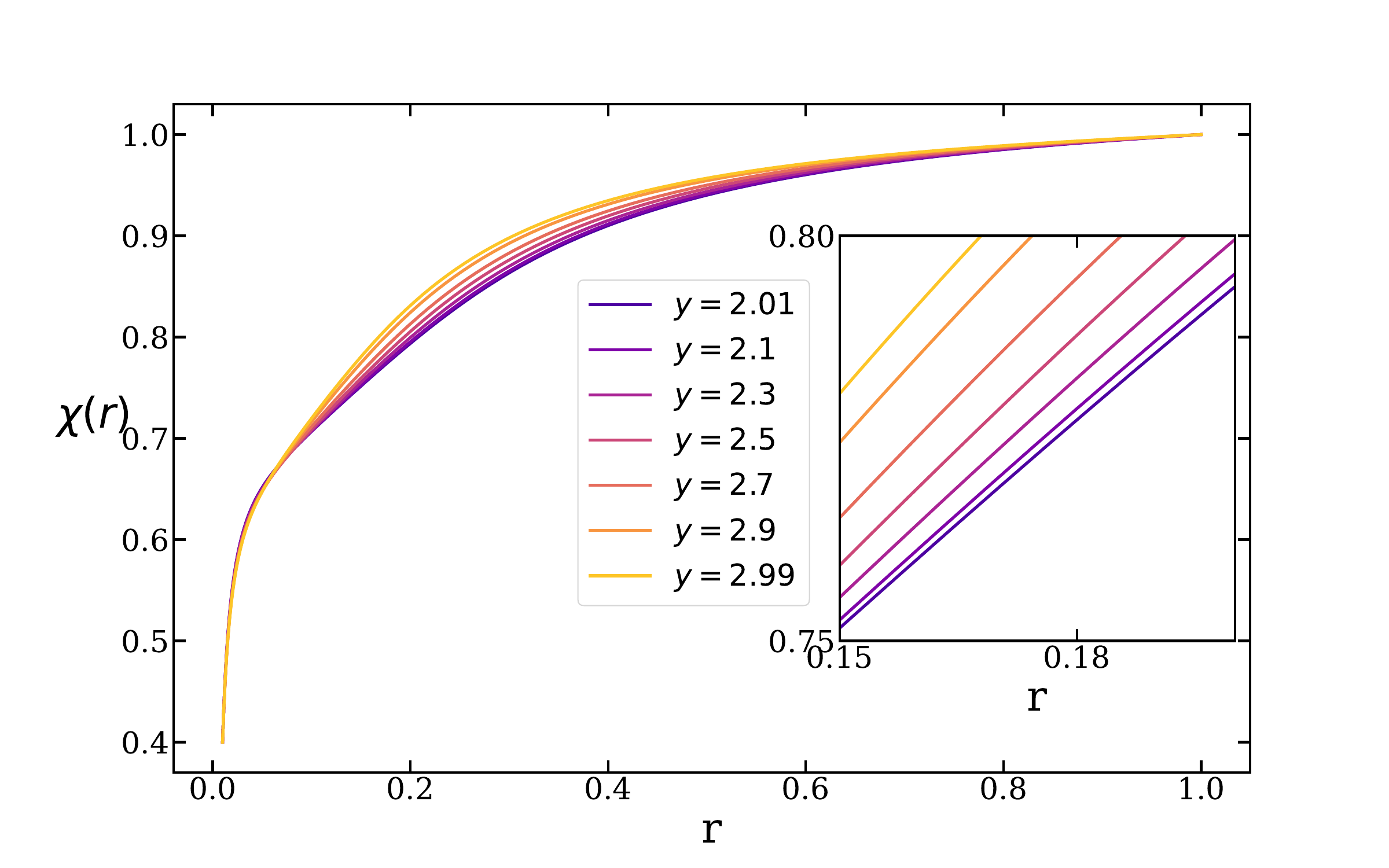} \includegraphics[width=0.49\textwidth]{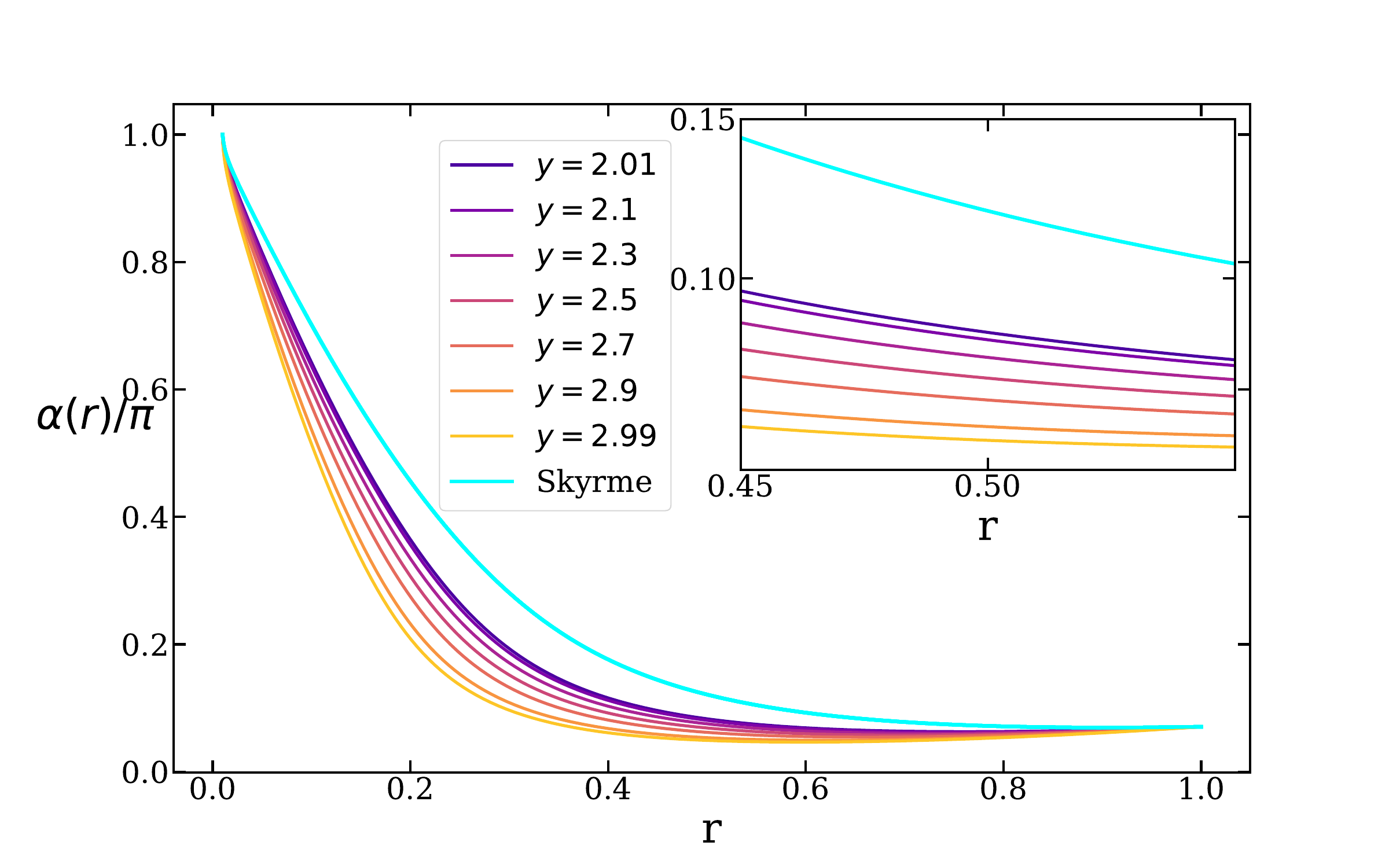} 
	\caption{Dilaton profile $\chi(r)$ (\emph{left}) and  $\alpha(r)/\pi$ (\emph{right}) for $m_\sigma=1$ and $\Delta=1$. The initial condition for the dilaton is $\chi(0)=2/5$.} 
	\label{solutionsmsigma1delta1}
\end{figure}

The time is ripe to discuss the impact of the anomalous dimension of the chiral condensate on the Skyrmion properties, i.e. the dependence on $y$. First, we note that all the conclusions drawn for the $y=5/2$ case in the previous sections hold for every $2<y<3$. The dilaton profile $\chi(r)$ becomes more flat and starts to converge earlier to its asymptotic value as $y$ increases. At the same time, the Skyrmion profile $\alpha(r)$ approaches the profile in the absence of the dilaton as $y$ decreases signaling the decoupling of $\sigma(r)$ from the dynamics. To illustrate this behavior we show the profiles obtained for $\Delta = m_\sigma = 1$ and different values of $y$ in Fig.\ref{solutionsmsigma1delta1}. 

\begin{figure}[t!]
\centering
\includegraphics[width=0.48\textwidth]{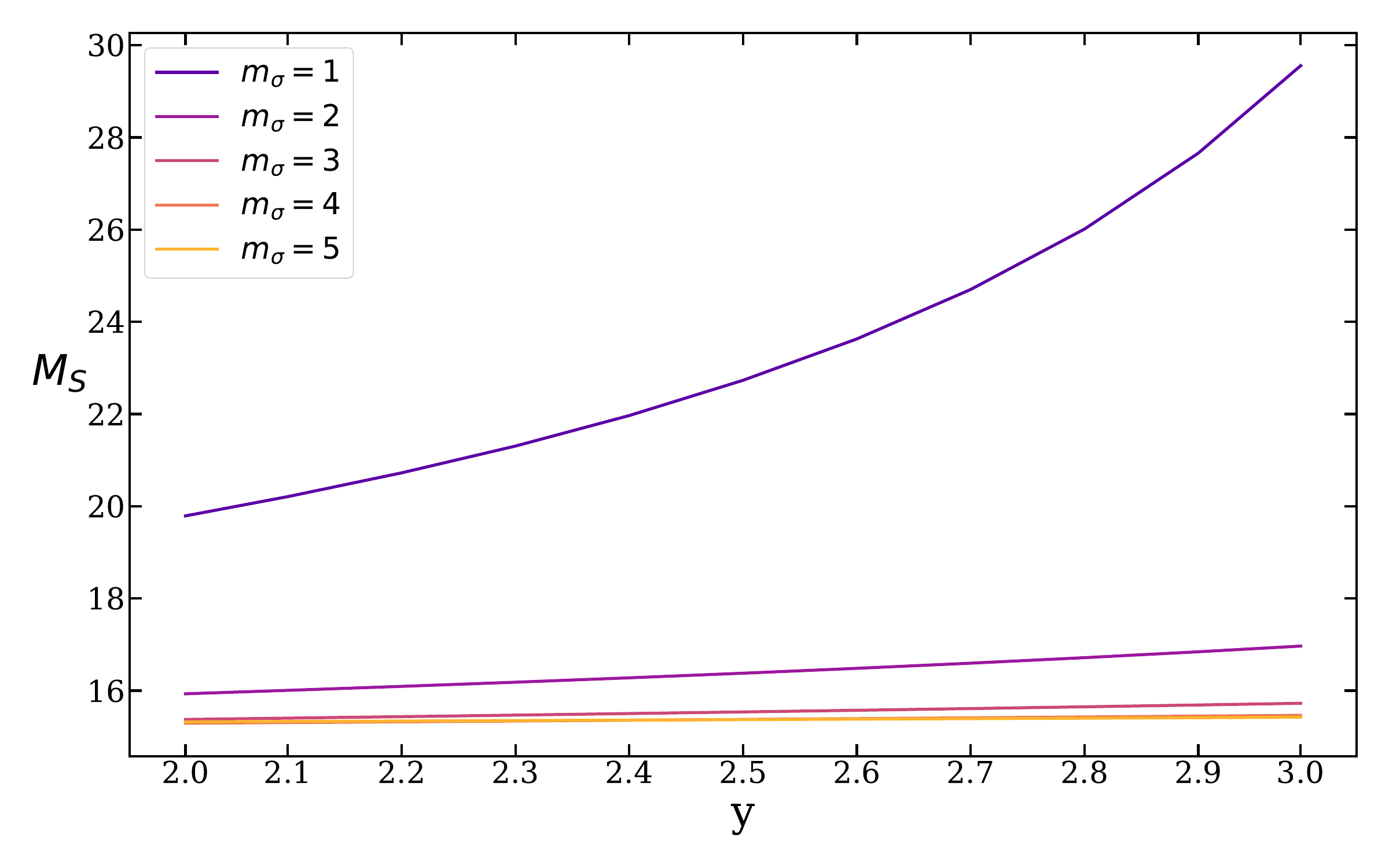} \includegraphics[width=0.48\textwidth]{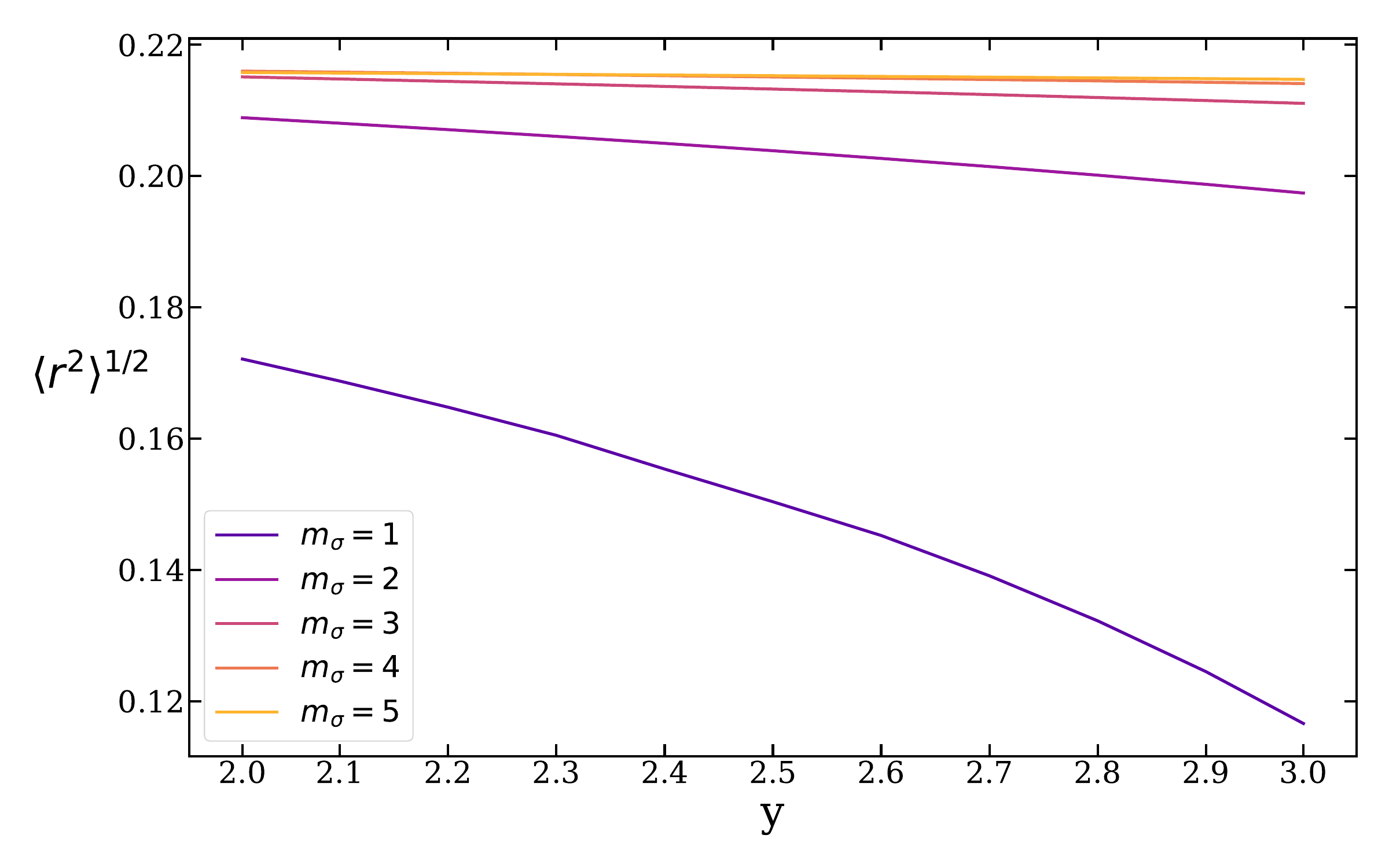} 
	\caption{$M_S$ (\emph{left}) and $\langle r^2 \rangle^{1/2}$ (\emph{right}) as a function of $y$ for $\Delta=1$ and $m_\sigma=1, 2, 3, 4, 5$.} 
	\label{delta1msigma1Msrquadro}
\end{figure}

\begin{figure}[t!]
\centering
\includegraphics[width=0.48\textwidth]{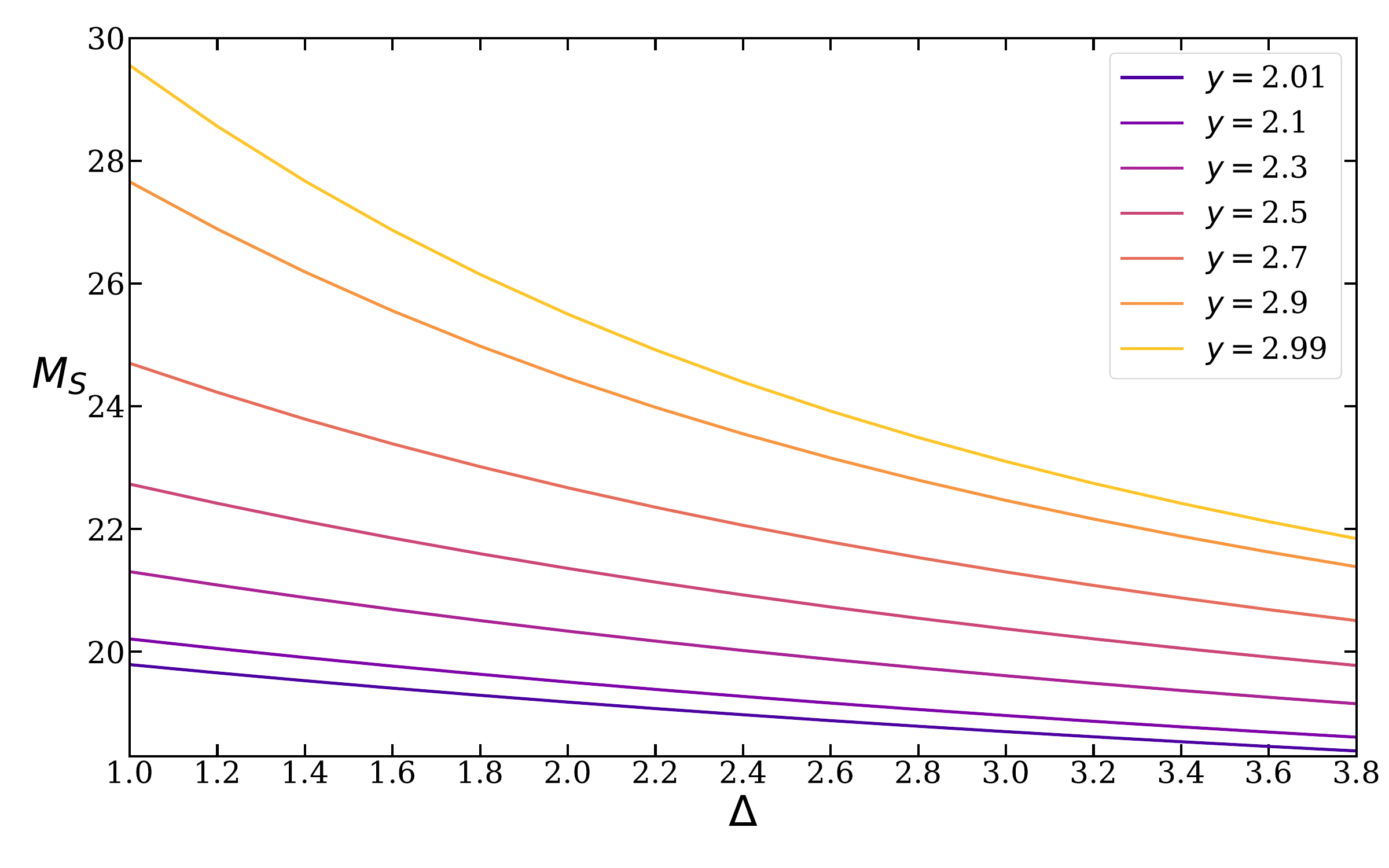}
\includegraphics[width=0.48\textwidth]{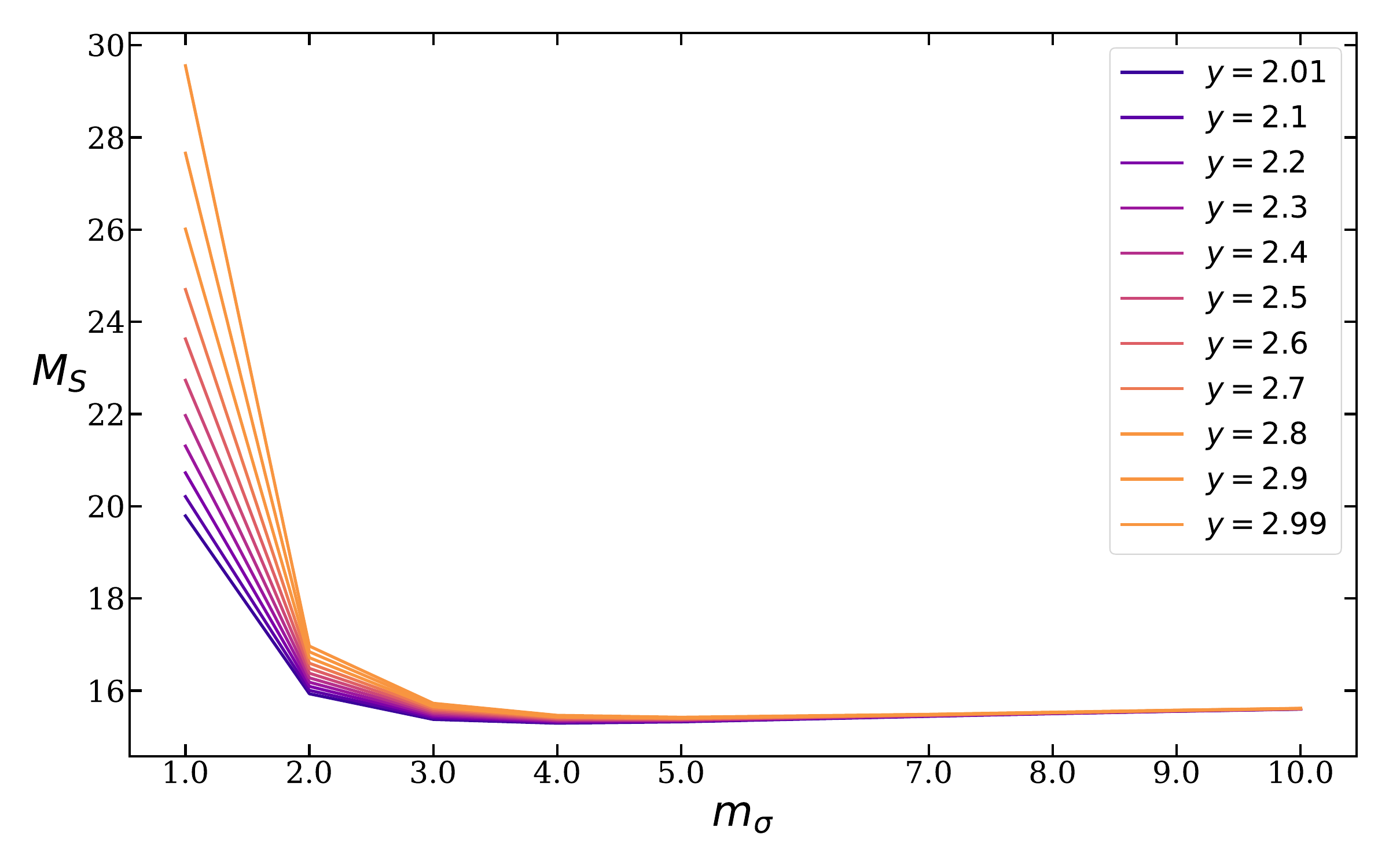} 
	\caption{The Skyrmion mass $M_S$ as a function of $\Delta$ for $m_\sigma=1$ and $y=2.01, 2.1, 2.3, 2.5, 2.7, 2.9, 2.99$ (\emph{left}) and as a function of $m_\sigma$ for $\Delta=1$ and  $y=2.01, 2.1, 2.3, 2.5, 2.7, 2.9, 2.99$ (\emph{right}).} 
	\label{massmsigma1y}
\end{figure}

$M_S$ and $\langle r^2 \rangle^{1/2}$ get, respectively, larger and smaller as $y$ varies from $y=2$ to $y=3$, as shown in Fig.\eqref{delta1msigma1Msrquadro}. From the figure is also evident that the dependence on $y$ gets weaker as the dilaton decouples for larger $m_\sigma$. Finally, our results for $M_S$ and $\langle r^2 \rangle^{1/2}$ confirm the observation that \emph{the dilaton decouples faster for smaller $y$}. For instance, this can be seen by looking at how the dependence of $M_S$ (and $\langle r^2 \rangle^{1/2}$) on both $\Delta$ and $m_\sigma$ becomes softer as $y$ decreases. This last point is exemplified in Fig.\ref{massmsigma1y}. 

\subsubsection{Numerical solutions in the chiral limit}

\begin{figure}[t!]
\centering
\includegraphics[width=0.43\textwidth]{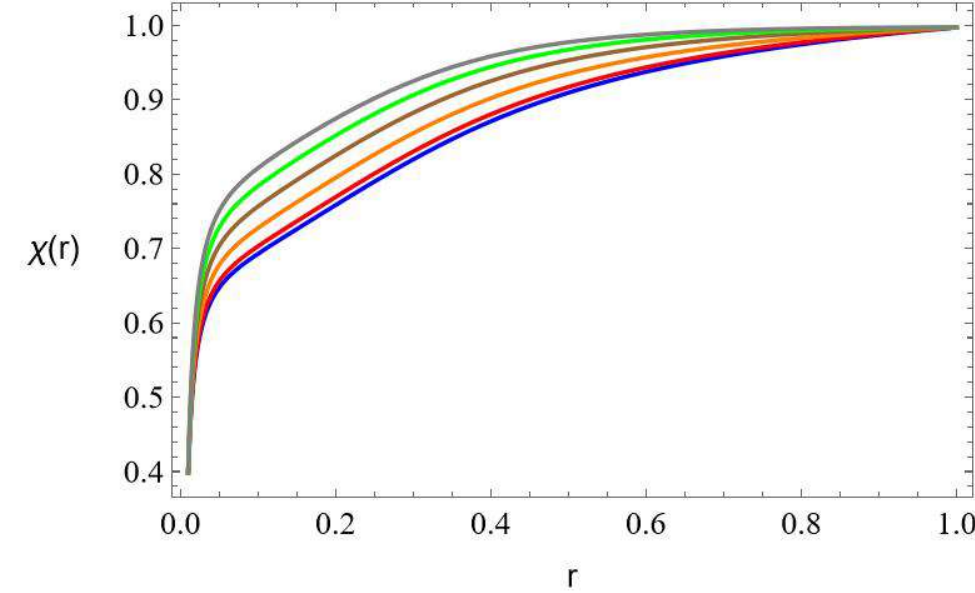} \includegraphics[width=0.56\textwidth]{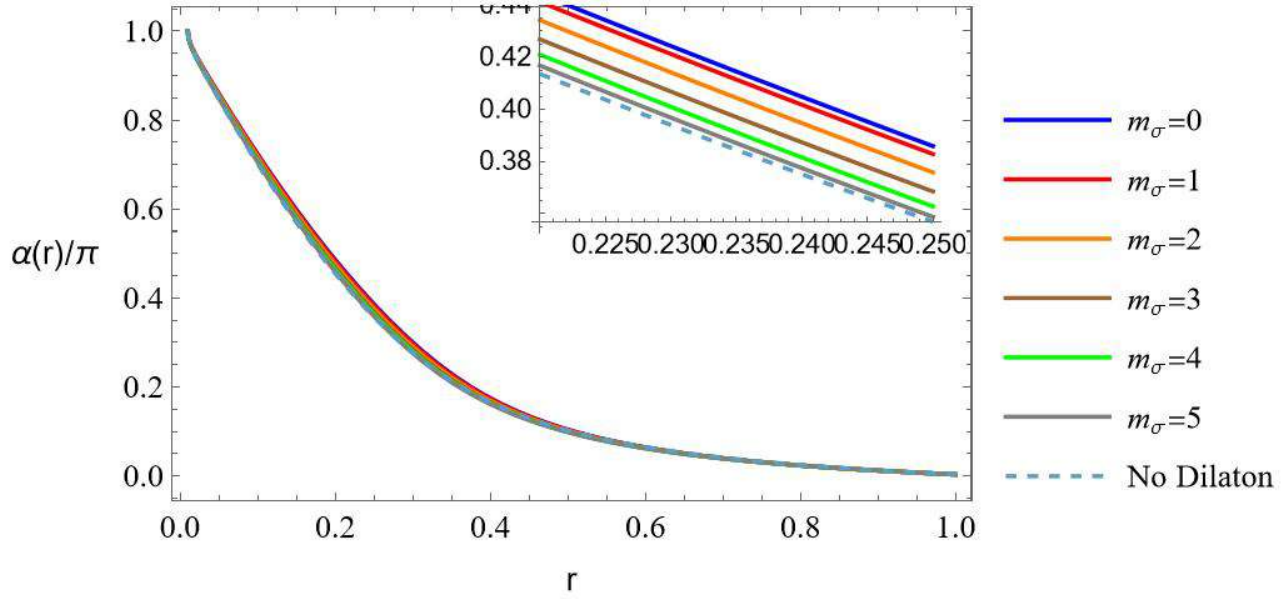} 
	\caption{Dilaton $\chi(r)$ (\emph{left}) and $\alpha(r)/\pi$ (\emph{right}) profiles for $\Delta=1$ and $m_\sigma=1,2,3,4,5$. The initial condition for the dilaton is $\chi(0)=2/5$.}
	\label{massless1}
\end{figure}

\begin{figure}[t!]
\centering
\includegraphics[width=0.48\textwidth]{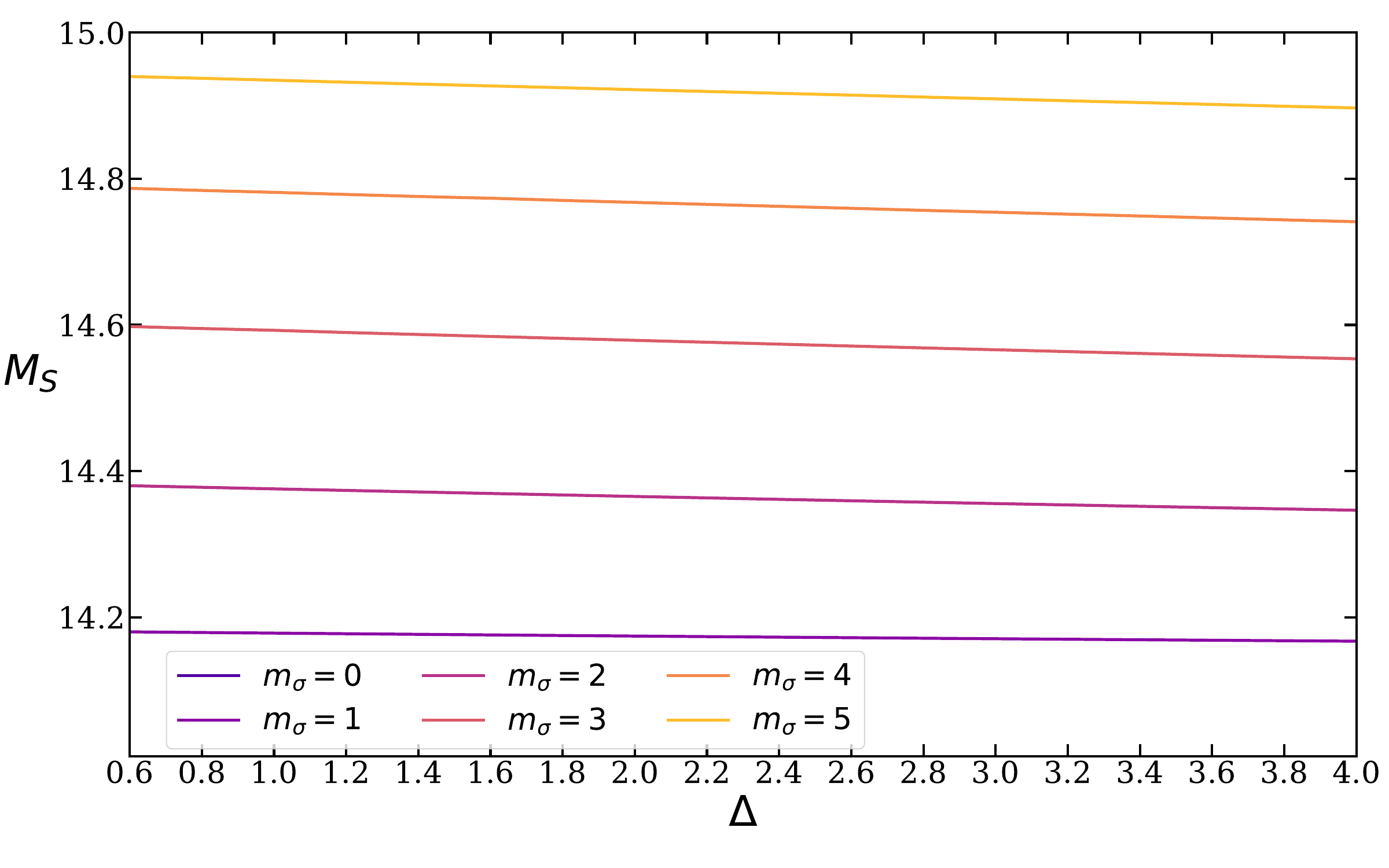} \includegraphics[width=0.48\textwidth]{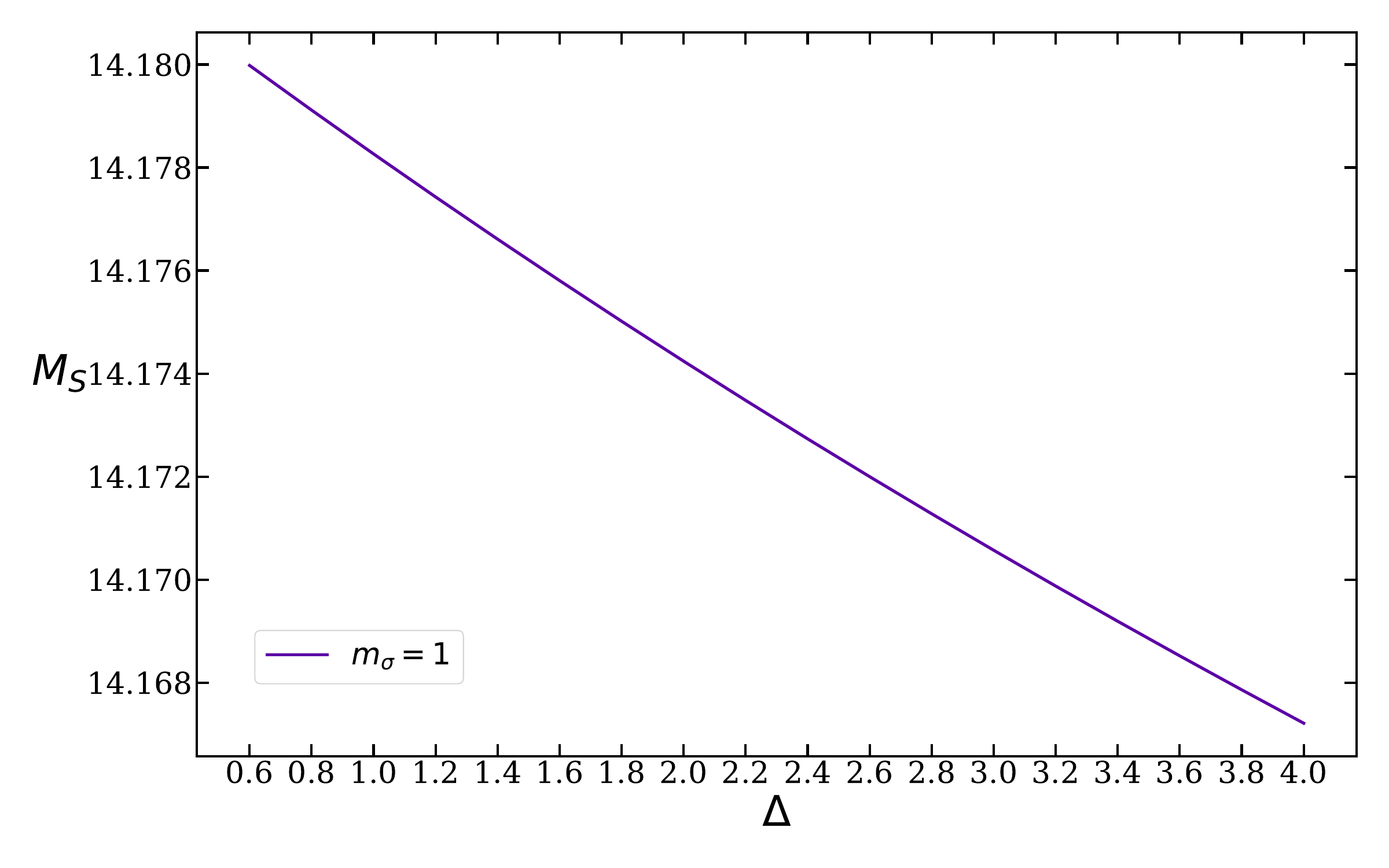} 
	\caption{Skyrmion mass $M_S$ as a function of $\Delta$ for $m_\sigma=0, 1,2,3,4,5$ (\emph{left}) and a detail of the $m_\sigma=1$ line (\emph{right}).} 
	\label{masslessmass}
\end{figure}

\begin{figure}[t!]
\centering
\includegraphics[width=0.5\textwidth]{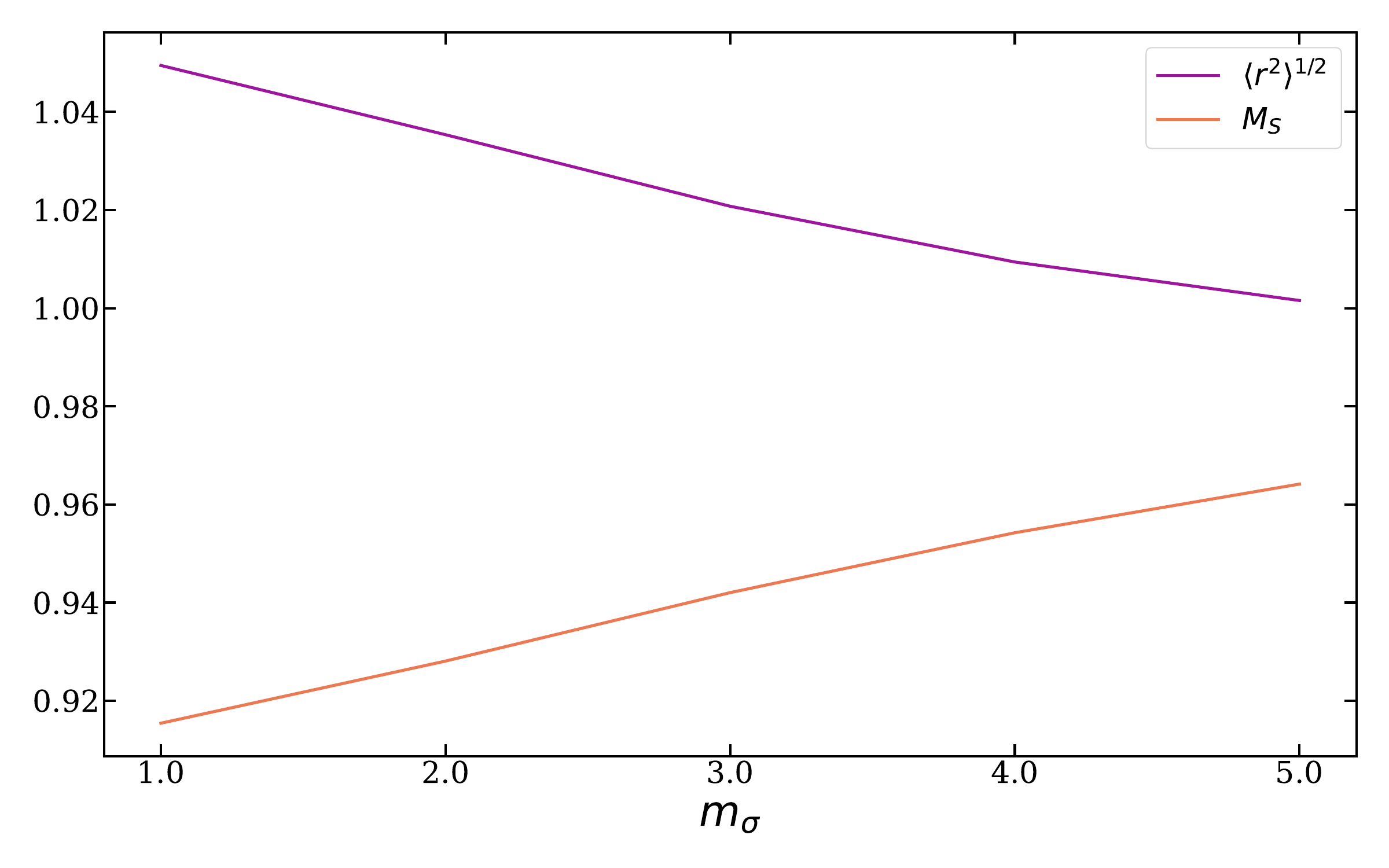} 
	\caption{Skyrmion mass $M_S$ and root mean square radius $\langle r^2 \rangle^{1/2}$ normalized at their values in the absence of the dilaton ($M_\text{Skyrme}= 15.49$ and ($\langle r^2 \rangle^{1/2}_\text{Skyrme}= 0.213$) as a function of $m_\sigma$.} 
	\label{masserad}
\end{figure}

We conclude our numerical analysis with a brief discussion of the solitonic solution in the chiral limit $m_\pi = 0$. Here the vacuum state occurs for $U=\mathbb{1}$ and $\sigma=0$. Therefore, the BC for $\chi(r)$ at infinity no longer depends on the value of $m_\sigma$ and $\Delta$.  Keeping the same numerical values adopted in Sec.\ref{numeri} we solve the EOMs for $\Delta=0.6,0.8, \dots, 4$ and $m_\sigma=1,2,3,4,5$. The first observation is that all the solitonic properties are nearly independent on $\Delta$ for any value of $m_\sigma$. This is a direct consequence of the fact that the dilaton ground state vanishes for any $\Delta$. The dilaton $\chi(r)$ and  $\alpha(r)/\pi$ profiles are displayed in Fig.\ref{massless1} for $\Delta=1$ and $m_\sigma=1,2,3,4,5$. Analogously to $m_\pi \neq 0$ case the dilaton profile progressively flattens as we increase its mass exhibiting a faster convergence to unity. At the same time, the associated $\alpha(r)/\pi$  profile shows almost no dependence on $m_\sigma$. The baryon mass $M_S$ is shown in Fig.\ref{masslessmass}. The dependence on $\Delta$ is similar ($M_S$ gets smaller as $\Delta$ increases) but much weaker than in the $m_\pi \neq 0$ case for all values of $m_\sigma$. For instance, for $m_\sigma=1$ the relative variation of the solitonic mass along the whole range of values of $\Delta$ is only $\sim 0.1\%$, to be compared with $\sim 13\%$ obtained for $m_\pi=140/93$ in Sec.\ref{numeri}. Moreover, the dependence on $m_\sigma $ of both $M_S$ and $\langle r^2 \rangle^{1/2}$ is such that now the solitonic mass increases and the radius decreases towards the dilaton-decoupled limit as shown in  Fig.\ref{masserad}. In the case of a non-vanishing $m_\pi$ it is the solitonic mass that decreases and the radius increases. 

\subsection{Analytical solution in the massless case}

In the case of the chiral Lagrangian alone, the spherical hedgehog ansatz leads to a differential equation that can only be solved using numerical methods. Quite remarkably, when the coupling with the dilaton is not neglected it is possible to find an analytical solution for a special value of the couplings $f$ and $K$ due to an interesting phenomenon, namely "BPS-like equations without a BPS bound". In fact, when $f^2 K=7/8$, the second order field equations can be reduced to a first order system even though there is no BPS bound on the energy of the system. 
The equations of motion \eqref{EOM1} and \eqref{EOM2} in the absence of mass and Skyrme term reduce to $2$ non-linear coupled ODEs
\begin{align}
r^2 \alpha ''    -2 r \alpha '\left(f r \sigma'-1\right)-\sin (2 \alpha)&=0  \ ,\label{ANWalpha}\\
r^2\sigma'' 
+r \left(f K r \alpha '^2+\sigma' \left(2-f r \sigma '\right)\right)
+2 f K \sin ^2(\alpha )&=0  \ .\label{ANWdilaton}
\end{align}
For later convenience, we change variables as $\rho=\log r$ and $\frac{\partial \sigma}{\partial \rho} \to \frac{\partial \sigma}{\partial \rho} +\frac{1}{2 f}$ and rewrite the EOMs as 
\begin{align} \label{rewri}
   \Ddot{\alpha} -2 f \dot \alpha \dot \sigma - \sin (2 \alpha ) &=0   \ ,\\ \label{rewri2}
\Ddot{\sigma}-f \dot \sigma ^2+f K \dot \alpha ^2+2 f K \sin ^2(\alpha)+ \frac{1}{4 f}&=0  \ ,
\end{align}
where the dot denotes the derivative with respect to $t$. Noticeably, when $f^2K=\frac{7}{8}$ we can find a first order system implying the EOMs
\begin{align}
   \alpha'&=\frac{\sqrt{2} }{r}\cos (\alpha )  \ , \\
   \sigma'&=-\frac{\sqrt{2} }{f r}\sin (\alpha (r))+\frac{1}{2 f r}  \ .
\end{align}
We can therefore solve for the analytical solution of the EOMs that read
\begin{align}
   \alpha_\pm(r) &= n \pi  +2 \tan ^{-1}\left(1-\frac{2 c_1}{c_1\pm\left(\frac{r}{r_0}\right)^{\sqrt{2}}}\right) \,, \qquad n \in \mathbb{Z}  \ , \\
   \sigma_\pm(r) &=\frac{1}{f}\left\{c_2+\frac{1}{2} \log \left(\frac{r}{r_0}\right)+\log \left[\cos \left(2 \tan ^{-1}\left(1-\frac{2 c_1}{c_1\pm\left(\frac{r}{r_0}\right)^{\sqrt{2}}}\right)\right)\right]\right\}  \ ,
\end{align}
where the plus and minus signs apply for $c_1 >0$ and $c_1<0$, respectively. The topological charge associated with this solution is $B=-1$. Conventional solitonic solutions of the chiral Lagrangian with negative winding numbers are called \emph{antiskyrmions}. However, the energy density of the solution reads
\begin{equation}
    \epsilon = \frac{e^{-2c_2}}{8 c_1^2 f^2 r^2} \left(\frac{r}{r_0}\right)^{-2 \sqrt{2}-1} \left[\left(\sqrt{2}+4\right) c_1^4-\left(\sqrt{2}-4\right) \left(\frac{r}{r_0}\right)^{4 \sqrt{2}}\right]  \ ,
\end{equation}
and diverges at the origin as $r^{-2 \sqrt{2}-3}$ due to the dilaton profile $\chi(r)$ being singular in $r=0$. Moreover, even excluding the singularity at $r=0$, the total energy would still diverge since the energy density does not decrease fast enough.


\section{Dilaton augmented Crystals} \label{boxer}
In this section, we first review the strategies to find analytical solutions describing arrays of tubes of baryons with non-trivial topological charge on flat space-time \cite{Canfora:2018rdz, Barriga:2021eki,Barriga:2022izc,Canfora:2023pkx} and then we will generalize the solutions when the dilaton is present. 

\subsection{Crystals without the dilaton}
\begin{figure}[t!]
     \centering
     \begin{subfigure}[h!]{0.4\textwidth}
         \centering
         \includegraphics[width=1\textwidth]{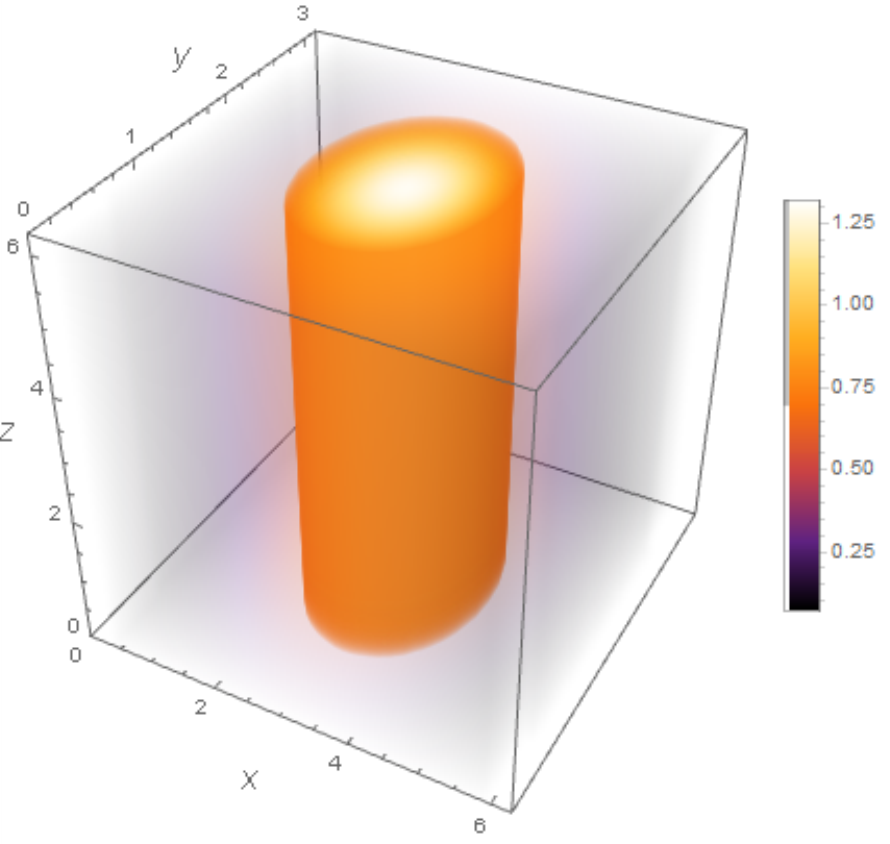}
         \label{fig:B=1}
     \end{subfigure}
     \begin{subfigure}[h!]{0.4\textwidth}
         \centering  \includegraphics[width=1\textwidth]{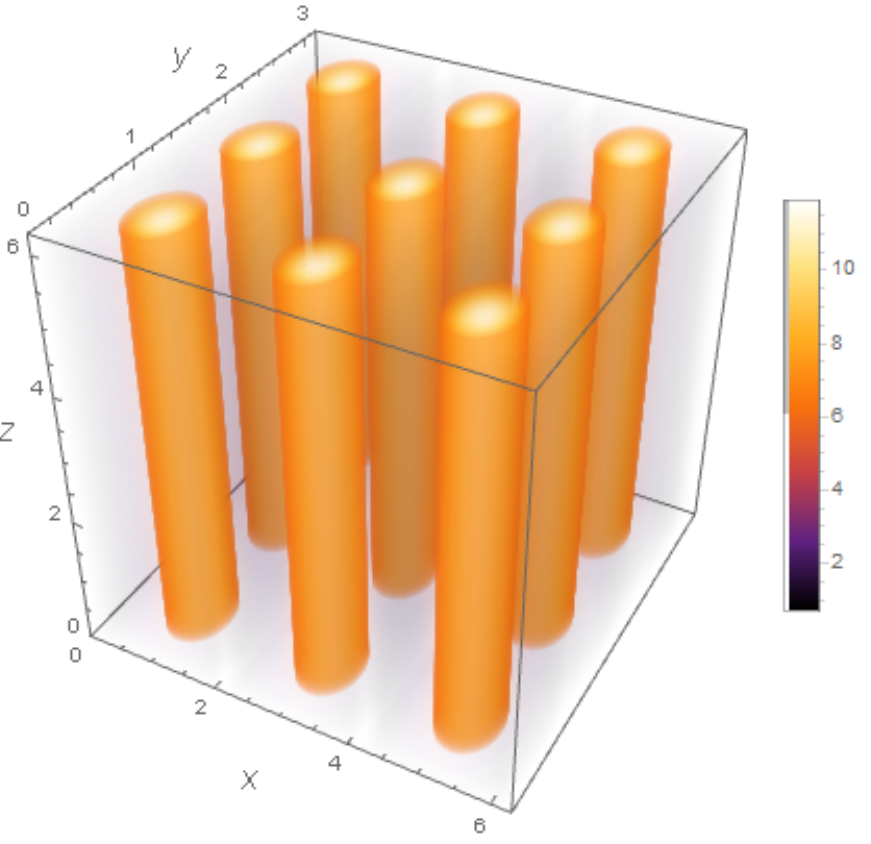}
         \label{fig:B=9}
     \end{subfigure}
        \caption{Energy density of the solution \eqref{cifamo2spaghi} for $p=q=n=1$ (\emph{left}) and $p=q=n=3$ (\emph{right}). In both cases, we have assumed $K=L_x=L_z=1$ and $L_y=2$.}
        \label{fig:three graphs}
\end{figure}

We consider the action \eqref{NLSMaction} in the absence of the Skyrme and mass terms and retain the same general parameterization for $U(x)$ given in \eqref{U}. The equations of motions \eqref{NLSMeq} constitute the following set of $3$ non-linear coupled PDEs  
\begin{align}
    -\square \alpha+\sin (\alpha) \cos (\alpha)\left(\nabla_\mu \Theta \nabla^\mu \Theta+\sin ^2 (\Theta) \nabla_\mu \Phi \nabla^\mu \Phi\right)&=0  \ ,\\
\sin ^2(\alpha) \square \Theta+2 \sin (\alpha) \cos (\alpha) \nabla_\mu \alpha \nabla^\mu \Theta-\sin ^2(\alpha) \sin (\Theta) \cos (\Theta) \nabla_\mu \Phi \nabla^\mu \Phi&=0  \ ,\\
\sin ^2(\alpha) \sin ^2(\Theta) \square \Phi+2 \sin (\alpha) \cos (\alpha) \sin ^2(\Theta) \nabla_\mu \alpha \nabla^\mu \Phi+2 \sin ^2(\alpha) \sin (\Theta) \cos (\Theta) \nabla_\mu \Theta \nabla^\mu \Phi&=0\ ,
\end{align}
where we have not yet made any ansatz for the parameters of $U$ in terms of the coordinates. Here the topological charge density $\rho_{B} \equiv J^{0}$ takes the following form
\begin{align}
    \rho_{B}=12 \sin ^2 (\alpha) \sin (\Theta) d \alpha \wedge d \Theta \wedge d \Phi\ .
\end{align}
From this expression, we observe that in order to have $d \alpha \wedge d \Theta \wedge d \Phi \neq 0 $ the fields $\alpha$, $\Theta$, and $\Phi$ must be independent. Additionally, in order to decouple the previous set of equations we impose
\begin{align} \label{condor}
    \nabla_\mu \Phi \nabla^\mu \alpha=\nabla_\mu \alpha \nabla^\mu \Theta=\nabla_\mu \Phi \nabla^\mu \Phi=\nabla_\mu \Theta \nabla^\mu \Phi=0  \ .
\end{align}
We proceed by confining the system into a box described by the following line element
\begin{align}
    ds^2=-dt^2+L_x^2 dx^2+L_{y}^2 dy^2+L_{z}^2 dz^2 \,,\label{metric}
\end{align}
where $\{x,y,z\}$ are dimensionless coordinates having the range
\begin{align}
    0 \leq x \leq 2 \pi, \quad 0 \leq y \leq \pi, \quad 0 \leq z \leq 2 \pi  \ .
\end{align}
The condition \eqref{condor} is then realized by considering the following ansatz 
\begin{align}
    \alpha=\alpha(x), \quad \Theta=q y, \quad \Phi=p\left(\frac{t}{L_z}-z\right), \quad q=2 v+1, \quad v,p \in \mathbb{Z}   \ ,\label{Ufields}
\end{align}
such that the equations of motion reduce to a single integrable ODE for the profile $\alpha(x)$
\begin{equation} \label{cifamo2spaghi}
   \partial_{x}\left[\frac{1}{2}(\alpha')^2-V(\alpha)-E_{0}\right]=0,\quad V(\alpha)=-\frac{q^2 L_{x}^2}{2 L_{y}^2}\cos(2\alpha)  \ ,
\end{equation}
where $E_0$ is an integration constant that depends on the boundary conditions for $\alpha$. The explicit solution can be written as the amplitude of a certain Jacobi elliptic function. 
The corresponding topological charge density reads
\begin{align} \label{topdensity}
    \rho_B=\frac{3q p \sin (qy  ) }{L_x L_{y} L_{z}}\partial_{x} (2 \alpha- \sin (2 \alpha ) )  \ .
\end{align}
By considering the boundary conditions for $\alpha$
\begin{equation}
    \alpha(2\pi)- \alpha(0) = n \pi \,, \qquad n\in \mathbb{Z}  \ ,
\end{equation}
then the topological charge is given by  $B= n p$. Therefore the baryon charge can be an arbitrary integer number. This solution describes stationary soliton crystals with the shape of ordered arrays of baryonic tubes carrying topological charge. Here the integer $p$ can be interpreted as the baryonic charge per unit of length in the $z$-direction per tube. This can be seen by looking at the energy density of the solution $\epsilon \equiv T_{00}$ 
\begin{equation} \label{Edensnodil}
    \epsilon =\frac{1}{2} K \left(\frac{\alpha '^2}{L_{x}^2}+\sin ^2(\alpha)  \left(\frac{q^2}{L_{y}^2}+\frac{2 p^2 \sin ^2(q y )}{L_{z}^2}\right)\right)  \ ,
\end{equation}
which is depicted in Fig.\ref{fig:three graphs}. 

\subsection{Crystals with the dilaton}

\begin{figure}[t!]
\centering
\includegraphics[width=0.48\textwidth]{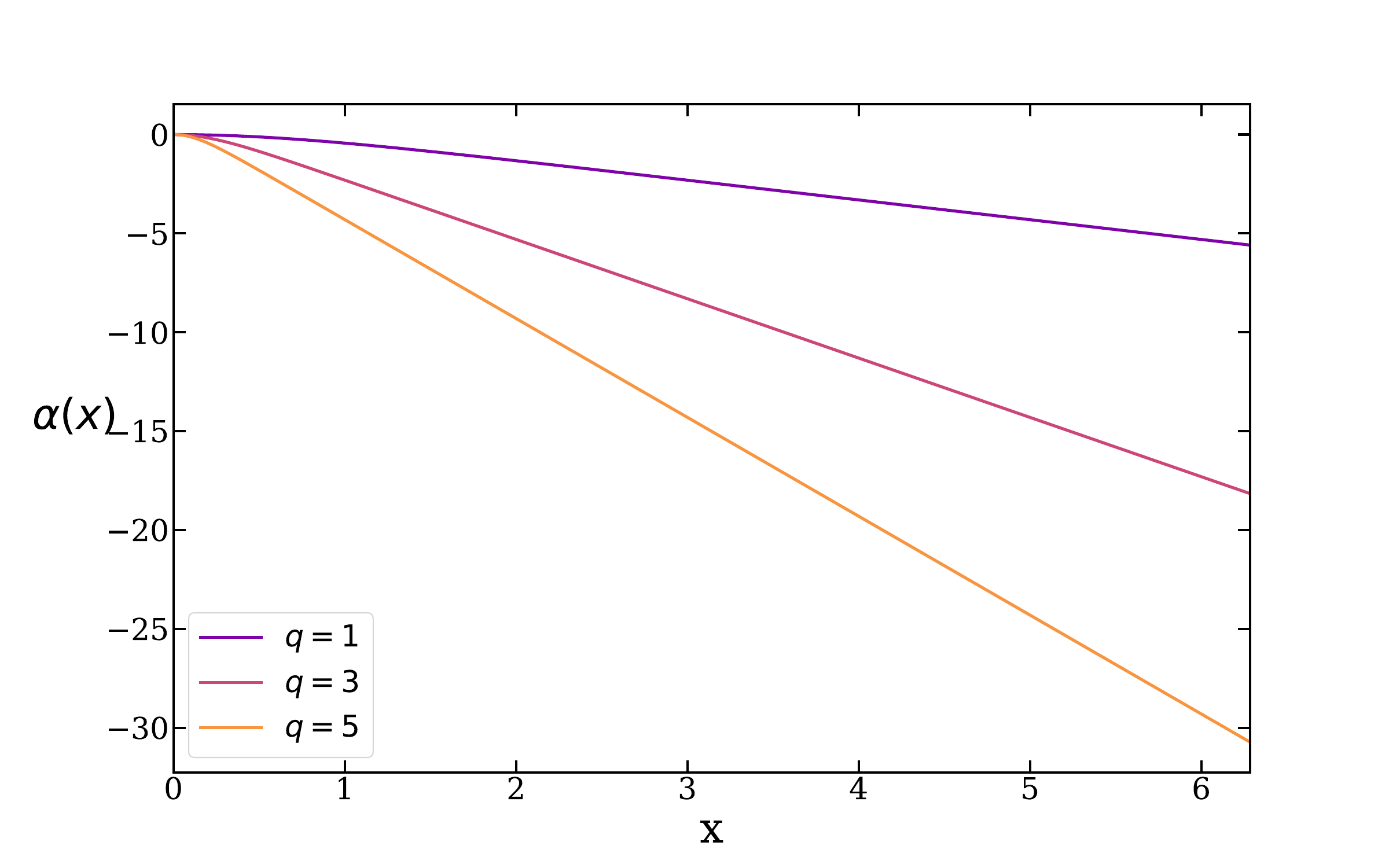} \includegraphics[width=0.48\textwidth]{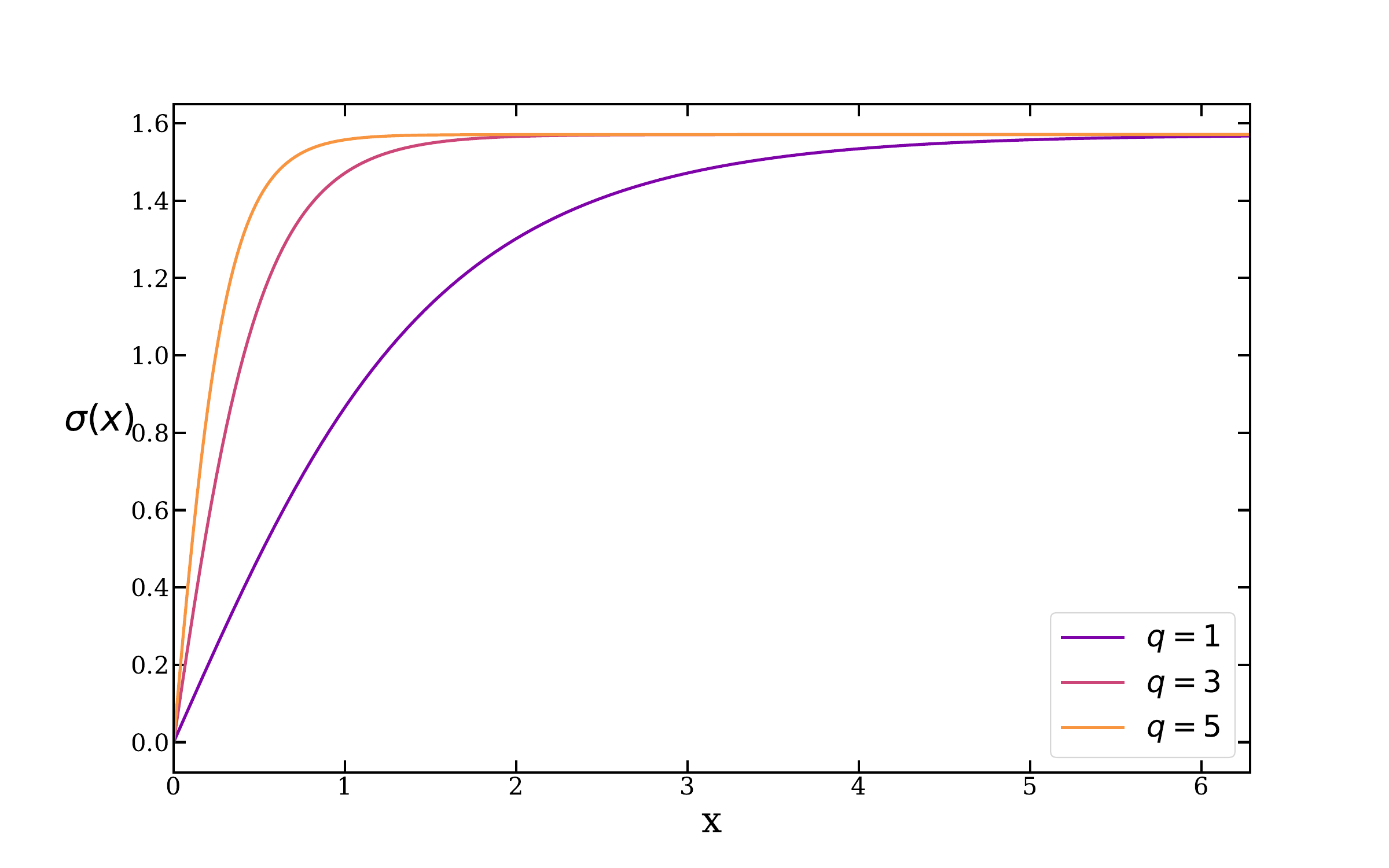} 
	\caption{The profiles $\alpha(x)$ (\emph{left}) and $\sigma(x)$ (\emph{right}) for $q=1,3,5$.} 
	\label{profilespasta}
\end{figure}

Here we study the effect of the dilaton on the solutions considered in the previous section. In particular, we consider $U$ given by eqs.\eqref{U} and \eqref{Ufields} and we assume that the dilaton field depends only on the $x$ coordinate, i.e. $\sigma=\sigma(x)$. Consequently, the field equations reduce to two coupled non-linear ODE for $\alpha(x)$ and $\sigma(x)$
\begin{align} \label{ecrist1}
   \alpha ''-2 f \alpha ' \sigma '-\frac{L^2_x}{2 L^2_y} q^2 \sin (2 \alpha )-  L^2_x m_{\pi }^2  \sin (\alpha) e^{-f (y-2) \sigma }&=0  \ ,\\ \label{ecrist2}
\sigma ''+f K \alpha '^2-f \sigma '^2+\frac{L^2_x}{L^2_y}f K q^2 \sin ^2(\alpha )-L^2_x e^{2 f \sigma} \partial_\sigma V(\sigma) -K f L^2_x m_{\pi }^2 y \cos (\alpha ) e^{-f (y-2) \sigma }&=0 \ .
\end{align}
The topological charge is given by eq.\eqref{topdensity} while the energy density evaluated on this ansatz reads
\begin{equation}
\epsilon =\frac{1}{2} K e^{-2 f \sigma} \left(\sin ^2(\alpha)  \left(\frac{q^2}{L_{y}^2}+\frac{2 p^2 \sin ^2(q y)}{L_{z}^2}\right)+\frac{\alpha '^2}{L_{x}^2}\right)+\frac{e^{-2 f \sigma} \sigma'^2}{2 L_{x}^2} -K m_{\pi }^2 \cos (\alpha ) e^{-f y \sigma}+ V(\sigma)  \,, 
\end{equation}
and reduces to Eq.\eqref{Edensnodil} for $\sigma=m_\pi=0$.
We proceed by focusing on the massless case $m_{\pi}=m_{\sigma}=0$, which unlike the massive case can be partially addressed analytically. 
In fact, it is quite remarkable that at the special parameter point, $f^2 K=1$ the field equations \eqref{ecrist1} and \eqref{ecrist2} can be reduced to a system of first order ODEs given by
\begin{align} \label{FO}
    \alpha'=q \cos (\alpha)  \ , \qquad
    \sigma'=-\frac{q}{f} \sin (\alpha)  \ ,
\end{align}
where for the sake of simplicity we have chosen $L_{x}=L_{y}$.
This can also be seen by noting that the EOMs \eqref{ecrist1} and \eqref{ecrist2} coincide with the EOMs \eqref{rewri} and \eqref{rewri2} for the hedgehog ansatz up to a constant term $\frac{1}{4 f}$ in eq.\eqref{rewri2}. The latter shifts the special combination of parameters from $f^2K=7/8$ to $f^2K=1$  reducing the second order differential equations of motion to the above set of first order ones yielding the solutions 
\begin{align} \label{linguine}
    \alpha(x) & =2 \arctan \left(\tanh \left(\frac{1}{2} (c_1+q x)\right)\right) \ ,  \\ 
   \sigma(x) & =-\frac{1}{f} \left[ c_2 +2 \arctanh \left(\tanh ^2\left(\frac{1}{2} (c_1+q x)\right)\right)\right] \ ,
\end{align}
where $c_{1}$ and $c_{2}$ are the integration constants. 
The behavior of these solutions is shown in Fig.\ref{profilespasta}.
We can restore all the integration constants stemming from the second order original EOMs and write the solution as 
\begin{align}
   \alpha(x) &= n \pi + \arctan \left(d_1 e^{q x}+d_2 e^{-q x}\right) \,, \qquad n \in \mathbb{Z}  \ ,\\ \sigma(x) &= -\frac{1}{f}\left(c_2 +\frac12 \log\left(\left(d_1 e^{q x}+d_2 e^{-q x}\right)^2+1\right)\right)\ .
\end{align}

When the parameter $p$ is an integer it is possible to have an integer topological charge $B$ only by considering a box that has an infinite length in the $x$ direction. In such a case we have $B=p$ but the total energy carried by the solution diverges. This is in net contrast with the solution in the absence of the dilaton field, where for $p \in \mathbb{Z}$ one can confine an arbitrary amount of topological charge in a finite-size box and the corresponding energy is finite. On the other hand, we can relax the condition  $p \in \mathbb{Z}$ and determine the value of $p$ such that a certain amount of topological charge is confined in a finite-size box. Without losing generality we consider $0 \le x \le 2 \pi$ and impose BCs $\alpha(0)=0$ and $\alpha(2\pi)=s$ with  $0<s<\pi/2$. The topological charge reads

\begin{equation}
  B =  -2\int_{\alpha (0)}^{\alpha (2 \pi )} \frac{ p }{\pi } \sin ^2(\alpha)\, d\alpha   \ .
\end{equation}
Hence $B \in \mathbb{Z}$ can be achieved when 
\begin{equation}
    p=\frac{2 \pi  B }{\sin(2 s)-2 s}\ .
\end{equation}

\begin{figure}[t!]
\centering
\includegraphics[width=0.32\textwidth]{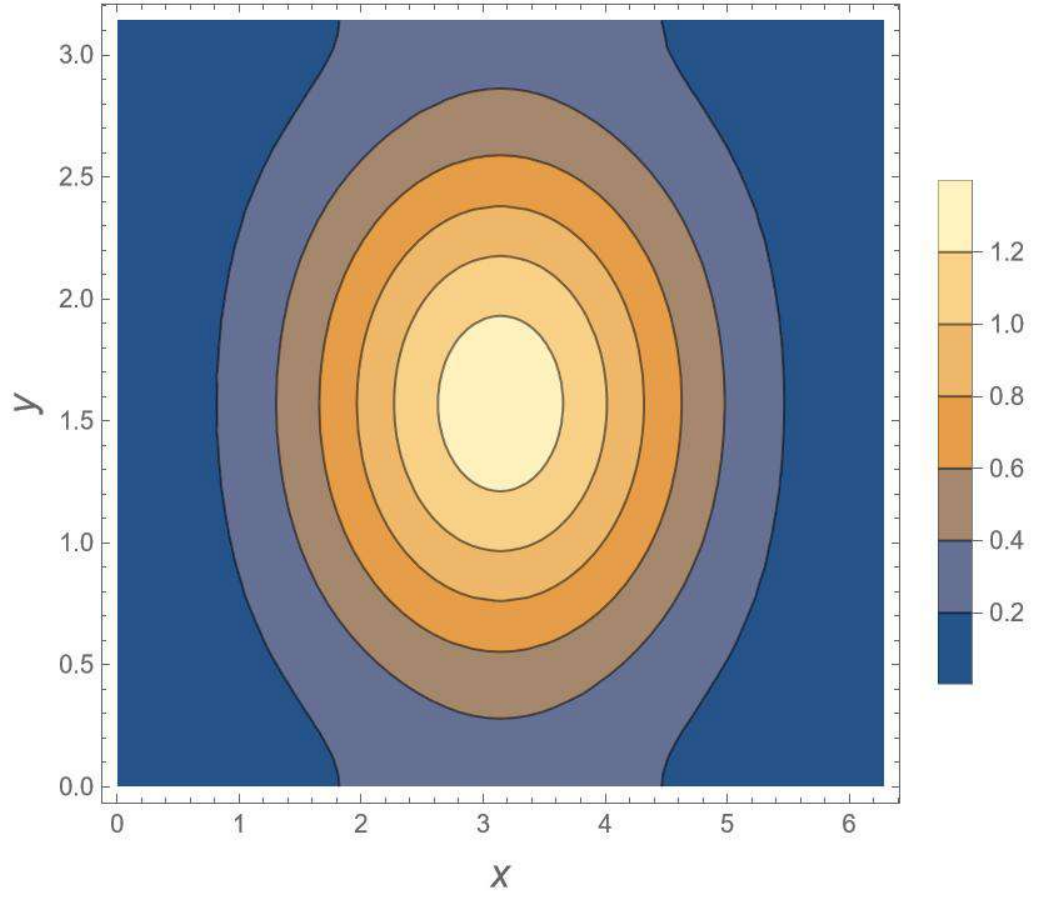} \includegraphics[width=0.32\textwidth]{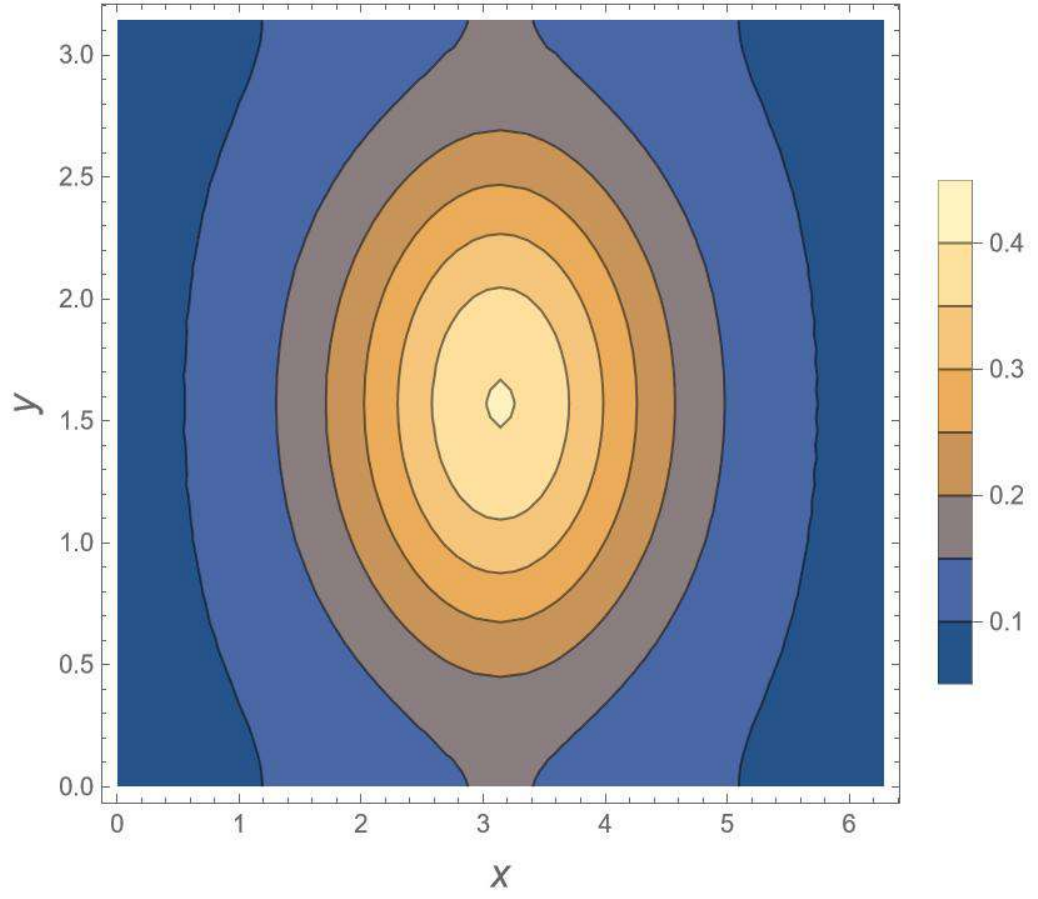} \includegraphics[width=0.325\textwidth]{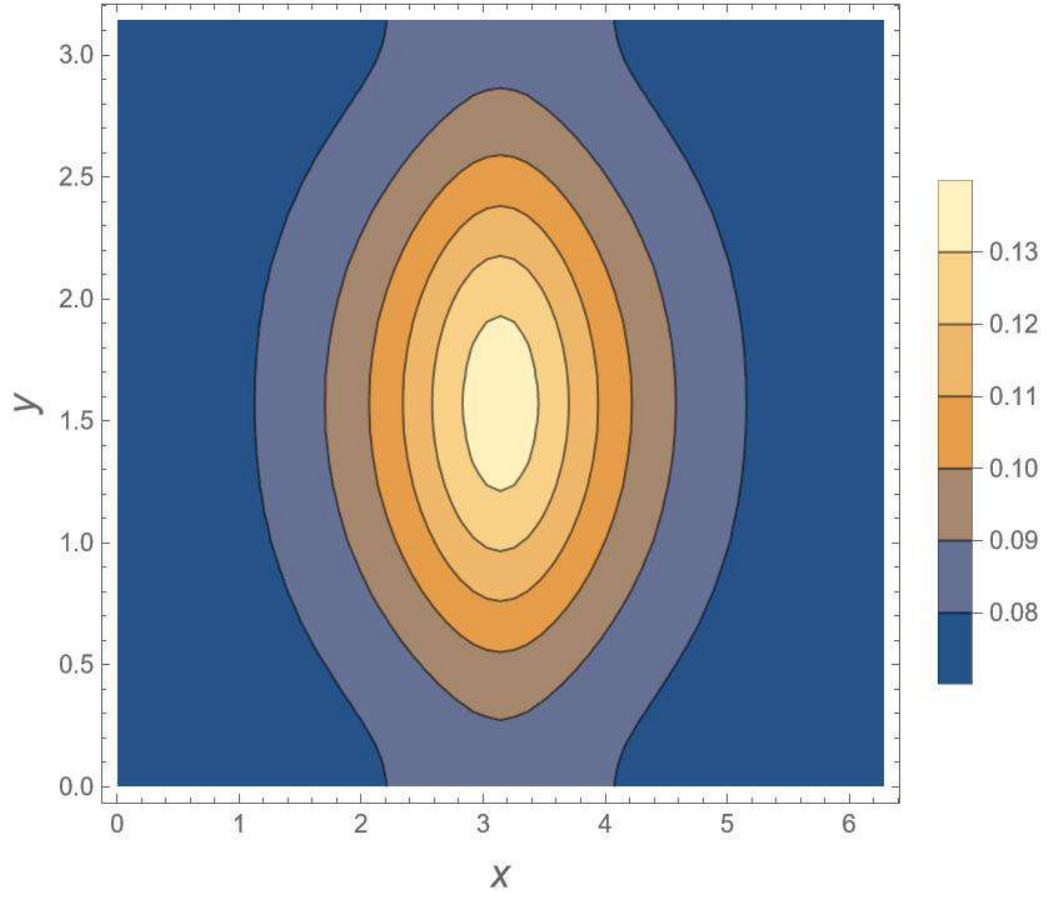}  \newline
\includegraphics[width=0.32\textwidth]{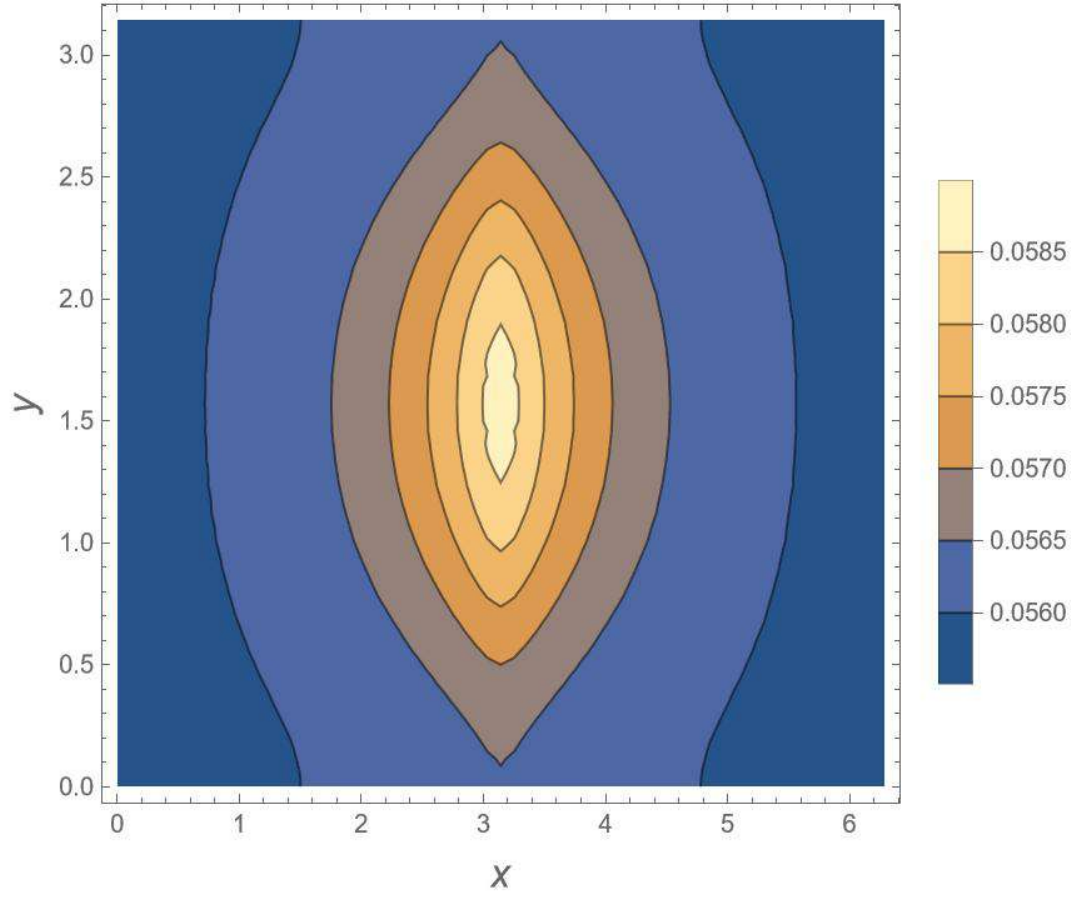} \includegraphics[width=0.35\textwidth]{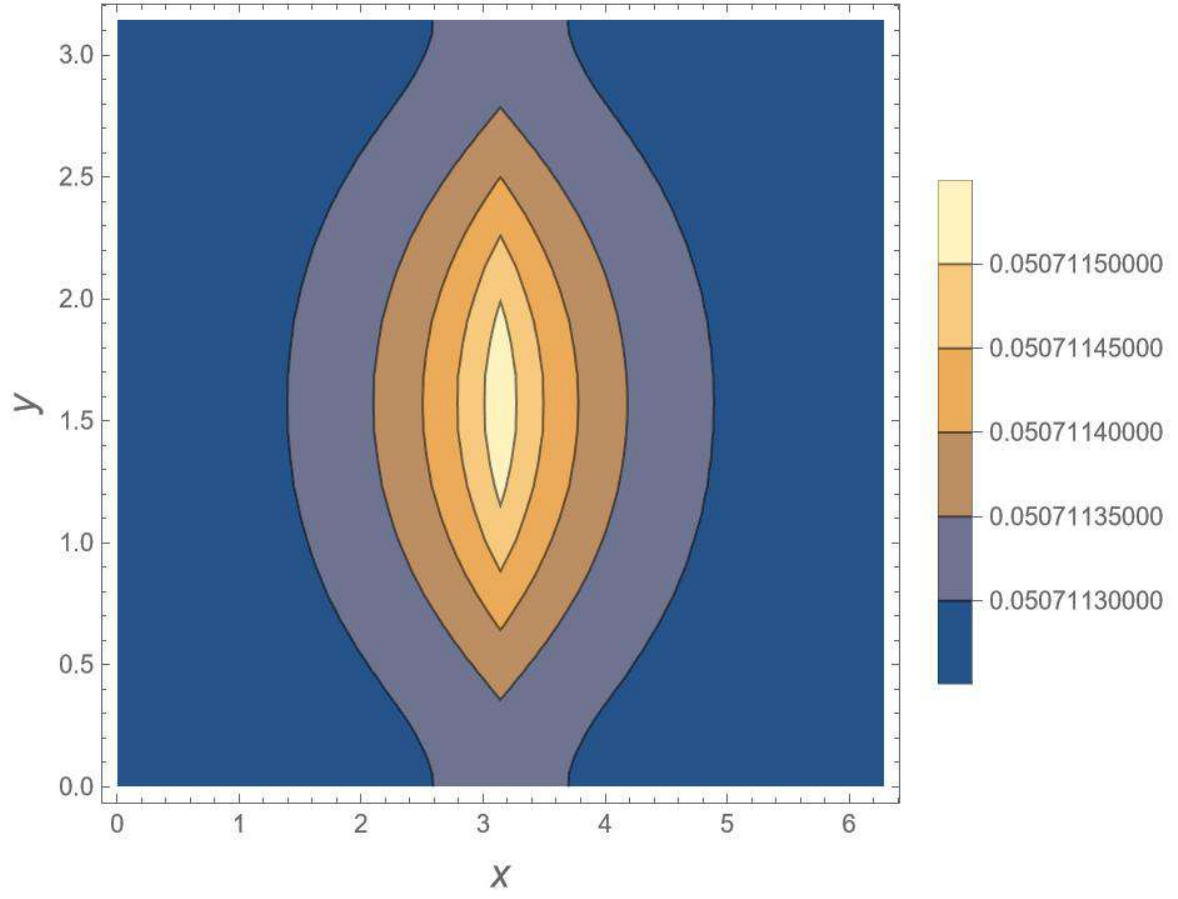} 
	\caption{Contour plot of the energy density in the $x-y$ plane for $m_\pi = m_\sigma= 0$ varying $X= f^2 K$. From left to right and top to bottom we have $X=0, 0.3, 0.6, 0.9, 1$. As $X$ goes from $0$ to $1$ the results interpolate between the analytical solution found in \cite{Canfora:2018rdz, Barriga:2021eki, Barriga:2022izc,Canfora:2023pkx} in the absence of the dilaton field and the solution \eqref{linguine}.}
	\label{1Dplot}
\end{figure}

We conclude this section by numerically investigating the behavior of the solution for different values of $X \equiv f^2 K$. The corresponding energy density is shown in Fig.\ref{1Dplot}. At $x=0$ the EOMs decouple; the dilaton solution is $\sigma(x)= c_1-\log (c_2+x)$, $\{c_1, c_2\} \in \mathbb{R}$ while $\alpha(x)$ solves eq.\eqref{cifamo2spaghi} yielding a field configuration with localized energy and integer topological charge. On the other hand, for $X=1$ we have the analytical solution \eqref{linguine}, whose energy grows indefinitely for increasing $x$. The physical impact of this solution will be discussed below.

\subsection{Isospin-Charge Separation} 
\begin{figure}[t!]
    \centering
    \includegraphics[ width=0.5\textwidth]{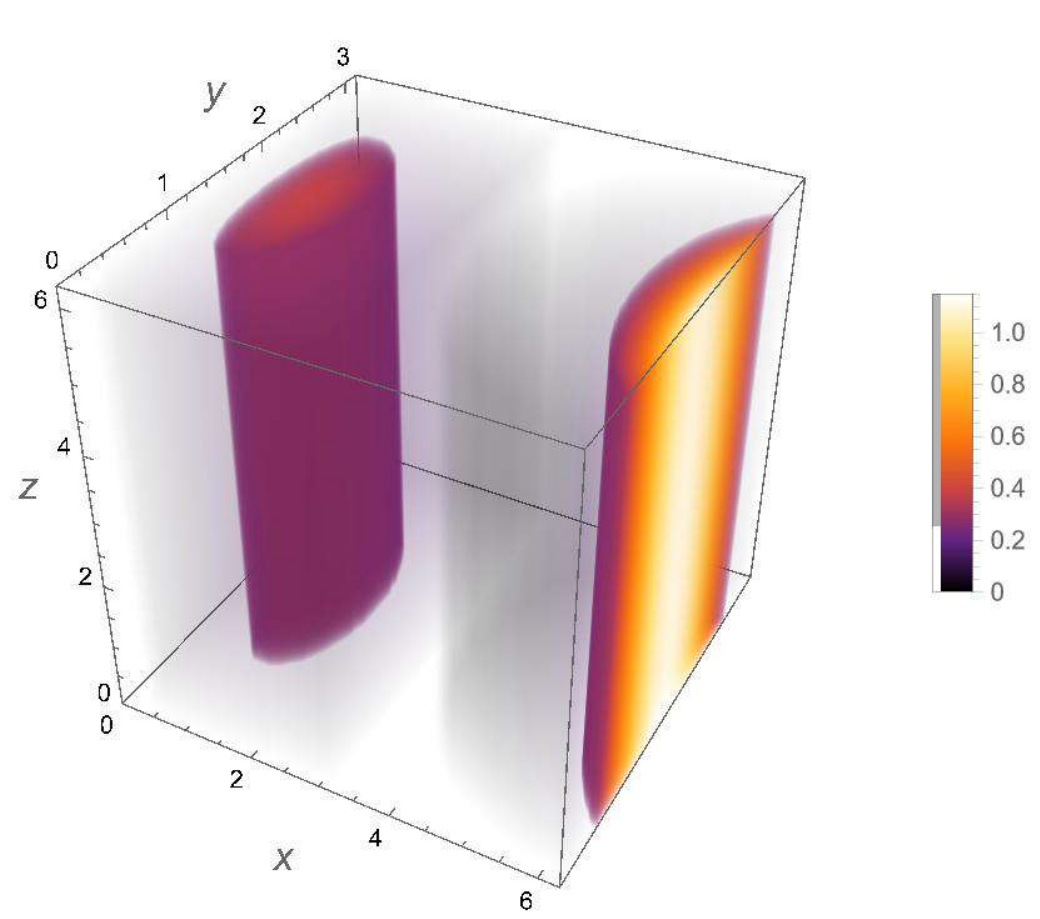}
    \caption{Topological  (\emph{left}) and isospin  (\emph{right}) charge densities for $L=B=q=1$ and $\alpha(2 \pi)=s=\frac{\pi}{4}$.}
    \label{fig:TCspaghetti1}
\end{figure}

A further intriguing feature of the current effective chiral Lagrangian dressed with the dilaton is the following. In the solutions without dilaton \cite{Canfora:2018rdz, Barriga:2021eki, Barriga:2022izc,Canfora:2023pkx} the peaks corresponding to the local maxima of the topological charge density are in the same locations as the peaks corresponding to the local maxima of the isospin charge density (as one can also check via a direct computation). On the other hand,   the isospin current (which is the Noether current associated with the $SU(2)$ isospin rotations of the theory) is deformed by the presence of the dilaton  while the definition of the topological charge density does not change (being the usual baryon charge). This implies that the local maxima of these two relevant quantities no longer coincide. Indeed, by setting for simplicity $L_x=L_y=L_z\equiv L$, the topological charge density reads
\begin{equation}
    \rho_B= \frac{12 p q}{L^3}\sin (y  q) \alpha '(x) \sin ^2(\alpha (x))  \ ,
\end{equation}
while the isospin charge density is  
\begin{equation}
    \rho_I=K \operatorname{Tr}(R^{0}\tau_3) e^{-2 f \sigma(x)}=\frac{2 K p}{L}e^{-2 f \sigma(x)} \sin ^2(y q) \sin ^2(\alpha (x))  \ .
\end{equation}
In Fig.\ref{fig:TCspaghetti1} we compare the two charge densities to illustrate the different positions of the peaks.

 This "separation" of the peaks of Baryonic and isospin charge densities could open the interesting possibility of defining the analogue of the well-known phenomenon in condensed matter physics called spin-charge separation \cite{PhysRevLett.125.190401, Lerda:1994kb} and references therein. In the condensed matter version of this phenomenon, due to the presence of strong correlations, spin and charge may cease to be tied generating independent branches of excitations. In the present case, it could be possible to observe isospin-baryon charge separation phenomenology thanks to the presence of the dilaton which can support excitations that change directly the isospin and only indirectly the baryon charge density.  
 
\section{Dilaton augmented Branes}

\begin{figure}[t!]
\centering
\includegraphics[ width=0.5\textwidth]{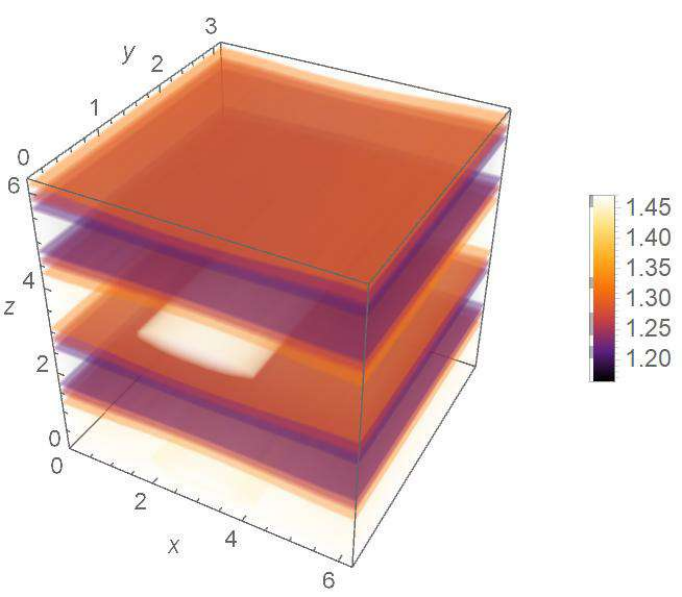}
	\caption{Energy density of a solution of the EOMs \eqref{confo} exhibiting a layer structure. We choose $F=F_+(\frac{t}{L_z}+z) = \cos(\frac{t}{L_z}+z)$.}
	\label{tortelliplot}
\end{figure}

In \cite{Alvarez:2017cjm,Ayon-Beato:2015eca,Canfora:2022jmh,Canfora:2023zmt,Canfora:2023pkx}, another family of analytical solutions of the massless chiral Lagrangian has been discovered. These are again  obtained by confining the theory in a box according to the metric eq.\eqref{metric} and considering the following ansatz
\begin{equation} \label{lasa}
    U=e^{\frac{1}{2} p y   \tau_3}e^{\frac{1}{4} x \tau_2}e^{\frac{1}{2} \tau_3 F(t,z )} \,, \qquad p \in \mathbb{Z} \ .
\end{equation}
The above represents a particular instance of Euler angles parametrization for $SU(2)$ which implies the following range for $x$ and $y$ \cite{Bertini:2005rc, Cacciatori:2012qi}
\begin{equation}
    0 \le y \le  \pi  \,,  \qquad  0 \le x \le  2\pi  \ .
\end{equation}
Without losing generality we can also consider 
\begin{equation}
 0 \le z \le  2\pi   \ .
\end{equation}
The topological charge density is given by
\begin{align} \label{chart}
   \rho_B &=\frac{3 p}{4 L_x L_y L_z} \sin \left( \frac{x}{2}\right) \partial_z F  \ .
\end{align}
The periodicity of physical observables along with the properties of the Euler angles parameterization fixes the boundary condition satisfied by $ F(t, z)$ as
\begin{equation} \label{beforec}
    F(t, z=0)-  F(t, z=2 \pi ) = \pm 8 \pi q \,,  \qquad q \in \mathbb{Z}  \ .
\end{equation}
As a consequence, the topological charge reads 
\begin{align} \label{chart2}
 B &= -\frac{p}{8 \pi} ( F(t, z=0)-  F(t, z=2 \pi ) ) = p q  \ .
\end{align}
Noticeably, for $m_\pi=0$ the EOMs \eqref{NLSMeq} evaluated on the ansatz \eqref{lasa} reduce to the field equation of a free massless scalar in $d=2$ dimensions \cite{Canfora:2022jmh}
\begin{align} \label{confo}
   \left( \partial_{t}^2 - \frac{1}{L_z^2} \partial_{z}^2 \right) F(t, z) = 0  \ .
\end{align}
The general solution to the previous equations can be decomposed as the sum of two independent modes $F=F_{+}\left(\frac{t}{L_{z}}+z\right) + F_{-} \left(\frac{t}{L_{z}}-z\right)$. The latter can be generically written using the following representation
\begin{align}
F_+&=\phi_+^0+v_+\left(\frac{t}{L_{z}}+z\right)+\sum_{n\neq 0}\left[a_n^+\sin\left(\frac{t}{L_{z}}+z\right)+b_n^+\cos\left(\frac{t}{L_{z}}+z\right)\right]  \ ,\label{rightmode}\\
F_-&=\phi_-^0+v_-\left(\frac{t}{L_{z}}-z\right)+\sum_{n\neq 0}\left[a_n^-\sin\left(\frac{t}{L_{z}}-z\right)+b_n^-\cos\left(\frac{t}{L_{z}}-z\right)\right]  \ ,\label{leftmode}
\end{align}
with coefficients $\phi^0_{\pm}, v_{\pm}, a_n^{\pm}, b_n^{\pm}$.
In this way the topological charge $\eqref{chart2}$ is non zero when $v_{+}-v_{-}\neq 0$, additionally the coefficients $v_{\pm}$ have to be chosen such that 
\begin{align}
F(t, \phi=0)&=\phi_0^{+}+\phi_0^{-}+\left(v_{+}+v_{-}\right) \frac{t}{L_z} \quad \Rightarrow \quad v_{+}+v_{-}=0  \ , \\
F(t, \phi=2 \pi)&=\phi_0^{+}+\phi_0^{-}+\left(v_{+}-v_{-}\right) 2 \pi \quad \Rightarrow \quad v_{+}-v_{-}=4 q  \ .
\end{align}
The energy density of the solution reads
\begin{equation}
    \epsilon = \frac{K}{8}\left[\frac{1}{4}\left(\frac{1}{L_x^2}+\frac{4 p^2}{L_y^2}\right)+\left(\partial_{t}F\right)^2+\frac{1}{L_z^2}\left(\partial_{z}F\right)^2\right]  \ ,
\end{equation}
and is illustrated in Fig.\ref{tortelliplot}. One can see that this solution describes modulated layers of nuclear matter. 

The inclusion of the Skyrme term yields an additional equation
\begin{align}
   \left( \partial_{t}F \right)^2 - \frac{1}{L_{z}^2} \left(\partial_{z}F\right)^2  =\left(\partial_{t}F-\frac{1}{L_{z}}\partial_{z}F\right) \left(\partial_{t}F+\frac{1}{L_{z}}\partial_{z}F\right)=0  \ ,
\end{align}
with solution $F(t, z) = F\left(z\pm\frac{t}{L_z}\right)$. On the other hand, the equations of motion do not admit any solution of the form \eqref{lasa} when $m_\pi \neq 0$.

Equipped with the above we now investigate the impact of the dilaton on this class of topologically non-trivial solutions of the chiral and Skyrme models in the chiral limit $m_\pi = 0$.  Assuming $\sigma= \sigma (t,z)$, the EOMs are
 
\begin{align} \label{laseom}
 \partial_{t}^2F - \frac{1}{L_z^2} \partial_{z}^2F &=0  \ ,\\
  (\partial_{t}F)^2 - \frac{1}{L_z^2} (\partial_{z}F)^2 &=0  \ ,\\
\partial_{z}F\ \partial_z \sigma-L_{z}^2\partial_t F\ \partial_{t}\sigma &=0  \ ,\\
\partial_{t}^2\sigma - \frac{1}{L_z^2} \partial_{z}^2\sigma-f \left(  (\partial_{t}\sigma)^2 - \frac{1}{L_z^2} (\partial_{z}\sigma)^2\right)-\frac{f k}{16}  \left(\frac{4 p^2}{L_{y}^2}+\frac{1}{L_{x}^2}\right) + e^{2 f \sigma} \partial_\sigma V(\sigma)&=0  \ .
\end{align} 
The baryon charge does not depend on the dilaton and, therefore, it is still given by eq.\eqref{chart2}. The energy density reads 
\begin{equation}
\begin{split}
 \epsilon &= \frac{K e^{-2 f \sigma} \left(4 L_{x}^2L_{y}^2 \left(L_{z}^2 \partial_{t}F^2+\partial_{z}F^2\right)+L_{z}^2 \left(L_{y}^2+4 L_{z}^2 p^2\right)\right)}{32 L_{x}^2L_{y}^2L_{z}^2}+\frac{e^{-2 f \sigma} \left(L_{z}^2 \partial_{t}\sigma^2+\partial_{z}\sigma^2\right)}{2 L_{z}^2}\\
 &+\frac{\lambda  K \left(\left((L_{z}^2 \partial_{t}F^2+\partial_{z}F^2\right) \left(L_{y}^2+4 L_{x}^2 p^2 \sin ^2\left(\frac{r}{2}\right)\right)+L_{z}^2 p^2\right)}{128 L_{x}^2L_{y}^2L_{z}^2}+V(\sigma)\ .
\end{split}
\end{equation}

\subsection{Branes in the conformal limit} 

\begin{figure}[t!]
\centering
\includegraphics[ width=0.5\textwidth]{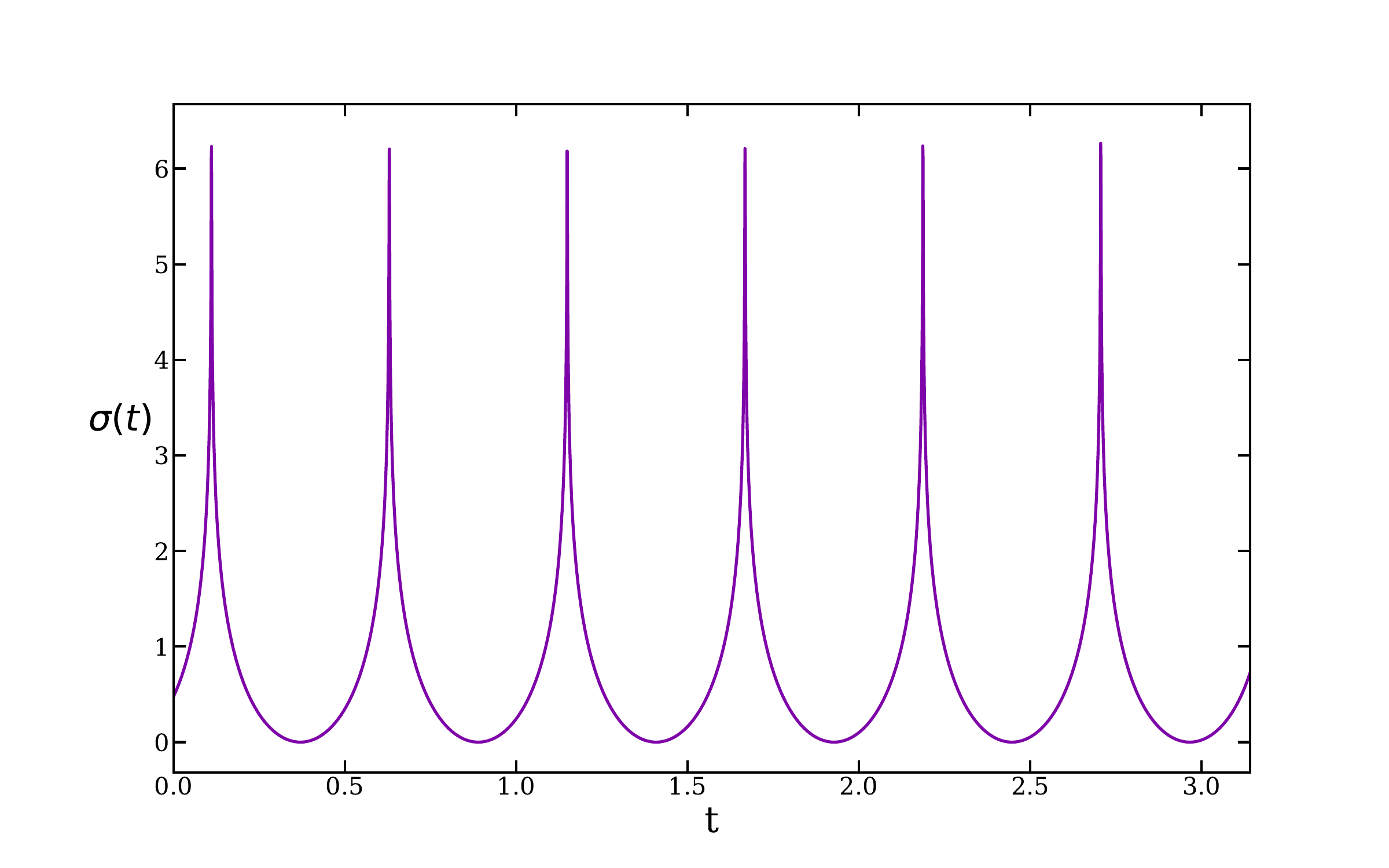}
	\caption{Time evolution of the dilaton solution eq.\eqref{static}.}
	\label{devolution}
\end{figure}

We start by setting to zero both the Skyrme term and the dilaton potential.
In this case the EOMs for the brane ansatz \eqref{lasa} admits an analytical solution carrying topological charge where the $U$ and  $\sigma$ fields are, respectively, static and homogeneous. It reads
\begin{align} \label{static}
F(z)= c_1 + c_2 z \,, \qquad  \sigma(t) =\frac{1}{f} \left\{d_1-\log \abs{\cos \left(\frac{\sqrt{K} \sqrt{4 c_2^2 L_y^2 L_x^2+L_z^2 \left(L_y^2+4 L_x^2 p^2\right)} \left(d_2 +f t\right)}{4 L_y L_x L_z}\right)}\right\}  \ ,
\end{align}
where $c_1$, $c_2$, $d_1$, and $d_2$ are arbitrary integration constants.
The boundary conditions \eqref{beforec} for $F(z)$ impose $c_2 = 4 q$ whereas the energy density evaluated on the solution is both homogeneous and static
\begin{equation} \label{halfo}
    \epsilon = \frac{K}{32 L_y^2 L_x^2 L_z^2} e^{-2 d_1} \left[L_y^2 \left(64 L_x^2 q^2+L_z^2\right)+4 L_x^2 L_z^2 p^2\right]  \ .
\end{equation}
Moreover, since the $U$ field is static, the isospin charge density of this solution vanishes identically. The time evolution of the dilaton field is depicted in Fig.\ref{devolution}. Note that this solution is non-analytic in time due to a series of branch-cut singularities lying at
\begin{equation}
    f  t = \frac{2 L_y L_x L_z\  \pi(2  n+ 1)}{\sqrt{K} \sqrt{L_y^2 \left(64 L_x^2 q^2+L_z^2\right)+4 L_x^2 L_z^2 p^2}}- d_2  \,, \qquad n \in \mathbb{Z} \ .
\end{equation}

It is possible to construct another analytical solution to the EOMs of the massless chiral Lagrangian where, akin to the case without the dilaton,  $F(z, t)$ can be expressed as a particular linear combination of left- and right-moving modes
\begin{align} \label{solb}
 F(t,z)  &=c_1+ c_2 \left[\left(z +\frac{t}{L_z}\right)- A \left(z -\frac{t}{L_z}\right)\right]  \ , \nonumber \\
 \frac{\sigma(t,z)}{\sqrt{k}} & = d_1+d_2  \left[\left(z+\frac{t}{L_z} \right)+  A \left(z -\frac{t}{L_z}\right)\right]  \ ,
\end{align}
where
\begin{equation}
   A = \frac{L_z^2 \left(L_y^2+4L_x^2 p^2\right)}{16 L_t^2L_x^2 \left(c_2^2+4 d_2^2\right)}  \ .
\end{equation}
However, for $d_2 \neq 0$ the energy of this solution either grows or decays exponentially in time depending on the boundary conditions. Noticeably, when $d_2 = 0$ the solution for the dilaton is a constant and the energy of the solution is exactly twice of \eqref{halfo}.

There are no nontrivial solutions with $U$ given by eq.\eqref{lasa} for $m_\pi \neq 0$.  Moreover, when $m_\sigma=0$ there are no solutions of the EOMs when adding the Skyrme term together with dilatonic dynamics.
However, as we will see in the next section, for non-vanishing dilaton mass we have been able to find analytical solutions with the ansatz eq.\eqref{lasa} including the Skyrme term.

\subsection{Branes in the near-conformal Skyrme model}
 
Our goal in this section is to study the most general solution of the EOMs \eqref{laseom} of the dilaton augmented Skyrme model including the Skyrme term. This is of the form
\begin{equation} \label{manuto}
F(t,z ) \equiv F\left(\frac{t}{L_z}\pm z \right)  \,, \qquad  \sigma(t,z ) = c  \ , \end{equation}
where for a dilaton potential of the form \eqref{sigmapotDelta} the dilaton vev $c$ is determined by the following equation 
\begin{equation} \label{dilvev}
    \chi_0 ^{\Delta -2}+\frac{\beta  (\Delta -4)}{m_\sigma^2}-\chi_0 ^2=0 \,, \qquad \chi_0 \equiv e^{-c f} \,, 
\end{equation} 
where
\begin{equation}
     \beta =\frac{1}{16} f^2 K \left(\frac{4 p^2}{L_y^2}+\frac{1}{L_x^2}\right)  \ .
\end{equation}
The properties of the soliton affect the dilaton vev through the value of the parameter $p$ which is related to the topological charge via eq.\eqref{chart2}.
For $m_\sigma \to \infty$ the solution is $c=0$. On the other hand, when $\Delta \to 4$ or $m_\sigma \to 0$ the vev becomes undetermined since the potential vanishes. In the special cases $\Delta \to 0,4$ and $\Delta = 2$ the dilaton vev reads
\begin{align} \label{voiello}
2 f  c = \begin{cases}
\sinh ^{-1}\left(\frac{2 \beta }{m_\sigma^2}\right) \qquad \quad \Delta \to 0 \\     -\log (1-\frac{2 \beta }{m_\sigma^2} )  \qquad \  \Delta = 2  \\ -W(-\frac{2 \beta }{m_\sigma^2} ) \qquad \qquad \Delta \to 4  \ ,  \end{cases}
\end{align}
where $W(x)$ denotes the Lambert W function. Only in the $\Delta=0$ case $c$ is real for arbitrary values of the dilaton mass. For $\Delta = 2$ the vev exhibits a branch cut singularity at $m_\sigma^2=2 \beta$ and the solution is complex for smaller masses. Analogously, in the $\Delta \to 4$ case the solution is real only for  $m_\sigma^2>2 e \beta$.

\begin{figure}[t!]
\centering
\includegraphics[ width=0.4\textwidth]{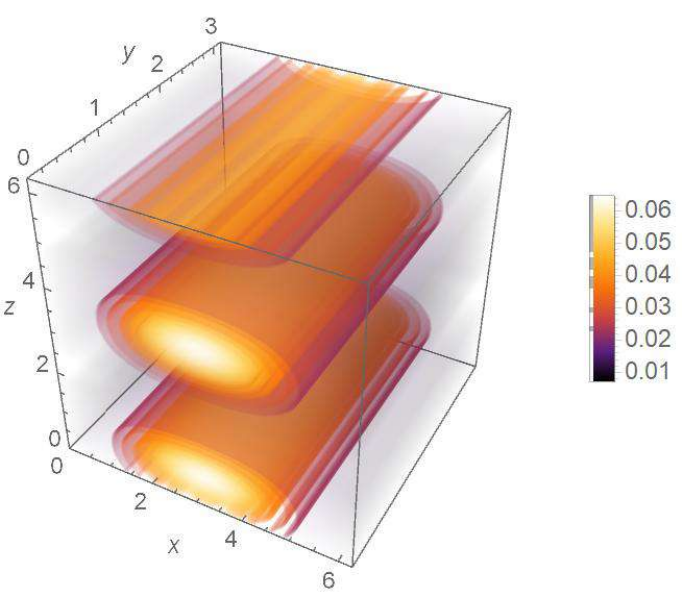} \includegraphics[width=0.4\textwidth]{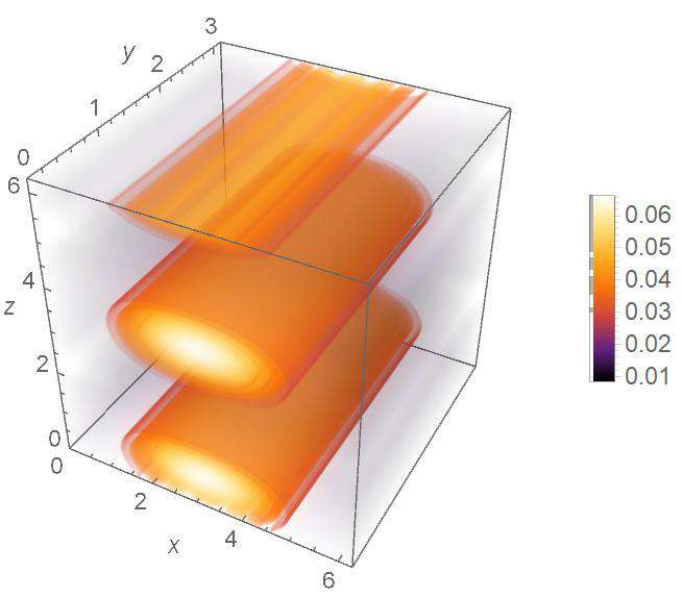}  \newline \includegraphics[width=0.4\textwidth]{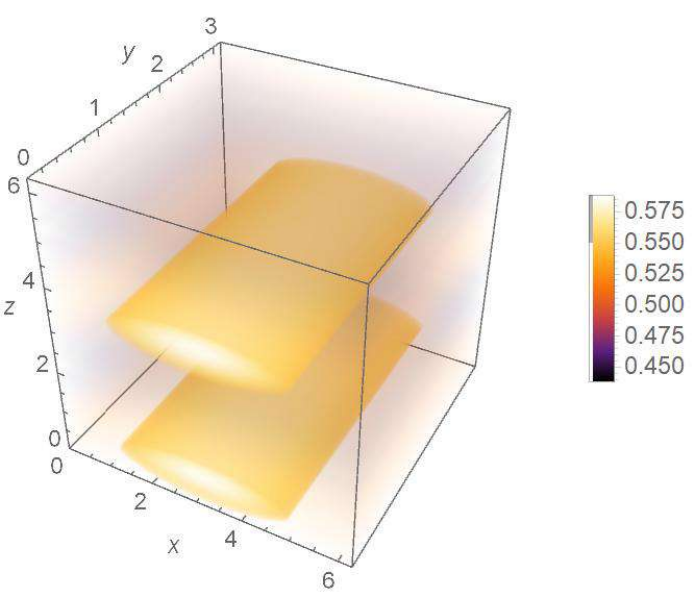} 
\includegraphics[width=0.4\textwidth]{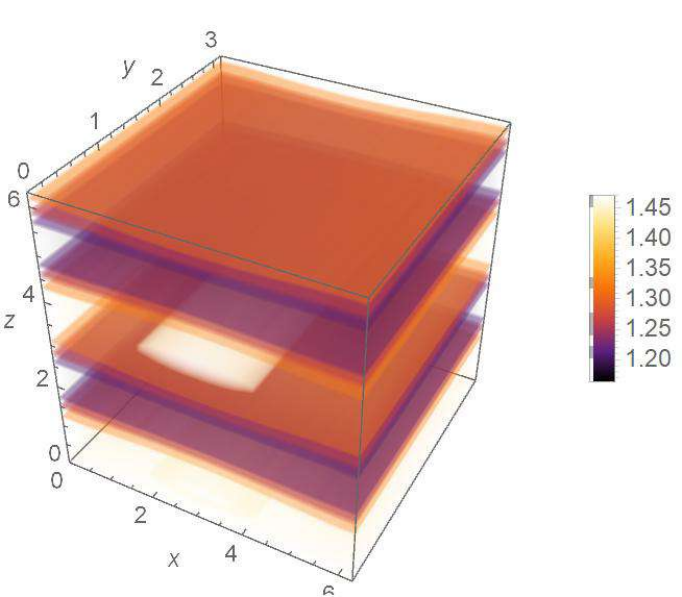}
	\caption{Energy density of the solution \eqref{manuto} for $\Delta \to 0$ and different values of the dilaton mass. We consider $m_\sigma/\sqrt{K} = 0.001$ (\emph{top-left}), $0.05$ (\emph{top-right}), $1$ (\emph{bottom-left}), and $20000$ (\emph{bottom-right}). As we increase $m_\sigma$ the dilaton decouples from the dynamics as can be seen by comparing the $m_\sigma/\sqrt{K} = 20000$ case to Fig.\ref{tortelliplot}. We choose $F(\frac{t}{L_z}+z) = \cos(\frac{t}{L_z}+z)$.}
	\label{tortelliplot2}
\end{figure}

The energy density carried by the solution for a generic $\Delta$ is
\begin{align}
 \epsilon =  \frac{K}{32} e^{-2 c f} \left(\frac{8 F'\left(\frac{t}{L_{z}}+\phi \right)^2}{L_{z}^2}+\frac{4 p^2}{L_{y}^2}+\frac{1}{L_{x}^2}\right) +\frac{K\lambda  \left(2 F'\left(\frac{t}{L_{z}}+\phi \right)^2 \left(L_{y}^2+4 L_{x}^2 p^2 \sin ^2\left(\frac{r}{2}\right)\right)+L_{z}^2 p^2\right)}{128 L_{x}^2L_{y}^2L_{z}^2}+V(c) \ .
\end{align}
As shown in Fig.\ref{tortelliplot2}, this solution describes ordered layers of matter analogously to the case without dilaton previously described. The dilaton vev suppresses the energy density of the $U$ field while adding a homogeneous contribution proportional to $m_\sigma^2$. As a consequence, the overall effect of the dilaton is to smooth out the layer structure.

Analogous solutions exist in the absence of the Skyrme term. In fact, we can now look for solutions of the form
\begin{equation}
  F(t,z ) \equiv F_+\left(\frac{t}{L_z}+ z \right) + F_-\left(\frac{t}{L_z}- z \right)  \,, \qquad  \sigma(t,z ) = c  \ .\end{equation}
Note that differently from eq.\eqref{manuto}, this is not the most general solution to the EOM.
By plugging the above into the EOMs \eqref{laseom} with $\lambda=0$ we have
\begin{equation}
   f K L_z^2 \left(L_y^2+4 L_x^2 p^2\right)-16 L_y^2 L_x^2 \left[L_z^2 e^{2 c f}\frac{\partial V(\sigma)}{\partial \sigma} \Bigg \rvert_{\sigma = c}-f K F_+'\left(\frac{t}{L_z}+\phi \right)  F_-'\left(\phi -\frac{t}{L_z}\right) \right] = 0  \ ,
\end{equation}
which admits solutions if and only if
\begin{enumerate}
    \item Either $F_+$ or $F_-$ vanishes.

    \item Both $F_+$ and $F_-$ are linear functions of their argument.
\end{enumerate}
In the first case the solution is equal to the one in the presence of the Skyrme term with the dilaton vev given by eq.\eqref{voiello}. In the second case, we have
\begin{equation}
F(t,z ) = a_0 + a_1 \left(\frac{t}{L_z}+\phi \right) + a_2 \left(\frac{t}{L_z}-\phi \right)  \ ,
\end{equation}
and the dilaton vev is again determined by eq.\eqref{dilvev} with $\beta$ now given by
\begin{equation}
  \beta =  \frac{f^2 K \left(16 a_1a_2L_y^2 L_x^2+L_z^2 \left(L_y^2+4 L_x^2 p^2\right)\right)}{16 L_y^2 L_x^2 L_z^2}  \ .
\end{equation}
For $m_\sigma=0$ this reduces to the solution \eqref{solb} with $d_2$ =0.

\section{Conclusions} \label{conclusion}
 We fused numerical and analytical methods to perform a thorough investigation of Skyrmion solutions featuring dilatonic dynamics. Employing the dilaton dressed chiral Lagrangian we addressed the impact on relevant physical quantities such as the traditional Skyrmion profile, mass, and size stemming from varying the near-conformal controlling parameters.  We discovered an intriguing dependence of these physical properties on the dilaton mass. A notable result is the observation of a quick dilaton decoupling once its mass surpasses the pion one. The decoupling is further enhanced at large values of the anomalous dimension of the chiral condensate.  
 Another interesting feature is the occurrence of a special value of the ratio of the dilaton to pion decay constants where the coupled second order field equations of the dilaton augmented EFT reduce to a first order solvable system. In this case, one arrives at analytic solitonic solutions even though this phenomenon is not related to a BPS bound. The resulting special solutions have negative topological charge $B=-1$  and are singular at the origin. 
 We further presented the impact of dilatonic dynamics on Skyrmion crystals and have identified a special parameter point, linked to the case of the traditional Skyrmion above, for which the second order differential equations also reduces to a set of first order ones. Here another observed feature,  stemming from the presence of the dilaton, is that the latter spatially separates the baryon and isospin charge distributions similarly to the spin-charge separation phenomenon.  
  Lastly, we considered dilaton-modified solitonic branes and showed that the soliton profile is unaltered by the presence of the dilaton while the latter acquires a homogeneous vacuum expectation value.
  Finally, we presented in which way the dilaton smooths out the brane configurations. We expect our results to have broad implications ranging from the near-conformal dynamics of  QCD-like theories to nuclear matter in extreme regimes.

\section*{Acknowledgments}
We are indebted to Fabrizio Canfora for a fruitful collaboration at the early stage of the project and precious suggestions. 
M. T. was supported by Agencia Nacional de Investigación y Desarrollo (ANID) grant 7221039. The work of J.B. was supported by the World Premier International Research Center Initiative (WPI Initiative), MEXT, Japan; and also supported in part by the JSPS KAKENHI Grant Number JP23K19047. The work of FS is partially supported by the Carlsberg Foundation, grant CF22-0922.

 \newpage
\begin{appendices}
    \section{Tables of the Skyrmion mass and the root mean square radius of the baryon charge}

In this appendix, we list in a series of tables our results for the Skyrmion mass $M_S$ and the root mean square radius of the baryon charge $\langle r^2 \rangle^{1/2}$ in the single spherical Skyrmion case studied in Sec.\ref{dahs}. We estimated an error not larger than the $0.5\%$ for all the entries. All the dimensionful quantities are measured in units of $\sqrt{K}$. $M_\text{Skyrme} = 15.67$ and $\langle r^2 \rangle^{1/2}_\text{Skyrme} = 0.215$ denote the Skyrmion mass and the root mean square radius in the absence of the dilaton, respectively.

\subsection{General case} \label{app1}

\FloatBarrier 
\begin{table}[h!]
\centering
\resizebox{\textwidth}{!}{
\begin{tabular}{@{}ccccccccccc@{}}
\toprule
\multicolumn{11}{c}{\textbf{Skyrmion mass}}                                                                   \\ \midrule
    & \multicolumn{5}{c}{$\mathbf{M_S}$}              & \multicolumn{5}{c}{$\mathbf{M_S/M_\text{Skyrme}}$} \\ \midrule
$\mathbf{\Delta}$ &
  $m_\sigma=1$ &
  $m_\sigma=2$ &
  $m_\sigma=3$ &
  $m_\sigma=4$ &
  \multicolumn{1}{c|}{$m_\sigma=5$} &
  $m_\sigma=1$ &
  $m_\sigma=2$ &
  $m_\sigma=3$ &
  $m_\sigma=4$ &
  $m_\sigma=5$  \\ \cmidrule(l){2-11} 
$1$  & $22.73$ & $16.38$ & $15.54$ & $15.38$ & \multicolumn{1}{c|}{$15.37$} & $1.45$ &  $1.05$ & $0.992$ & $0.979$ & $0.981$ \\ \cmidrule{2-11}
$1.2$  & $22.42$ & $16.36$ & $15.53$ & $15.37$ & \multicolumn{1}{c|}{$15.37$} & $1.43$ &  $1.04$ & $0.991$ & $0.979$ & $0.981$  \\
\cmidrule{2-11}
$1.4$  & $22.13$ & $16.34$ & $15.53$ & $15.37$ & \multicolumn{1}{c|}{$15.37$} & $1.41$ &  $1.04$ & $0.991$ & $0.980$ & $0.981$ \\ \cmidrule(l){2-11} 
$1.6$  & $21.85$ & $16.32$ & $15.53$ & $15.37$ & \multicolumn{1}{c|}{$15.37$} & $1.39$ &  $1.04$ & $0.991$ & $0.980$ & $0.981$ \\ \cmidrule(l){2-11} 
$1.8$  & $21.60$ & $16.31$ & $15.52$ & $15.37$ & \multicolumn{1}{c|}{$15.36$} & $1.38$ & $1.04$ & $0.991$ & $0.980$ & $0.980$ \\ \cmidrule(l){2-11} 
$2$  & $21.36$ & $16.29$ & $15.52$ & $15.36$ & \multicolumn{1}{c|}{$15.36$} & $1.36$ &  $1.04$ & $0.990$ & $0.980$ & $0.980$ \\ \cmidrule(l){2-11} 
$2.2$  & $21.14$ & $16.27$ & $15.52$ & $15.36$ & \multicolumn{1}{c|}{$15.36$} & $1.35$ &  $1.04$ & $0.990$ & $0.980$ & $0.980$ \\ \cmidrule(l){2-11} 
$2.4$  & $20.93$ & $16.26$ & $15.51$ & $15.36$ & \multicolumn{1}{c|}{$15.36$} & $1.34$ &  $1.04$ & $0.990$  & $0.980$ & $0.980$ \\ \cmidrule(l){2-11} 
$2.6$  & $20.73$ & $16.24$ & $15.51$ & $15.36$ & \multicolumn{1}{c|}{$15.36$} & $1.32$ & $1.04$ & $0.990$ & $0.980$ & $0.980$ \\ \cmidrule(l){2-11} 
$2.8$  & $20.55$ & $16.23$ & $15.51$ & $15.36$ & \multicolumn{1}{c|}{$15.35$} & $1.31$ & $1.04$ & $0.989$ & $0.980$ & $0.980$ \\ \cmidrule(l){2-11} 
$3$  & $20.37$ & $16.21$ & $15.50$ & $15.35$ & \multicolumn{1}{c|}{$15.35$} & $1.30$ & $1.03$ & $0.989$ & $0.981$ & $0.980$ \\ \cmidrule(l){2-11} 
$3.2$  & $20.21$ & $16.20$ & $15.50$ & $15.35$ & \multicolumn{1}{c|}{$15.35$} & $1.29$ & $1.03$ & $0.989$ & $0.981$ & $0.980$ \\ \cmidrule(l){2-11} 
$3.4$  & $20.06$ & $16.18$ & $15.50$ & $15.35$ & \multicolumn{1}{c|}{$15.35$} & $1.28$ & $1.03$ & $0.989$ & $0.981$ & $0.979$ \\ \cmidrule(l){2-11} 
$3.6$  & $19.91$ & $16.17$ & $15.49$ & $15.35$ & \multicolumn{1}{c|}{$15.35$} & $1.27$ & $1.03$ & $0.989$ & $0.981$ & $0.979$ \\ \cmidrule(l){2-11} 
$3.8$  & $19.78$ & $16.16$ & $15.49$ & $15.35$ & \multicolumn{1}{c|}{$15.35$} & $1.26$ & $1.03$ & $0.988$ & $0.981$ & $0.979$ \\ \cmidrule(l){2-11} 
$4$  & $19.65$ & $16.14$ & $15.49$ & $15.34$ & \multicolumn{1}{c|}{$15.34$} & $1.23$ & $1.03$ & $0.99$ & $0.988$ & $0.979$ \\ \bottomrule
\end{tabular}%
}
\caption{\em Numerical values of the Skyrmion mass for different values of the conformal dimension $\Delta$ and dilaton mass $m_\sigma=1,2,3,4,5$. $M_\text{Skyrme} = 15.67$ denotes the Skyrmion mass in the absence of the dilaton.}
\label{tab:bigtable1b}
\end{table}

\begin{table}[h!]
\centering
\resizebox{\textwidth}{!}{
\begin{tabular}{@{}ccccccccccc@{}}
\toprule
\multicolumn{11}{c}{\textbf{Root mean square radius of the baryon charge}}                                                                   \\ \midrule
    & \multicolumn{5}{c}{$\mathbf{\langle r^2\rangle ^{1/2}}\ $ \textbf{in units of} $\mathbf{10^{-1}}$} &\multicolumn{5}{c}{$\mathbf{\langle r^2\rangle ^{1/2}/\langle r^2\rangle_{Skyrme} ^{1/2}}$ \textbf{in units of} $\mathbf{10^{-1}}$} \\ \midrule
$\mathbf{\Delta}$ &
  $m_\sigma=1$ &
  $m_\sigma=2$ &
  $m_\sigma=3$ &
  $m_\sigma=4$ &
  \multicolumn{1}{c|}{$m_\sigma=5$} &
  $m_\sigma=1$ &
  $m_\sigma=2$ &
  $m_\sigma=3$ &
  $m_\sigma=4$ &
  $m_\sigma=5$  \\ \cmidrule(l){2-11} 
$1$  & $1.50$ & $2.04$ & $2.13$ & $2.151$ & \multicolumn{1}{c|}{$2.153$} & $6.99$ & $9.47$ & $9.91$ & $9.996$ & $10.01$ \\
\cmidrule{2-11}
$1.2$  & $1.52$ & $2.04$ & $2.13$ & $2.151$ & \multicolumn{1}{c|}{$2.153$} & $7.08$ & $9.48$ & $9.91$ & $9.997$ & $10.01$ \\
\cmidrule{2-11}
$1.4$  & $1.54$ & $2.04$ & $2.13$ & $2.151$ & \multicolumn{1}{c|}{$2.153$} & $7.17$ & $9.49$ & $9.91$ & $9.998$ & $10.01$ \\
\cmidrule{2-11}
$1.6$  & $1.56$ & $2.04$ & $2.13$ & $2.151$ & \multicolumn{1}{c|}{$2.153$} & $7.25$ & $9.50$ & $9.92$ & $9.999$ & $10.01$ \\
\cmidrule{2-11}
$1.8$  & $1.58$ & $2.05$ & $2.13$ & $2.152$ & \multicolumn{1}{c|}{$2.153$} & $7.33$ & $9.51$ & $9.92$ & $10.00$ & $10.01$ \\
\cmidrule{2-11}
$2$  & $1.60$ & $2.05$ & $2.13$ & $2.152$ & \multicolumn{1}{c|}{$2.153$} & $7.43$ & $9.52$ & $9.92$ & $10.00$ & $10.01$ \\
\cmidrule{2-11}
$2.2$  & $1.61$ & $2.05$ & $2.13$ & $2.152$ & \multicolumn{1}{c|}{$2.153$} & $7.50$ & $9.53$ & $9.92$ & $10.00$ & $10.01$ \\
\cmidrule{2-11}
$2.4$  & $1.63$ & $2.05$ & $2.14$ & $2.152$ & \multicolumn{1}{c|}{$2.153$} & $7.57$ & $9.54$ & $9.92$  & $10.00$ & $10.01$ \\
\cmidrule{2-11}
$2.6$  & $1.64$ & $2.05$ & $2.14$ & $2.152$ & \multicolumn{1}{c|}{$2.153$} & $7.64$ & $9.54$ & $9.93$ & $10.00$ & $10.01$ \\
\cmidrule{2-11}
$2.8$  & $1.66$ & $2.06$ & $2.14$ & $2.152$ & \multicolumn{1}{c|}{$2.153$} & $7.70$ & $9.55$ & $9.93$ & $10.00$ & $10.01$ \\
\cmidrule{2-11}
$3$  & $1.67$ & $2.06$ & $2.14$ & $2.153$ & \multicolumn{1}{c|}{$2.154$} & $7.76$ & $9.56$ & $9.93$ & $10.00$ & $10.01$ \\
\cmidrule{2-11}
$3.2$  & $1.68$ & $2.06$ & $2.14$ & $2.153$ & \multicolumn{1}{c|}{$2.154$} & $7.82$ & $9.57$ & $9.93$ & $10.01$ & $10.01$ \\
\cmidrule{2-11}
$3.4$  & $1.70$ & $2.06$ & $2.14$ & $2.153$ & \multicolumn{1}{c|}{$2.154$} & $7.88$ & $9.57$ & $9.93$ & $10.01$ & $10.01$ \\
\cmidrule{2-11}
$3.6$  & $1.71$ & $2.06$ & $2.14$ & $2.153$ & \multicolumn{1}{c|}{$2.154$} & $7.93$ & $9.58$ & $9.93$ & $10.01$ & $10.01$ \\
\cmidrule{2-11}
$3.8$  & $1.72$ & $2.06$ & $2.14$ & $2.153$ & \multicolumn{1}{c|}{$2.154$} & $7.98$ & $9.59$ & $9.94$ & $10.01$ & $10.01$ \\
\cmidrule{2-11}
$4$  & $1.73$ & $2.06$ & $2.14$ & $2.153$ & \multicolumn{1}{c|}{$2.154$} & $8.03$ & $9.59$ & $9.94$ & $10.01$ & $10.01$ \\
\bottomrule
\end{tabular}
}
\caption{\em Numerical values of the root mean square radius $\langle r^2\rangle ^{1/2}$ of the baryon charge for different values of the conformal dimension $\Delta$ and dilaton mass $m_\sigma=1,2,3,4,5$. $\langle r^2\rangle_{Skyrme} ^{1/2} = 0.215$ denotes the root mean square radius of the baryon charge for the Skyrme model in the absence of the dilaton.}
\label{tab:bigtable2b}
\end{table}%

\FloatBarrier
\begin{table}[h!]
\centering
\begin{tabular}{@{}ccccccccc@{}}
\toprule
\multicolumn{9}{c}{\textbf{Skyrmion mass}}                                                                   \\ \midrule
    & \multicolumn{4}{c}{$\mathbf{M_S}$}              & \multicolumn{4}{c}{$\mathbf{M_S/M_\text{Skyrme}}$} \\ \midrule
$\mathbf{\Delta}$ &
  $m_\sigma=7$ &
  $m_\sigma=8$ &
  $m_\sigma=9$ &
  \multicolumn{1}{c|}{$m_\sigma=10$} &
  $m_\sigma=7$ &
  $m_\sigma=8$ &
  $m_\sigma=9$ &
  $m_\sigma=10$  \\ \cmidrule(l){2-9} 
				$1$  & $15.46$ & $15.52$ & $15.56$ & \multicolumn{1}{c|}{$15.61$} & $0.987$ & $0.990$ & $0.993$ & $0.996$ \\
    \cmidrule{2-9}
				$1.2$  & $15.46$ & $15.51$ & $15.56$ & \multicolumn{1}{c|}{$15.61$} & $0.987$ & $0.990$ & $0.993$ & $0.996$ \\
    \cmidrule{2-9}
				$1.4$  & $15.46$ & $15.51$ & $15.56$ & \multicolumn{1}{c|}{$15.60$} & $0.987$ & $0.990$ & $0.993$ & $0.996$ \\
    \cmidrule{2-9}
				$1.6$  & $15.46$ & $15.51$ & $15.56$ & \multicolumn{1}{c|}{$15.60$} & $0.986$ & $0.990$ & $0.993$ & $0.996$ \\
    \cmidrule{2-9}
				$1.8$  & $15.46$ & $15.51$ & $15.56$ & \multicolumn{1}{c|}{$15.60$}  & $0.986$ & $0.990$ & $0.993$ & $0.996$ \\
    \cmidrule{2-9}
				$2$  & $15.45$ & $15.51$ & $15.56$ & \multicolumn{1}{c|}{$15.60$} & $0.986$ & $0.990$ & $0.993$ & $0.995$ \\
    \cmidrule{2-9}
				$2.2$  & $15.45$ & $15.51$ & $15.56$ & \multicolumn{1}{c|}{$15.60$} & $0.986$ & $0.990$ & $0.993$ & $0.995$ \\
   \cmidrule{2-9}
				$2.4$  & $15.45$ & $15.51$ & $15.55$ & \multicolumn{1}{c|}{$15.60$} & $0.986$ & $0.989$ & $0.993$  & $0.995$ \\
    \cmidrule{2-9}
				$2.6$  & $15.45$ & $15.50$ & $15.55$ & \multicolumn{1}{c|}{$15.60$} & $0.986$ & $0.989$ & $0.992$ & $0.995$ \\
    \cmidrule{2-9}
				$2.8$  & $15.45$ & $15.50$ & $15.55$ & \multicolumn{1}{c|}{$15.60$} & $0.986$ & $0.989$ & $0.992$ & $0.995$ \\
    \cmidrule{2-9}
				$3$  & $15.45$ & $15.50$ & $15.55$ & \multicolumn{1}{c|}{$15.59$} & $0.986$ & $0.989$ & $0.992$ & $0.995$ \\
    \cmidrule{2-9}
				$3.2$  & $15.45$ & $15.50$ & $15.55$ & \multicolumn{1}{c|}{$15.59$} & $0.986$ & $0.989$ & $0.992$ & $0.995$ \\
   \cmidrule{2-9}
				$3.4$  & $15.44$ & $15.50$ & $15.55$ & \multicolumn{1}{c|}{$15.59$} & $0.986$ & $0.989$ & $0.992$ & $0.995$ \\
   \cmidrule{2-9}
				$3.6$  & $15.44$ & $15.50$ & $15.55$ & \multicolumn{1}{c|}{$15.59$} & $0.985$ & $0.989$ & $0.992$ & $0.995$ \\
    \cmidrule{2-9}
				$3.8$  & $15.44$ & $15.49$ & $15.54$ & \multicolumn{1}{c|}{$15.59$} & $0.985$ & $0.989$ & $0.992$ & $0.995$ \\
    \cmidrule{2-9}
				$4$  & $15.44$ & $15.49$ & $15.54$ & \multicolumn{1}{c|}{$15.59$} & $0.985$ & $0.989$ & $0.992$ & $0.995$ \\
    \bottomrule
			\end{tabular}%
			\caption{\em Numerical values of the Skyrmion mass for different values of the conformal dimension $\Delta$ and dilaton mass $m_\sigma=7,8,9,10$. $M_\text{Skyrme} = 15.67$ denotes the Skyrmion mass in the absence of the dilaton.}
  \label{tab:bigtable1a}
	\end{table}%


\begin{table}[h!]
\centering
\begin{tabular}{@{}ccccccccc@{}}
\toprule
\multicolumn{9}{c}{\textbf{Root mean square radius of the baryon charge}}                                                                   \\ \midrule
    & \multicolumn{4}{c}{$\mathbf{\langle r^2\rangle ^{1/2}}\ $ \textbf{in units of} $\mathbf{10^{-1}}$} &\multicolumn{4}{c}{$\mathbf{\langle r^2\rangle ^{1/2}/\langle r^2\rangle_{Skyrme} ^{1/2}}$ \textbf{in units of} $\mathbf{10^{-1}}$} \\ \midrule
$\mathbf{\Delta}$ &
  $m_\sigma=7$ &
  $m_\sigma=8$ &
  $m_\sigma=9$ &
  \multicolumn{1}{c|}{$m_\sigma=10$} &
  $m_\sigma=7$ &
  $m_\sigma=8$ &
  $m_\sigma=9$ &
$m_\sigma=10$  \\ \cmidrule(l){2-9} 
$1$  & $2.150$ & $2.149$ & $2.148$ & \multicolumn{1}{c|}{$2.149$} & $9.991$ & $9.986$ & $9.985$ & $9.986$ \\
\cmidrule{2-9}
$1.2$  & $2.150$ & $2.149$ & $2.148$ & \multicolumn{1}{c|}{$2.149$} & $9.990$ & $9.986$ & $9.984$ & $9.986$ \\
\cmidrule{2-9}
$1.4$  & $2.150$ & $2.149$ & $2.148$ & \multicolumn{1}{c|}{$2.148$} & $9.990$ & $9.985$ & $9.984$ & $9.985$ \\
\cmidrule{2-9}
$1.6$  & $2.149$ & $2.148$ & $2.148$ & \multicolumn{1}{c|}{$2.148$} & $9.990$ & $9.985$ & $9.983$ & $9.985$ \\
\cmidrule{2-9}
$1.8$  & $2.149$ & $2.148$ & $2.148$ & \multicolumn{1}{c|}{$2.148$}  & $9.990$ & $9.984$ & $9.983$ & $9.984$ \\
\cmidrule{2-9}
$2$  & $2.149$ & $2.148$ & $2.148$ & \multicolumn{1}{c|}{$2.148$} & $9.989$ & $9.984$ & $9.982$ & $9.983$ \\
\cmidrule{2-9}
$2.2$  & $2.149$ & $2.148$ & $2.148$ & \multicolumn{1}{c|}{$2.148$} & $9.989$ & $9.984$ & $9.982$ & $9.983$ \\
\cmidrule{2-9}
$2.4$  & $2.149$ & $2.148$ & $2.148$ & \multicolumn{1}{c|}{$2.148$} & $9.989$ & $9.983$ & $9.981$  & $9.982$ \\
\cmidrule{2-9}
$2.6$  & $2.149$ & $2.148$ & $2.148$ & \multicolumn{1}{c|}{$2.148$} & $9.989$ & $9.983$ & $9.981$ & $9.982$ \\
\cmidrule{2-9}
$2.8$  & $2.149$ & $2.148$ & $2.147$ & \multicolumn{1}{c|}{$2.148$} & $9.989$ & $9.983$ & $9.980$ & $9.981$ \\
\cmidrule{2-9}
$3$  & $2.149$ & $2.148$ & $2.147$ & \multicolumn{1}{c|}{$2.147$} & $0.988$ & $9.982$ & $9.980$ & $9.981$ \\
\cmidrule{2-9}
$3.2$  & $2.149$ & $2.148$ & $2.147$ & \multicolumn{1}{c|}{$2.147$} & $0.988$ & $9.982$ & $9.979$ & $9.980$ \\
\cmidrule{2-9}
$3.4$  & $2.149$ & $2.148$ & $2.147$ & \multicolumn{1}{c|}{$2.147$} & $0.988$ & $9.981$ & $9.979$ & $9.979$ \\
\cmidrule{2-9}
$3.6$  & $2.149$ & $2.148$ & $2.147$ & \multicolumn{1}{c|}{$2.147$} & $0.988$ & $9.981$ & $0.978$ & $9.979$ \\
\cmidrule{2-9}
$3.8$  & $2.149$ & $2.148$ & $2.147$ & \multicolumn{1}{c|}{$2.147$} & $0.988$ & $9.981$ & $0.978$ & $9.978$ \\
\cmidrule{2-9}
$4$  & $2.149$ & $2.147$ & $2.147$ & \multicolumn{1}{c|}{$2.147$} & $0.989$ & $9.980$ & $0.978$ & $9.978$ \\
\bottomrule
\end{tabular}
\caption{\em Numerical values of the root mean square radius $\langle r^2\rangle ^{1/2}$ of the baryon charge for different values of the conformal dimension $\Delta$ and dilaton mass $m_\sigma=7,8,9,10$. $\langle r^2\rangle_{Skyrme} ^{1/2} = 0.215$ denotes the root mean square radius of the baryon charge for the Skyrme model in the absence of the dilaton.}
\label{tab:bigtable1}
\end{table}%

\FloatBarrier
\subsection{$M_S$ and  $\langle r^2\rangle ^{1/2}$ in the $\Delta \to 0$ limit} \label{app2}

\begin{table}[h!]
\centering
\begin{tabular}{@{}ccc@{}}
\toprule
\multicolumn{3}{c}{\textbf{Skyrmion mass}}                                                                   \\ \midrule
    & $\mathbf{M_S}$             & $\mathbf{M_S/M_\text{Skyrme}}$ \\ 
     \cmidrule{2-3}    
				$m_\sigma=1$  & $10.45$ & $0.67$  \\ 
     \cmidrule{2-3}    
				$m_\sigma=2$  & $13.31$ & $0.45$  \\ 
     \cmidrule{2-3}    
				$m_\sigma=3$  & $14.17$ & $0.90$  \\ 
     \cmidrule{2-3}    
				$m_\sigma=4$  & $14.57$ & $0.93$  \\ 
     \cmidrule{2-3}    
				$m_\sigma=5$  & $14.82$ & $0.95$  \\ 
    \toprule
				& \multicolumn{2}{c}{\textbf{Root mean square radius of the baryon charge}}  \\  
				\midrule
      &  $\mathbf{\langle r^2\rangle ^{1/2}}$ & $\mathbf{\langle r^2\rangle ^{1/2}/\langle r^2\rangle_{Skyrme} ^{1/2}}$  \\ 
      \cmidrule{2-3}    
				$m_\sigma=1$ & $0.296$ & $1.38$  \\ 
     \cmidrule{2-3}    
				$m_\sigma=2$  & $0.236$ & $1.09$  \\ 
     \cmidrule{2-3}    
				$m_\sigma=3$  & $0.223$ & $1.04$  \\ 
     \cmidrule{2-3}    
				$m_\sigma=4$  & $0.219$ & $1.02$  \\ 
     \cmidrule{2-3}    
				$m_\sigma=5$  & $0.217$ & $1.01$  \\ 
    \bottomrule
			\end{tabular}
		\caption{{\em {Numerical values of the Skyrmion mass $M_S$ and the root mean square radius of the baryon charge $\langle r^2\rangle ^{1/2}$ for $m_\sigma =0,1,2,3,4,5$.}}}
  \label{tab:bigtableCOLEMANa}
	\end{table}%

\FloatBarrier

\begin{table}[h!]
\centering
\begin{tabular}{@{}ccc@{}}
\toprule
\multicolumn{3}{c}{\textbf{Skyrmion mass}}                                                                   \\ \midrule
    & $\mathbf{M_S}$             & $\mathbf{M_S/M_\text{Skyrme}}$ \\ 
     \cmidrule{2-3}    
				$m_\sigma=7$  & $15.13$ & $0.966$  \\ 
     \cmidrule{2-3}    
				$m_\sigma=8$  & $15.24$ & $0.973$  \\ 
     \cmidrule{2-3}    
				$m_\sigma=9$  & $15.33$ & $0.978$  \\ 
     \cmidrule{2-3}    
				$m_\sigma=10$  & $15.41$ & $0.983$  \\ 
    \toprule
				& \multicolumn{2}{c}{\textbf{Root mean square radius of the baryon charge}}  \\  
				\midrule
      &  $\mathbf{\langle r^2\rangle ^{1/2}}$ & $\mathbf{\langle r^2\rangle ^{1/2}/\langle r^2\rangle_{Skyrme} ^{1/2}}$  \\ 
      \cmidrule{2-3}     
				$m_\sigma=7$ & $0.215$ & $1.001$  \\ 
     \cmidrule{2-3}    
				$m_\sigma=8$  & $0.215$ & $1.000$  \\ 
     \cmidrule{2-3}    
				$m_\sigma=9$  & $0.215$ & $1.000$  \\ 
     \cmidrule{2-3}    
				$m_\sigma=10$  & $0.215$ & $1.000$  \\ 
    \bottomrule
			\end{tabular}
		\caption{{\em {Numerical values of the Skyrmion mass and the root mean square radius $\langle r^2\rangle ^{1/2}$ of the baryon charge for different values of the dilaton mass measured in units of $\sqrt{K}$. }}}
  \label{tab:bigtableCOLEMANb}
	\end{table}%

\FloatBarrier
\end{appendices}

 \newpage
\printbibliography

@article{Lee:2003eg,
    author = "Lee, Hee-Jung and Park, Byung-Yoon and Rho, Mannque and Vento, Vicente",
    title = "{Sliding vacua in dense skyrmion matter}",
    eprint = "hep-ph/0304066",
    archivePrefix = "arXiv",
    reportNumber = "KIAS-P03026",
    doi = "10.1016/S0375-9474(03)01626-9",
    journal = "Nucl. Phys. A",
    volume = "726",
    pages = "69--92",
    year = "2003"
}

@article{brown1991scaling,
  title={Scaling effective Lagrangians in a dense medium},
  author={Brown, GE and Rho, Mannque},
  journal={Physical review letters},
  volume={66},
  number={21},
  pages={2720},
  year={1991},
  publisher={APS}
}

@article{Bersini:2022bnx,
    author = "Bersini, Jahmall and D'Alise, Alessandra and Sannino, Francesco and Torres, Matías",
    title = "{Charging the conformal window at nonzero \ensuremath{\theta} angle}",
    eprint = "2208.09227",
    archivePrefix = "arXiv",
    primaryClass = "hep-th",
    reportNumber = "RBI-ThPhys-2022-32",
    doi = "10.1103/PhysRevD.107.125024",
    journal = "Phys. Rev. D",
    volume = "107",
    number = "12",
    pages = "125024",
    year = "2023"
}

@article{Ma:2023ugl,
    author = "Ma, Yong-Liang and Yang, Wen-Cong",
    title = "{Topology and Emergent Symmetries in Dense Compact Star Matter}",
    eprint = "2301.02105",
    archivePrefix = "arXiv",
    primaryClass = "nucl-th",
    doi = "10.3390/sym15030776",
    journal = "Symmetry",
    volume = "15",
    number = "3",
    pages = "776",
    year = "2023"
}

@article{Ma:2018xjw,
    author = "Ma, Yong-Liang and Rho, Mannque",
    title = "{Pseudoconformal structure in dense baryonic matter}",
    eprint = "1810.06062",
    archivePrefix = "arXiv",
    primaryClass = "nucl-th",
    doi = "10.1103/PhysRevD.99.014034",
    journal = "Phys. Rev. D",
    volume = "99",
    number = "1",
    pages = "014034",
    year = "2019"
}

@article{Shao:2022njr,
    author = "Shao, Long-Qi and Ma, Yong-Liang",
    title = "{Scale symmetry and composition of compact star matter}",
    eprint = "2202.09957",
    archivePrefix = "arXiv",
    primaryClass = "nucl-th",
    doi = "10.1103/PhysRevD.106.014014",
    journal = "Phys. Rev. D",
    volume = "106",
    number = "1",
    pages = "014014",
    year = "2022"
}

@article{Paeng:2017qvp,
    author = "Paeng, Won-Gi and Kuo, Thomas T. S. and Lee, Hyun Kyu and Ma, Yong-Liang and Rho, Mannque",
    title = "{Scale-invariant hidden local symmetry, topology change, and dense baryonic matter. II.}",
    eprint = "1704.02775",
    archivePrefix = "arXiv",
    primaryClass = "nucl-th",
    doi = "10.1103/PhysRevD.96.014031",
    journal = "Phys. Rev. D",
    volume = "96",
    number = "1",
    pages = "014031",
    year = "2017"
}

@article{Holt:2014hma,
    author = "Holt, Jeremy W. and Rho, Mannque and Weise, Wolfram",
    title = "{Chiral symmetry and effective field theories for hadronic, nuclear and stellar matter}",
    eprint = "1411.6681",
    archivePrefix = "arXiv",
    primaryClass = "nucl-th",
    doi = "10.1016/j.physrep.2015.10.011",
    journal = "Phys. Rept.",
    volume = "621",
    pages = "2--75",
    year = "2016"
}

@article{Ma:2013ooa,
    author = "Ma, Yong-Liang and Harada, Masayasu and Lee, Hyun Kyu and Oh, Yongseok and Park, Byung-Yoon and Rho, Mannque",
    title = "{Dense baryonic matter in the hidden local symmetry approach: Half-skyrmions and nucleon mass}",
    eprint = "1304.5638",
    archivePrefix = "arXiv",
    primaryClass = "hep-ph",
    doi = "10.1103/PhysRevD.88.014016",
    journal = "Phys. Rev. D",
    volume = "88",
    number = "1",
    pages = "014016",
    year = "2013",
    note = "[Erratum: Phys.Rev.D 88, 079904 (2013)]"
}

@article{Ma:2021nuf,
    author = "Ma, Yong-Liang and Rho, Mannque",
    title = "{Topology change, emergent symmetries and compact star matter}",
    eprint = "2103.00744",
    archivePrefix = "arXiv",
    primaryClass = "nucl-th",
    doi = "10.1007/s43673-021-00016-1",
    journal = "AAPPS Bull.",
    volume = "31",
    number = "1",
    pages = "16",
    year = "2021"
}

@article{Cata:2018wzl,
    author = "Catá, O. and Crewther, R. J. and Tunstall, Lewis C.",
    title = "{Crawling technicolor}",
    eprint = "1803.08513",
    archivePrefix = "arXiv",
    primaryClass = "hep-ph",
    reportNumber = "SI-HEP-2018-09, QFET-2018-05, ADP-18-3/T1051, ADP-18-3-T1051",
    doi = "10.1103/PhysRevD.100.095007",
    journal = "Phys. Rev. D",
    volume = "100",
    number = "9",
    pages = "095007",
    year = "2019"
}

@article{Crewther:2013vea,
    author = "Crewther, R. J. and Tunstall, Lewis C.",
    title = "{$\Delta I=1/2$ rule for kaon decays derived from QCD infrared fixed point}",
    eprint = "1312.3319",
    archivePrefix = "arXiv",
    primaryClass = "hep-ph",
    reportNumber = "ADP-12-09-T776",
    doi = "10.1103/PhysRevD.91.034016",
    journal = "Phys. Rev. D",
    volume = "91",
    number = "3",
    pages = "034016",
    year = "2015"
}

@article{Paeng:2015noa,
    author = "Paeng, Won-Gi and Kuo, Thomas T. S. and Lee, Hyun Kyu and Rho, Mannque",
    title = "{Scale-Invariant Hidden Local Symmetry, Topology Change and Dense Baryonic Matter}",
    eprint = "1508.05210",
    archivePrefix = "arXiv",
    primaryClass = "hep-ph",
    doi = "10.1103/PhysRevC.93.055203",
    journal = "Phys. Rev. C",
    volume = "93",
    number = "5",
    pages = "055203",
    year = "2016"
}

@article{Brown:2001nh,
    author = "Brown, G. E. and Rho, Mannque",
    title = "{On the manifestation of chiral symmetry in nuclei and dense nuclear matter}",
    eprint = "hep-ph/0103102",
    archivePrefix = "arXiv",
    doi = "10.1016/S0370-1573(01)00084-9",
    journal = "Phys. Rept.",
    volume = "363",
    pages = "85--171",
    year = "2002"
}

@article{Cohen:1988sq,
    author = "Cohen, Andrew G. and Georgi, Howard",
    title = "{Walking Beyond the Rainbow}",
    reportNumber = "HUTP-88/A007",
    doi = "10.1016/0550-3213(89)90109-0",
    journal = "Nucl. Phys. B",
    volume = "314",
    pages = "7--24",
    year = "1989"
}

@article{Jaffe:1976yi,
    author = "Jaffe, Robert L.",
    title = "{Perhaps a Stable Dihyperon}",
    reportNumber = "SLAC-PUB-1828",
    doi = "10.1103/PhysRevLett.38.195",
    journal = "Phys. Rev. Lett.",
    volume = "38",
    pages = "195--198",
    year = "1977",
    note = "[Erratum: Phys.Rev.Lett. 38, 617 (1977)]"
}

@article{Eichten:1980mw,
    author = "Eichten, E. and Feinberg, F.",
    title = "{Spin Dependent Forces in QCD}",
    reportNumber = "HUTP-80-A053",
    doi = "10.1103/PhysRevD.23.2724",
    journal = "Phys. Rev. D",
    volume = "23",
    pages = "2724",
    year = "1981"
}

@article{Isgur:1989vq,
    author = "Isgur, Nathan and Wise, Mark B.",
    title = "{Weak Decays of Heavy Mesons in the Static Quark Approximation}",
    reportNumber = "UTPT-89-27, CALT-68-1585",
    doi = "10.1016/0370-2693(89)90566-2",
    journal = "Phys. Lett. B",
    volume = "232",
    pages = "113--117",
    year = "1989"
}

@article{Miransky:1984ef,
    author = "Miransky, V. A.",
    title = "{Dynamics of Spontaneous Chiral Symmetry Breaking and Continuum Limit in Quantum Electrodynamics}",
    reportNumber = "ITF-84-152E",
    doi = "10.1007/BF02724229",
    journal = "Nuovo Cim. A",
    volume = "90",
    pages = "149--170",
    year = "1985"
}

@article{Harada:1996wb,
    author = "Harada, M. and Sannino, F. and Schechter, J. and Weigel, H.",
    title = "{Scaling behavior in soliton models}",
    eprint = "nucl-th/9605001",
    archivePrefix = "arXiv",
    reportNumber = "UNITU-THEP-7-1996, SU-4240-631",
    doi = "10.1016/0370-2693(96)00801-5",
    journal = "Phys. Lett. B",
    volume = "384",
    pages = "5--12",
    year = "1996"
}

@article{Sannino:2016sfx,
    author = "Sannino, Francesco and Strumia, Alessandro and Tesi, Andrea and Vigiani, Elena",
    title = "{Fundamental partial compositeness}",
    eprint = "1607.01659",
    archivePrefix = "arXiv",
    primaryClass = "hep-ph",
    reportNumber = "CP3-ORIGINS-2016-027, EFI-16-15, CERN-TH-2016-148",
    doi = "10.1007/JHEP11(2016)029",
    journal = "JHEP",
    volume = "11",
    pages = "029",
    year = "2016"
}

@article{Holdom:1988gr,
    author = "Holdom, Bob",
    title = "{Continuum Limit of Quenched Theories}",
    reportNumber = "UTPT-88-15",
    doi = "10.1103/PhysRevLett.62.997",
    journal = "Phys. Rev. Lett.",
    volume = "62",
    pages = "997",
    year = "1989"
}

@article{Holdom:1988gs,
    author = "Holdom, Bob",
    title = "{Raising Condensates Beyond the Ladder}",
    reportNumber = "UTPT-88-13",
    doi = "10.1016/0370-2693(88)91776-5",
    journal = "Phys. Lett. B",
    volume = "213",
    pages = "365--369",
    year = "1988"
}

@article{HeavyFlavorAveragingGroup:2022wzx,
    author = "Amhis, Yasmine Sara and others",
    collaboration = "Heavy Flavor Averaging Group, HFLAV",
    title = "{Averages of b-hadron, c-hadron, and \ensuremath{\tau}-lepton properties as of 2021}",
    eprint = "2206.07501",
    archivePrefix = "arXiv",
    primaryClass = "hep-ex",
    doi = "10.1103/PhysRevD.107.052008",
    journal = "Phys. Rev. D",
    volume = "107",
    number = "5",
    pages = "052008",
    year = "2023"
}

@article{Bigi:2016mdz,
    author = "Bigi, Dante and Gambino, Paolo",
    title = "{Revisiting $B\to D \ell \nu$}",
    eprint = "1606.08030",
    archivePrefix = "arXiv",
    primaryClass = "hep-ph",
    doi = "10.1103/PhysRevD.94.094008",
    journal = "Phys. Rev. D",
    volume = "94",
    number = "9",
    pages = "094008",
    year = "2016"
}

@article{Detmold:2015aaa,
    author = "Detmold, William and Lehner, Christoph and Meinel, Stefan",
    title = "{$\Lambda_b \to p \ell^- \bar{\nu}_\ell$ and $\Lambda_b \to \Lambda_c \ell^- \bar{\nu}_\ell$ form factors from lattice QCD with relativistic heavy quarks}",
    eprint = "1503.01421",
    archivePrefix = "arXiv",
    primaryClass = "hep-lat",
    reportNumber = "RBRC-1111",
    doi = "10.1103/PhysRevD.92.034503",
    journal = "Phys. Rev. D",
    volume = "92",
    number = "3",
    pages = "034503",
    year = "2015"
}

@article{Falk:1990pz,
    author = "Falk, Adam F. and Grinstein, Benjamin and Luke, Michael E.",
    title = "{Leading mass corrections to the heavy quark effective theory}",
    reportNumber = "HUTP-90-A044",
    doi = "10.1016/0550-3213(91)90464-9",
    journal = "Nucl. Phys. B",
    volume = "357",
    pages = "185--207",
    year = "1991"
}

@article{Falk:1992ws,
    author = "Falk, Adam F. and Neubert, Matthias",
    title = "{Second order power corrections in the heavy quark effective theory. 2. Baryon form-factors}",
    eprint = "hep-ph/9209269",
    archivePrefix = "arXiv",
    reportNumber = "SLAC-PUB-5898",
    doi = "10.1103/PhysRevD.47.2982",
    journal = "Phys. Rev. D",
    volume = "47",
    pages = "2982--2990",
    year = "1993"
}

@article{Zwicky:2023bzk,
    author = "Zwicky, Roman",
    title = "{QCD with an Infrared Fixed Point -- Pion Sector}",
    eprint = "2306.06752",
    archivePrefix = "arXiv",
    primaryClass = "hep-ph",
    reportNumber = "CERN-TH-2023-100",
    month = "6",
    year = "2023"
}

@article{Luke:1992cs,
    author = "Luke, Michael E. and Manohar, Aneesh V.",
    title = "{Reparametrization invariance constraints on heavy particle effective field theories}",
    eprint = "hep-ph/9205228",
    archivePrefix = "arXiv",
    reportNumber = "UCSD-PTH-92-15",
    doi = "10.1016/0370-2693(92)91786-9",
    journal = "Phys. Lett. B",
    volume = "286",
    pages = "348--354",
    year = "1992"
}

@article{Neubert:1993mb,
    author = "Neubert, Matthias",
    title = "{Heavy quark symmetry}",
    eprint = "hep-ph/9306320",
    archivePrefix = "arXiv",
    reportNumber = "SLAC-PUB-6263",
    doi = "10.1016/0370-1573(94)90091-4",
    journal = "Phys. Rept.",
    volume = "245",
    pages = "259--396",
    year = "1994"
}

@inproceedings{Wise:1993wa,
    author = "Wise, Mark B.",
    title = "{Combining chiral and heavy quark symmetry}",
    booktitle = "{CCAST Symposium on Particle Physics at the Fermi Scale}",
    eprint = "hep-ph/9306277",
    archivePrefix = "arXiv",
    reportNumber = "CALT-68-1860",
    pages = "71--114",
    month = "5",
    year = "1993"
}

@article{Voloshin:1986dir,
    author = "Voloshin, M. B. and Shifman, Mikhail A.",
    title = "{On the annihilation constants of mesons consisting of a heavy and a light quark, and $B^0 \leftrightarrow \overline{B}^{-0}$ oscillations}",
    reportNumber = "ITEP-54-1986",
    journal = "Sov. J. Nucl. Phys.",
    volume = "45",
    pages = "292",
    year = "1987"
}

@article{Harada:1997we,
    author = "Harada, Masayasu and Qamar, Asif and Sannino, Francesco and Schechter, Joseph and Weigel, Herbert",
    title = "{Hyperfine splitting of low lying heavy baryons}",
    eprint = "hep-ph/9703234",
    archivePrefix = "arXiv",
    reportNumber = "SU-4240-658, UNITU-THEP-3-1997",
    doi = "10.1016/S0375-9474(97)00400-4",
    journal = "Nucl. Phys. A",
    volume = "625",
    pages = "789",
    year = "1997"
}

@article{Harada:1997bk,
    author = "Harada, Masayasu and Sannino, Francesco and Schechter, Joseph and Weigel, Herbert",
    title = "{Generalization of the bound state model}",
    eprint = "hep-ph/9704358",
    archivePrefix = "arXiv",
    reportNumber = "SU-4240-659, UNITU-THEP-6-1997",
    doi = "10.1103/PhysRevD.56.4098",
    journal = "Phys. Rev. D",
    volume = "56",
    pages = "4098--4114",
    year = "1997"
}

@article{Meissner:1987ge,
    author = "Meissner, Ulf G.",
    title = "{Low-Energy Hadron Physics from Effective Chiral Lagrangians with Vector Mesons}",
    reportNumber = "MIT-CTP-1471",
    doi = "10.1016/0370-1573(88)90090-7",
    journal = "Phys. Rept.",
    volume = "161",
    pages = "213",
    year = "1988"
}

@article{Jain:1987sz,
    author = "Jain, P. and Johnson, R. and Meissner, Ulf G. and Park, N. W. and Schechter, J.",
    title = "{Realistic Pseudoscalar Vector Chiral Lagrangian and Its Soliton Excitations}",
    reportNumber = "MIT-CTP-1532, SU-4228-371",
    doi = "10.1103/PhysRevD.37.3252",
    journal = "Phys. Rev. D",
    volume = "37",
    pages = "3252",
    year = "1988"
}

@article{Schechter:1995vr,
    author = "Schechter, J. and Subbaraman, A. and Vaidya, S. and Weigel, H.",
    title = "{Heavy quark solitons: Towards realistic masses}",
    eprint = "hep-ph/9503307",
    archivePrefix = "arXiv",
    reportNumber = "SU-4240-606, UCI-TR-95-10, UNITU-THEP-6-1995",
    doi = "10.1016/0375-9474(95)00182-Z",
    journal = "Nucl. Phys. A",
    volume = "590",
    pages = "655--679",
    year = "1995",
    note = "[Erratum: Nucl.Phys.A 598, 583--583 (1996)]"
}

@article{Gupta:1993kd,
    author = "Gupta, Kumar S. and Arshad Momen, M. and Schechter, J. and Subbaraman, A.",
    title = "{Heavy quark solitons}",
    eprint = "hep-ph/9303207",
    archivePrefix = "arXiv",
    reportNumber = "SU-4240-532, UR-1306, ER-40685-755",
    doi = "10.1103/PhysRevD.47.R4835",
    journal = "Phys. Rev. D",
    volume = "47",
    pages = "R4835--R4839",
    year = "1993"
}

@article{Oh:1994vd,
    author = "Oh, Yong-seok and Park, Byung-Yoon and Min, Dong-Pil",
    title = "{Heavy baryons as Skyrmion with 1/m(Q) corrections}",
    eprint = "hep-ph/9402205",
    archivePrefix = "arXiv",
    reportNumber = "SNUTP-93-80",
    doi = "10.1103/PhysRevD.49.4649",
    journal = "Phys. Rev. D",
    volume = "49",
    pages = "4649--4658",
    year = "1994"
}

@article{Jenkins:1992ic,
    author = "Jenkins, Elizabeth Ellen and Manohar, Aneesh V.",
    title = "{Hyperfine splittings of baryons containing a heavy quark in the Skyrme model}",
    eprint = "hep-ph/9208238",
    archivePrefix = "arXiv",
    reportNumber = "UCSD-PTH-92-26",
    doi = "10.1016/0370-2693(92)90694-Y",
    journal = "Phys. Lett. B",
    volume = "294",
    pages = "273--280",
    year = "1992"
}

@article{Guralnik:1992dj,
    author = "Guralnik, Zachary and Luke, Michael E. and Manohar, Aneesh V.",
    title = "{Properties of baryons containing a heavy quark in the Skyrme model}",
    eprint = "hep-ph/9208221",
    archivePrefix = "arXiv",
    reportNumber = "UCSD-PTH-92-24",
    doi = "10.1016/0550-3213(93)90465-2",
    journal = "Nucl. Phys. B",
    volume = "390",
    pages = "474--500",
    year = "1993"
}

@article{Orlando:2020yii,
    author = "Orlando, Domenico and Reffert, Susanne and Sannino, Francesco",
    title = "{Charging the Conformal Window}",
    eprint = "2003.08396",
    archivePrefix = "arXiv",
    primaryClass = "hep-th",
    doi = "10.1103/PhysRevD.103.105026",
    journal = "Phys. Rev. D",
    volume = "103",
    number = "10",
    pages = "105026",
    year = "2021"
}

@article{Weigel:1993zd,
    author = "Weigel, H. and Alkofer, Reinhard and Reinhardt, H.",
    title = "{Hyperons in the bound state approach to the Nambu-Jona-Lasinio chiral soliton}",
    eprint = "hep-ph/9310309",
    archivePrefix = "arXiv",
    reportNumber = "UNITU-THEP-14-1993",
    doi = "10.1016/0375-9474(94)90742-0",
    journal = "Nucl. Phys. A",
    volume = "576",
    pages = "477--524",
    year = "1994"
}

@article{Kondo:1991fc,
    author = "Kondo, Y. and Saito, S. and Otofuji, T.",
    title = "{Semileptonic hyperon decays in the Skyrme model with bound kaon approach to strangeness}",
    doi = "10.1016/0370-2693(91)91768-Q",
    journal = "Phys. Lett. B",
    volume = "256",
    pages = "316--320",
    year = "1991"
}

@article{Kaplan:1989fc,
    author = "Kaplan, David B. and Klebanov, Igor R.",
    title = "{The Role of a Massive Strange Quark in the Large $N$ Skyrme Model}",
    reportNumber = "SLAC-PUB-4964, UCSD-PTH-89-02",
    doi = "10.1016/0550-3213(90)90168-D",
    journal = "Nucl. Phys. B",
    volume = "335",
    pages = "45--66",
    year = "1990"
}

@article{Scoccola:1988wa,
    author = "Scoccola, N. N. and Nadeau, H. and Nowak, Maciej A. and Rho, Mannque",
    title = "{The Hyperons as Skyrmions With Vector Mesons}",
    reportNumber = "SACLAY-SPHT-87-116",
    doi = "10.1016/0370-2693(88)90595-3",
    journal = "Phys. Lett. B",
    volume = "201",
    pages = "425",
    year = "1988",
    note = "[Erratum: Phys.Lett.B 220, 658 (1989)]"
}

@article{Callan:1985hy,
    author = "Callan, Jr., Curtis G. and Klebanov, Igor R.",
    title = "{Bound State Approach to Strangeness in the Skyrme Model}",
    reportNumber = "Print-85-0733 (PRINCETON)",
    doi = "10.1016/0550-3213(85)90292-5",
    journal = "Nucl. Phys. B",
    volume = "262",
    pages = "365--382",
    year = "1985"
}

@article{Adkins:1983hy,
    author = "Adkins, Gregory S. and Nappi, Chiara R.",
    title = "{The Skyrme Model with Pion Masses}",
    reportNumber = "Print-83-0704 (IAS,PRINCETON)",
    doi = "10.1016/0550-3213(84)90172-X",
    journal = "Nucl. Phys. B",
    volume = "233",
    pages = "109--115",
    year = "1984"
}

@article{Weinberg:1975gm,
    author = "Weinberg, Steven",
    title = "{Implications of Dynamical Symmetry Breaking}",
    reportNumber = "PRINT-75-0804 (HARVARD)",
    doi = "10.1103/PhysRevD.19.1277",
    journal = "Phys. Rev. D",
    volume = "13",
    pages = "974--996",
    year = "1976",
    note = "[Addendum: Phys.Rev.D 19, 1277--1280 (1979)]"
}

@article{Appelquist:1999dq,
    author = "Appelquist, Thomas and Rodrigues da Silva, P. S. and Sannino, Francesco",
    title = "{Enhanced global symmetries and the chiral phase transition}",
    eprint = "hep-ph/9906555",
    archivePrefix = "arXiv",
    reportNumber = "YCTP-P14-99",
    doi = "10.1103/PhysRevD.60.116007",
    journal = "Phys. Rev. D",
    volume = "60",
    pages = "116007",
    year = "1999"
}

@article{Duan:2000dy,
    author = "Duan, Zhi-yong and Rodrigues da Silva, P. S. and Sannino, Francesco",
    title = "{Enhanced global symmetry constraints on epsilon terms}",
    eprint = "hep-ph/0001303",
    archivePrefix = "arXiv",
    reportNumber = "YCTP-P1-00",
    doi = "10.1016/S0550-3213(00)00550-2",
    journal = "Nucl. Phys. B",
    volume = "592",
    pages = "371--390",
    year = "2001"
}

@article{Kaplan:1983fs,
    author = "Kaplan, David B. and Georgi, Howard",
    title = "{SU(2) x U(1) Breaking by Vacuum Misalignment}",
    reportNumber = "HUTP-83/A069",
    doi = "10.1016/0370-2693(84)91177-8",
    journal = "Phys. Lett. B",
    volume = "136",
    pages = "183--186",
    year = "1984"
}

@article{Kaplan:1983sm,
    author = "Kaplan, David B. and Georgi, Howard and Dimopoulos, Savas",
    title = "{Composite Higgs Scalars}",
    reportNumber = "HUTP-83/A079",
    doi = "10.1016/0370-2693(84)91178-X",
    journal = "Phys. Lett. B",
    volume = "136",
    pages = "187--190",
    year = "1984"
}

@article{Miransky:1996pd,
    author = "Miransky, V. A. and Yamawaki, Koichi",
    title = "{Conformal phase transition in gauge theories}",
    eprint = "hep-th/9611142",
    archivePrefix = "arXiv",
    reportNumber = "DPNU-96-58",
    doi = "10.1103/PhysRevD.56.3768",
    journal = "Phys. Rev. D",
    volume = "55",
    pages = "5051--5066",
    year = "1997",
    note = "[Erratum: Phys.Rev.D 56, 3768 (1997)]"
}

@article{Cacciapaglia:2014uja,
    author = "Cacciapaglia, Giacomo and Sannino, Francesco",
    title = "{Fundamental Composite (Goldstone) Higgs Dynamics}",
    eprint = "1402.0233",
    archivePrefix = "arXiv",
    primaryClass = "hep-ph",
    reportNumber = "CP3-ORIGINS-2014-003, DIAS-2014-3, LYCEN-2014-02",
    doi = "10.1007/JHEP04(2014)111",
    journal = "JHEP",
    volume = "04",
    pages = "111",
    year = "2014"
}

@article{Gripaios:2009pe,
    author = "Gripaios, Ben and Pomarol, Alex and Riva, Francesco and Serra, Javi",
    title = "{Beyond the Minimal Composite Higgs Model}",
    eprint = "0902.1483",
    archivePrefix = "arXiv",
    primaryClass = "hep-ph",
    doi = "10.1088/1126-6708/2009/04/070",
    journal = "JHEP",
    volume = "04",
    pages = "070",
    year = "2009"
}

@article{Galloway:2010bp,
    author = "Galloway, Jamison and Evans, Jared A. and Luty, Markus A. and Tacchi, Ruggero Altair",
    title = "{Minimal Conformal Technicolor and Precision Electroweak Tests}",
    eprint = "1001.1361",
    archivePrefix = "arXiv",
    primaryClass = "hep-ph",
    doi = "10.1007/JHEP10(2010)086",
    journal = "JHEP",
    volume = "10",
    pages = "086",
    year = "2010"
}

@article{Barnard:2013zea,
    author = "Barnard, James and Gherghetta, Tony and Ray, Tirtha Sankar",
    title = "{UV descriptions of composite Higgs models without elementary scalars}",
    eprint = "1311.6562",
    archivePrefix = "arXiv",
    primaryClass = "hep-ph",
    doi = "10.1007/JHEP02(2014)002",
    journal = "JHEP",
    volume = "02",
    pages = "002",
    year = "2014"
}

@article{Chacko:2012sy,
    author = "Chacko, Zackaria and Mishra, Rashmish K.",
    title = "{Effective Theory of a Light Dilaton}",
    eprint = "1209.3022",
    archivePrefix = "arXiv",
    primaryClass = "hep-ph",
    doi = "10.1103/PhysRevD.87.115006",
    journal = "Phys. Rev. D",
    volume = "87",
    number = "11",
    pages = "115006",
    year = "2013"
}

@article{Li:2016uzn,
    author = "Li, Yan-Ling and Ma, Yong-Liang and Rho, Mannque",
    title = "{Chiral-scale effective theory including a dilatonic meson}",
    eprint = "1609.07014",
    archivePrefix = "arXiv",
    primaryClass = "hep-ph",
    doi = "10.1103/PhysRevD.95.114011",
    journal = "Phys. Rev. D",
    volume = "95",
    number = "11",
    pages = "114011",
    year = "2017"
}

@article{Kasai:2016ifi,
    author = "Kasai, Aya and Okumura, Ken-ichi and Suzuki, Hiroshi",
    title = "{A dilaton-pion mass relation}",
    eprint = "1609.02264",
    archivePrefix = "arXiv",
    primaryClass = "hep-lat",
    reportNumber = "KYUSHU-HET-168",
    month = "9",
    year = "2016"
}

@article{Appelquist:2017wcg,
    author = "Appelquist, Thomas and Ingoldby, James and Piai, Maurizio",
    title = "{Dilaton EFT Framework For Lattice Data}",
    eprint = "1702.04410",
    archivePrefix = "arXiv",
    primaryClass = "hep-ph",
    doi = "10.1007/JHEP07(2017)035",
    journal = "JHEP",
    volume = "07",
    pages = "035",
    year = "2017"
}

@article{Appelquist:2020bqj,
    author = "Appelquist, Thomas and Ingoldby, James and Piai, Maurizio",
    title = "{Nearly Conformal Composite Higgs Model}",
    eprint = "2012.09698",
    archivePrefix = "arXiv",
    primaryClass = "hep-ph",
    doi = "10.1103/PhysRevLett.126.191804",
    journal = "Phys. Rev. Lett.",
    volume = "126",
    number = "19",
    pages = "191804",
    year = "2021"
}

@article{Hansen:2016fri,
    author = "Hansen, Martin and Lang\ae{}ble, Kasper and Sannino, Francesco",
    title = "{Extending Chiral Perturbation Theory with an Isosinglet Scalar}",
    eprint = "1610.02904",
    archivePrefix = "arXiv",
    primaryClass = "hep-ph",
    reportNumber = "CP3-ORIGINS-2016-041",
    doi = "10.1103/PhysRevD.95.036005",
    journal = "Phys. Rev. D",
    volume = "95",
    number = "3",
    pages = "036005",
    year = "2017"
}

@article{Matsuzaki:2013eva,
    author = "Matsuzaki, Shinya and Yamawaki, Koichi",
    title = "{Dilaton Chiral Perturbation Theory: Determining the Mass and Decay Constant of the Technidilaton on the Lattice}",
    eprint = "1311.3784",
    archivePrefix = "arXiv",
    primaryClass = "hep-lat",
    doi = "10.1103/PhysRevLett.113.082002",
    journal = "Phys. Rev. Lett.",
    volume = "113",
    number = "8",
    pages = "082002",
    year = "2014"
}

@article{Cacciapaglia:2020kgq,
    author = "Cacciapaglia, Giacomo and Pica, Claudio and Sannino, Francesco",
    title = "{Fundamental Composite Dynamics: A Review}",
    eprint = "2002.04914",
    archivePrefix = "arXiv",
    primaryClass = "hep-ph",
    doi = "10.1016/j.physrep.2020.07.002",
    journal = "Phys. Rept.",
    volume = "877",
    pages = "1--70",
    year = "2020"
}

@article{Dietrich:2005jn,
    author = "Dietrich, Dennis D. and Sannino, Francesco and Tuominen, Kimmo",
    title = "{Light composite Higgs from higher representations versus electroweak precision measurements: Predictions for CERN LHC}",
    eprint = "hep-ph/0505059",
    archivePrefix = "arXiv",
    doi = "10.1103/PhysRevD.72.055001",
    journal = "Phys. Rev. D",
    volume = "72",
    pages = "055001",
    year = "2005"
}

@article{Golterman:2016lsd,
    author = "Golterman, Maarten and Shamir, Yigal",
    title = "{Low-energy effective action for pions and a dilatonic meson}",
    eprint = "1603.04575",
    archivePrefix = "arXiv",
    primaryClass = "hep-ph",
    doi = "10.1103/PhysRevD.94.054502",
    journal = "Phys. Rev. D",
    volume = "94",
    number = "5",
    pages = "054502",
    year = "2016"
}

@article{Golterman:2016cdd,
    author = "Golterman, Maarten and Shamir, Yigal",
    title = "{Effective pion mass term and the trace anomaly}",
    eprint = "1611.04275",
    archivePrefix = "arXiv",
    primaryClass = "hep-ph",
    doi = "10.1103/PhysRevD.95.016003",
    journal = "Phys. Rev. D",
    volume = "95",
    number = "1",
    pages = "016003",
    year = "2017"
}

@article{Golterman:2018mfm,
    author = "Golterman, Maarten and Shamir, Yigal",
    title = "{Large-mass regime of the dilaton-pion low-energy effective theory}",
    eprint = "1805.00198",
    archivePrefix = "arXiv",
    primaryClass = "hep-ph",
    doi = "10.1103/PhysRevD.98.056025",
    journal = "Phys. Rev. D",
    volume = "98",
    number = "5",
    pages = "056025",
    year = "2018"
}

@article{Golterman:2020tdq,
    author = "Golterman, Maarten and Neil, Ethan T. and Shamir, Yigal",
    title = "{Application of dilaton chiral perturbation theory to $N_f=8$, ${\rm SU}(3)$ spectral data}",
    eprint = "2003.00114",
    archivePrefix = "arXiv",
    primaryClass = "hep-ph",
    doi = "10.1103/PhysRevD.102.034515",
    journal = "Phys. Rev. D",
    volume = "102",
    number = "3",
    pages = "034515",
    year = "2020"
}

@article{Golterman:2021ohm,
    author = "Golterman, Maarten and Shamir, Yigal",
    title = "{Dilaton chiral perturbation theory and applications}",
    eprint = "2110.07930",
    archivePrefix = "arXiv",
    primaryClass = "hep-lat",
    doi = "10.22323/1.396.0372",
    journal = "PoS",
    volume = "LATTICE2021",
    pages = "372",
    year = "2022"
}

@article{Droz:1991cx,
    author = "Droz, Serge and Heusler, Markus and Straumann, Norbert",
    title = "{New black hole solutions with hair}",
    reportNumber = "ZU-TH-17-91",
    doi = "10.1016/0370-2693(91)91592-J",
    journal = "Phys. Lett. B",
    volume = "268",
    pages = "371--376",
    year = "1991"
}

@book{coleman1988aspects,
  title={Aspects of symmetry: selected Erice lectures},
  author={Coleman, Sidney},
  year={1988},
  publisher={Cambridge University Press}
}

@article{goldberger2007light,
    author = "Goldberger, Walter D. and Grinstein, Benjamin and Skiba, Witold",
    title = "{Distinguishing the Higgs boson from the dilaton at the Large Hadron Collider}",
    eprint = "0708.1463",
    archivePrefix = "arXiv",
    primaryClass = "hep-ph",
    reportNumber = "UCSD-PTH-07-09",
    doi = "10.1103/PhysRevLett.100.111802",
    journal = "Phys. Rev. Lett.",
    volume = "100",
    pages = "111802",
    year = "2008"
}

@article{rattazzi2001comments,
  title={Comments on the holographic picture of the Randall-Sundrum model},
  author={Rattazzi, Riccardo and Zaffaroni, Alberto},
  journal={Journal of High Energy Physics},
  volume={2001},
  number={04},
  pages={021},
  year={2001},
  publisher={IOP Publishing}
}

@article{Adkins:1983nw,
    author = "Adkins, Gregory S. and Nappi, Chiara R.",
    title = "{Stabilization of Chiral Solitons via Vector Mesons}",
    reportNumber = "Print-84-0053 (PRINCETON)",
    doi = "10.1016/0370-2693(84)90239-9",
    journal = "Phys. Lett. B",
    volume = "137",
    pages = "251--256",
    year = "1984"
}

@article{Zahed:1986qz,
    author = "Zahed, I. and Brown, G. E.",
    title = "{The Skyrme Model}",
    reportNumber = "PRINT-86-0160 (STONY-BROOK)",
    doi = "10.1016/0370-1573(86)90142-0",
    journal = "Phys. Rept.",
    volume = "142",
    pages = "1--102",
    year = "1986"
}

@article{chacko2013effective,
  title={Effective theory of a light dilaton},
  author={Chacko, Zackaria and Mishra, Rashmish K},
  journal={Physical Review D},
  volume={87},
  number={11},
  pages={115006},
  year={2013},
  publisher={APS}
}

@article{Sannino:2004qp,
    author = "Sannino, Francesco and Tuominen, Kimmo",
    title = "{Orientifold theory dynamics and symmetry breaking}",
    eprint = "hep-ph/0405209",
    archivePrefix = "arXiv",
    doi = "10.1103/PhysRevD.71.051901",
    journal = "Phys. Rev. D",
    volume = "71",
    pages = "051901",
    year = "2005"
}

@article{Appelquist:2022mjb,
    author = "Appelquist, Thomas and Ingoldby, James and Piai, Maurizio",
    title = "{Dilaton Effective Field Theory}",
    eprint = "2209.14867",
    archivePrefix = "arXiv",
    primaryClass = "hep-ph",
    doi = "10.3390/universe9010010",
    journal = "Universe",
    volume = "9",
    number = "1",
    pages = "10",
    year = "2023"
}

@article{Adkins:1983ya,
    author = "Adkins, Gregory S. and Nappi, Chiara R. and Witten, Edward",
    title = "{Static Properties of Nucleons in the Skyrme Model}",
    reportNumber = "PRINT-83-0493 (IAS,PRINCETON)",
    doi = "10.1016/0550-3213(83)90559-X",
    journal = "Nucl. Phys. B",
    volume = "228",
    pages = "552",
    year = "1983"
}

@article{Skyrme:1961vq,
    author = "Skyrme, T. H. R.",
    title = "{A Nonlinear field theory}",
    doi = "10.1098/rspa.1961.0018",
    journal = "Proc. Roy. Soc. Lond. A",
    volume = "260",
    pages = "127--138",
    year = "1961"
}

@article{Bertini:2005rc,
    author = "Bertini, Stefan and Cacciatori, S. L. and Cerchiai, Bianca L.",
    title = "{On the Euler angles for SU(N)}",
    eprint = "math-ph/0510075",
    archivePrefix = "arXiv",
    reportNumber = "IFUM-846-FT, LBNL-57265, UCB-PTH-05-05",
    doi = "10.1063/1.2190898",
    journal = "J. Math. Phys.",
    volume = "47",
    pages = "043510",
    year = "2006"
}

@article{Cacciatori:2012qi,
    author = "Cacciatori, S. L. and Dalla Piazza, F. and Scotti, A.",
    title = "{Compact Lie groups: Euler constructions and generalized Dyson conjecture}",
    eprint = "1207.1262",
    archivePrefix = "arXiv",
    primaryClass = "math.GR",
    doi = "10.1090/tran/6795",
    journal = "Trans. Am. Math. Soc.",
    volume = "369",
    number = "7",
    pages = "4709--4724",
    year = "2017"
}

@article{Canfora:2022jmh,
    author = "Canfora, Fabrizio and Hidalgo, Diego and Lagos, Marcela and Meneses, Enzo and Vera, Aldo",
    title = "{Infinite conformal symmetry and emergent chiral fields of topologically nontrivial configurations: From Yang-Mills-Higgs theory to the Skyrme model}",
    eprint = "2208.07925",
    archivePrefix = "arXiv",
    primaryClass = "hep-th",
    doi = "10.1103/PhysRevD.106.105016",
    journal = "Phys. Rev. D",
    volume = "106",
    number = "10",
    pages = "105016",
    year = "2022"
}

@article{Canfora:2023zmt,
    author = "Canfora, Fabrizio",
    title = "{Magnetized Baryonic layer and a novel BPS bound in the gauged-Non-Linear-Sigma-Model-Maxwell theory in (3+1)-dimensions through Hamilton-Jacobi equation}",
    eprint = "2309.03153",
    archivePrefix = "arXiv",
    primaryClass = "hep-th",
    month = "9",
    year = "2023"
}

@article{Alvarez:2017cjm,
    author = "Alvarez, P. D. and Canfora, F. and Dimakis, N. and Paliathanasis, A.",
    title = "{Integrability and chemical potential in the (3 + 1)-dimensional Skyrme model}",
    eprint = "1707.07421",
    archivePrefix = "arXiv",
    primaryClass = "hep-th",
    doi = "10.1016/j.physletb.2017.08.073",
    journal = "Phys. Lett. B",
    volume = "773",
    pages = "401--407",
    year = "2017"
}

@article{Ayon-Beato:2015eca,
    author = "Ayon-Beato, Eloy and Canfora, Fabrizio and Zanelli, Jorge",
    title = "{Analytic self-gravitating Skyrmions, cosmological bounces and AdS wormholes}",
    eprint = "1509.02659",
    archivePrefix = "arXiv",
    primaryClass = "gr-qc",
    doi = "10.1016/j.physletb.2015.11.065",
    journal = "Phys. Lett. B",
    volume = "752",
    pages = "201--205",
    year = "2016"
}

@article{Canfora:2018rdz,
    author = "Canfora, Fabrizio",
    title = "{Ordered arrays of Baryonic tubes in the Skyrme model in ( $3+1$ ) dimensions at finite density}",
    eprint = "1807.02090",
    archivePrefix = "arXiv",
    primaryClass = "hep-th",
    doi = "10.1140/epjc/s10052-018-6404-x",
    journal = "Eur. Phys. J. C",
    volume = "78",
    number = "11",
    pages = "929",
    year = "2018"
}

@article{Barriga:2021eki,
    author = "Barriga, Gonzalo and Canfora, Fabrizio and Torres, Matías and Vera, Aldo",
    title = "{Crystals of gauged solitons, force free plasma and resurgence}",
    eprint = "2105.01172",
    archivePrefix = "arXiv",
    primaryClass = "hep-th",
    doi = "10.1103/PhysRevD.103.096023",
    journal = "Phys. Rev. D",
    volume = "103",
    number = "9",
    pages = "096023",
    year = "2021"
}

@article{Barriga:2022izc,
    author = "Barriga, Gonzalo and Canfora, Fabrizio and Lagos, Marcela and Torres, Matías and Vera, Aldo",
    title = "{On the robustness of solitons crystals in the Skyrme model}",
    eprint = "2207.08712",
    archivePrefix = "arXiv",
    primaryClass = "hep-th",
    doi = "10.1016/j.nuclphysb.2022.115913",
    journal = "Nucl. Phys. B",
    volume = "983",
    pages = "115913",
    year = "2022"
}

@article{Canfora:2023pkx,
    author = "Canfora, Fabrizio and Rebolledo-Caceres, Scarlett C.",
    title = "{Skyrmions at finite density}",
    eprint = "2306.10226",
    archivePrefix = "arXiv",
    primaryClass = "hep-th",
    doi = "10.1142/S0217732323300021",
    journal = "Mod. Phys. Lett. A",
    volume = "38",
    number = "12n13",
    pages = "2330002",
    year = "2023"
}

@article{Dietrich:2006cm,
    author = "Dietrich, Dennis D. and Sannino, Francesco",
    title = "{Conformal window of SU(N) gauge theories with fermions in higher dimensional representations}",
    eprint = "hep-ph/0611341",
    archivePrefix = "arXiv",
    doi = "10.1103/PhysRevD.75.085018",
    journal = "Phys. Rev. D",
    volume = "75",
    pages = "085018",
    year = "2007"
}

@article{PhysRevLett.125.190401,
  title = {Emergence and Disruption of Spin-Charge Separation in One-Dimensional Repulsive Fermions},
  author = {He, Feng and Jiang, Yu-Zhu and Lin, Hai-Qing and Hulet, Randall G. and Pu, Han and Guan, Xi-Wen},
  journal = {Phys. Rev. Lett.},
  volume = {125},
  issue = {19},
  pages = {190401},
  numpages = {7},
  year = {2020},
  month = {Nov},
  publisher = {American Physical Society},
  doi = {10.1103/PhysRevLett.125.190401},
  url = {https://link.aps.org/doi/10.1103/PhysRevLett.125.190401}
}

@article{Lerda:1994kb,
    author = "Lerda, Alberto",
    title = "{A Field theory approach to the t-J model and the spin charge separation}",
    eprint = "hep-th/9403176",
    archivePrefix = "arXiv",
    reportNumber = "DFT-US-1-94",
    doi = "10.1016/0550-3213(94)90367-0",
    journal = "Nucl. Phys. B",
    volume = "428",
    pages = "629--654",
    year = "1994"
}

@article{appelquist2020dilaton,
  title={Dilaton potential and lattice data},
  author={Appelquist, Thomas and Ingoldby, James and Piai, Maurizio},
  journal={Physical Review D},
  volume={101},
  number={7},
  pages={075025},
  year={2020},
  publisher={APS}
}

@article{Park:2008zg,
    author = "Park, Byung-Yoon and Rho, Mannque and Vento, Vicente",
    title = "{The Role of the Dilaton in Dense Skyrmion Matter}",
    eprint = "0801.1374",
    archivePrefix = "arXiv",
    primaryClass = "hep-ph",
    doi = "10.1016/j.nuclphysa.2008.03.015",
    journal = "Nucl. Phys. A",
    volume = "807",
    pages = "28--37",
    year = "2008"
}

@article{Cata:2019edh,
    author = {Catá, Oscar and Müller, Christoph},
    title = "{Chiral effective theories with a light scalar at one loop}",
    eprint = "1906.01879",
    archivePrefix = "arXiv",
    primaryClass = "hep-ph",
    doi = "10.1016/j.nuclphysb.2020.114938",
    journal = "Nucl. Phys. B",
    volume = "952",
    pages = "114938",
    year = "2020"
}
\end{document}